\documentclass[aps,prd,twocolumn,floatfix,noshowpacs,tightenlines,noshowkeys,superscriptaddress,amsmath,amssymb,nofootinbib]{revtex4-1}
\usepackage{amssymb,amsbsy,epsfig,color,graphicx}
\usepackage{color}
\usepackage{[longtable}
\usepackage{array}
\usepackage{dcolumn}   
\usepackage{cellspace}
\usepackage{mathtools}
\usepackage{amstext}
\usepackage{amssymb}
\usepackage{stmaryrd}
\usepackage{stackrel}
\usepackage{graphicx}
\usepackage{esint}
\usepackage[utf8]{inputenc}
\usepackage{adjustbox}
\usepackage{multirow}
\usepackage{float}
\usepackage{booktabs}
\usepackage{enumitem}
\usepackage{slashed}
\usepackage{hyperref}
\usepackage{etoolbox} 
\usepackage{lipsum} 
\usepackage[capitalize]{cleveref}
\usepackage{lipsum,booktabs}
\usepackage{changepage}
\usepackage{multirow}
\usepackage[caption=false]{subfig}
\allowdisplaybreaks
\usepackage{soul}
\def\bea{\begin{eqnarray}}
	\def\eea{\end{eqnarray}}

\newcommand{\ba}{\begin{eqnarray}}
	\newcommand{\ea}{\end{eqnarray}}

\makeatletter

\appto{\appendix}{%
	\@ifstar{\def\theequation@prefix{A.}}%
	{}%
}
\makeatother

\hypersetup{pdfstartview=FitV,colorlinks=true,linkcolor=midblue,citecolor=midblue,filecolor=midblue,urlcolor=midblue}

\definecolor{midblue}{rgb}{0,0,0.5}
\definecolor{cadmiumorange}{rgb}{0.93, 0.53, 0.18}

\def\BL{Boyer-Lindquist}

\def\hJ{\hat{J_{z}}}
\def\hE{\hat{E}}
\def\hr{\hat{r}}
\def\ha{\hat{a}}
\def\hx{\hat{x}}
\def\hO{\hat{\Omega}}
\def\ho{\hat{\omega}}
\def\hht{\hat{t}}

\def\mode{\ell m \hat{\omega}}
\def\mOde{\ell m \hat{\Omega}}
\def\Ang{S_{\mode}^{\ha\ho}(\theta)}
\def\In{\textrm{in}}
\def\up{\textrm{up}}

\def\ta{(a)}
\def\tb{(b)}
\def\tc{(c)}
\def\td{(d)}

\usepackage{hyperref}
\hypersetup{colorlinks=true}

\def\equationautorefname~#1\null{%
	Eq.~(#1)\null
}
\def\figureautorefname~#1\null{%
	Fig.~#1\null
}
\def\tableautorefname~#1\null{%
	Table.~#1\null
}
\def\sectionautorefname~#1\null{%
	Section #1\null
}
\def\appendixautorefname~#1\null{%
	Appendix #1\null
}

\begin{document}

	\title{Prospects for determining the nature of the
secondaries of extreme mass-ratio inspirals using the spin-induced
quadrupole deformation}

	\author{Mostafizur Rahman}
	\email{mostafizur.r@iitgn.ac.in}
	\affiliation{Indian Institute of Technology, Gandhinagar, Gujarat-382355, India}
	
	\author{Arpan Bhattacharyya}
	\email{abhattacharyya@iitgn.ac.in}
	\affiliation{Indian Institute of Technology, Gandhinagar, Gujarat-382355, India}

	\begin{abstract}
The measurement of multipole moments of astrophysical objects through gravitational wave (GW) observations provides a novel way to distinguish black holes from other astrophysical objects. 
This paper studies the gravitational wave radiation from an extreme mass ratio inspiral (EMRI) system consisting of a supermassive Kerr black hole (the primary object) and a spinning stellar-mass compact object (the secondary object). The quadrupolar deformation induced by the spin of the secondary is different for different astrophysical objects. We compute the effect of the quadrupolar deformation on the GW phase and provide an order of magnitude estimate of whether LISA can distinguish different astrophysical objects through GW phase measurement. We find that although LISA can not distinguish between a black hole and a neutron star, it can distinguish black holes from a large variety of highly spinning astrophysical objects like superspinars and highly deformable exotic compact objects like boson stars for EMRI systems with relatively large mass ratio ($q\sim 10^{-4}$). Furthermore, we show that the effect of spin-induced quadrupolar deformation on the GW phase for white dwarf and brown dwarf-EMRI systems can be quite significant even for small values of mass ratio ($q\lesssim 10^{-6}$). 
	\end{abstract}
	
	\maketitle

	
	\section{Introduction}\label{Intro}
	
The detection of gravitational waves  \cite{LIGOScientific:2016aoc, LIGOScientific:2021djp} paved the way for a new era in observational astrophysics, which allows us to probe physics in the strong gravity regime for the very first time \cite{Carson:2020rea, Berti:2015itd, Perkins:2020tra, Barack:2018yly, LIGOScientific:2021sio}. The ground-based gravitational-wave detectors successfully observed the merger of stellar-mass black holes and neutron stars. Unfortunately, these detectors are only sensitive to frequencies above $\sim 10$ Hz due to the presence of seismic noise (future third-generation detectors like the Einstein Telescope hope to evade the seismic noise by going underground and can probe signals in the frequency band ranging from $\sim 3$ Hz to several kHz \cite{ET2020}). The future space-based gravitational wave detectors like the Laser Interferometer Space Antenna (LISA), on the other hand, will be unconstrained from such restrictions and can detect gravitational waves in the millihertz frequency band \cite{eLISA:2013xep}. It can detect gravitational wave signals from a wide variety of astrophysical and cosmological sources \cite{eLISA:2013xep, Amaro-Seoane:2007osp, Gair:2017ynp, Babak:2017tow, PhysRevD.93.024003, Tamanini_2016, Gair:2010yu, Caprini:2015zlo,Gair:2012nm}.\par
	One primary source for LISA observations is extreme mass ratio inspiral (EMRI), a binary system with a very small mass ratio ($q\equiv m_s/M\sim 10^{-7}-10^{-4}$), where a stellar-mass object inspirals into a supermassive compact object \cite{Amaro-Seoane:2007osp, Gair:2017ynp, Babak:2017tow,Gair:2010yu, PhysRevD.78.064028}. The stellar-mass object (hereafter, the secondary) completes $\sim 10^{4}-10^{6}$ orbits around the supermassive central object (hereafter, the primary) within the LISA frequency band before plunging \cite{PhysRevD.78.064028, PhysRev.136.B1224}. The gravitational waveforms from the system can be used to extract accurate information about the parameters of the binary system \cite{Gair:2017ynp, Berry:2019wgg} and the geometry surrounding the primary object \cite{PhysRevD.52.5707,PhysRevD.56.1845, PhysRevD.69.124022, Glampedakis:2005cf, PhysRevD.75.042003, PhysRevD.77.024035, PhysRevLett.126.141102, PhysRevLett.103.111101, PhysRevD.81.024030, PhysRevD.102.064041, PhysRevD.104.064023, PhysRevLett.123.101103,Gupta:2021cno, Gupta:2022jdt}. Recent studies have shown that LISA can measure the redshifted mass and spin of the primary with much better accuracy than current
ground-based detectors and X-ray measurements \cite{Gair:2017ynp, Berry:2019wgg}.\par
	Furthermore, the EMRI system is an ideal testbed to analyze the nature of the supermassive object \cite{PhysRevD.52.5707, PhysRevD.69.124022, Glampedakis:2005cf, PhysRevD.75.042003, PhysRevD.77.024035, PhysRevD.81.024030}. The uniqueness and no-hair theorems in the context of general relativity assert that the astrophysical objects beyond a certain mass limit are Kerr black holes. Their geometry and multipole moments depend only on their mass and angular momentum \cite{PhysRevLett.11.237, PhysRevLett.34.905, PhysRevD.52.5707, PhysRevD.75.042003}. 
	However, recently, black hole alternative models like gravastars \cite{Visser:2003ge, Mottola:2011ud}, boson stars \cite{Liebling:2012fv, PhysRevLett.115.111301,Siemonsen:2020hcg}, and fuzzballs \cite{PhysRevLett.88.211303, LUNIN2002342} have gained much attention. These objects are collectively known as exotic compact objects (ECOs) \cite{Cardoso:2019rvt}. They are slightly larger than the black holes with the same mass and angular momentum and have finite reflectivity. 
	Gravitational waves produced in the compact binary coalescence process provide a way to identify these objects as their ringdown signals differ from that of a black hole \cite{Kokkotas:1995av, PhysRevD.102.064053, PhysRevLett.116.171101, PhysRevD.96.084002, PhysRevD.101.024031, Cardoso:2019rvt, PhysRevD.104.044045, PhysRevD.100.062006}.
	Moreover, the multipolar structure of some of these objects is drastically different from that of the Kerr black holes  \cite{ PhysRevD.55.6081, Herdeiro:2014goa,  PhysRevLett.125.221602, Bianchi:2020bxa, Bah:2021jno, Uchikata:2015yma, PhysRevD.94.064015}, which left its imprint on the gravitational waveform. Thus, the measurement of higher order multipole moments presents an opportunity to distinguish ECOs from black holes through gravitational wave observations \cite{ Krishnendu:2017shb, PhysRevD.95.084014, PhysRevD.96.024002, Narikawa:2021pak, Saleem:2021vph}. 
	Since LISA can measure the quadrupolar moment of the primary with great precision (independent of its mass and angular momentum),  the emitted gravitational radiations from the system can testify for the ``Kerr-ness'' of the primary \cite{PhysRevD.52.5707, PhysRevD.75.042003,Bianchi:2020bxa}. Other notable EMRI based tests to identify the nature of primary include measuring the change in tidal heating \cite{PhysRevD.8.1010, PhysRevD.101.044004,PhysRevD.61.084004} and energy flux \cite{PhysRevD.104.104026, PhysRevD.104.064009} due to the presence of finite reflectivity and the measurement of tidal Love numbers \cite{Pani:2019cyc}.  \par
	Relatively less attention has been given to finding the nature of the secondary object. This is because the effect of the secondary's spin and higher-order multipole moments is expected to get suppressed by the system's tiny mass ratio. Several authors have recently considered the effect of secondary's spin on orbital dynamics and gravitational wave production \cite{PhysRevD.83.044044, PhysRevD.89.064011, PhysRevD.53.622, PhysRevD.54.3762, PhysRevD.103.104045, PhysRevD.69.044011, PhysRevD.58.064005, Pani, Piovano:2020ooe, PhysRevD.96.064051, PhysRevD.102.064013, Skoupy:2022adh,PhysRevD.102.024091, Timogiannis:2022bks}. In particular, Piovano et al. studied the adiabatic evolution of spinning secondary in circular and equatorial orbit \cite{Pani, Piovano:2020ooe}. Their study shows that the gravitational wave dephasing due to the secondary's spin could be large enough for detection. Moreover, LISA can detect a class of exotic compact object models called the superspinars that can breach the Kerr bound. 
	Interestingly, some recent studies also considered the effect of quadrupolar deformation of the secondary in an intermediate-mass ratio inspiral (IMRI) system \cite{Chen:2019hac} and Schwarzschild background \cite{PhysRevD.102.024091}. In this paper, we consider the secondary as a spinning object that inspirals into a supermassive Kerr black hole in a circular, equatorial orbit. Moreover, the rotation induces quadrupolar deformation in the secondary. Several authors have emphasized the importance of considering second-order effects like quadrupolar deformation for the correct modelling of EMRI waveform \cite{PhysRevD.73.044034,PhysRevD.103.064048}. The argument follows from the fact that over the long inspiral period ($T_i\sim M/q$) of an EMRI system, the second-order force terms $q^{2}f^{\alpha}_{(2)}$ have a considerable effect on orbital dynamics $\delta z^{\alpha}\sim q^{2}f^{\alpha}_{(2)} T_i^2\sim q^0$. Thus, one can not neglect the contribution of these terms.   
Since the quadrupolar moment carries information about the object's internal structure, it can help us identify the nature of the object. In this paper, we calculate the corrections in the gravitational wave phase due to the effect and show that those corrections can be large enough for LISA to detect and thus can distinguish between black holes and other astrophysical objects.  \par
	The paper is organized as follows: In \autoref{Sec_1_MPD}, we briefly describe the equation of motion of a deformed spinning object in curved spacetime. In \autoref{Sec_3_Orbital_motion}, we describe the orbital motion of the object in Kerr spacetime. \autoref{Sec_4_GW_flux} gives a brief review of the Teukolsky formalism and gravitational wave emission from the EMRI system. In \autoref{Sec_5_Result}, we present our main results. \autoref{Sec_Conclusion} contains our conclusion. The equations for circular orbits and orbital frequency of the secondary object are presented in  Appendix~\ref{App:Circular}. In Appendix~\ref{App:source_term}, we provide a detailed calculation for the Teukolsky source term for a spinning, deformed object. Finally, in Appendix~\ref{App:Comparison}, we compare our results with the ones existing in the literature.
\par
	
	\textit{Notation and Convention}: Throughout the paper, we adopt positive signature convention $(-,+,+,+)$ and geometrical unit $c=G=1$. Greek letters $\alpha,~\beta, \gamma,...$ are used to denote four-dimensional spacetime indices, whereas the bracketed lowercase roman letters $(a),(b),(c),...$ are used to denote tetrad indices. Round and square bracket around a pair of indices denote
 symmetrization and antisymmetrization respectively: $T^{(\mu\nu)}=(T^{\mu\nu}+T^{\nu\mu})/2$, $T^{[\mu\nu]}=(T^{\mu\nu}-T^{\nu\mu})/2$.  
	\section{Dynamics of extended objects}\label{Sec_1_MPD}
	\subsection{Equation of motion}
	
	The dynamics of the stellar mass object, immersed in the gravitational field of the supermassive black hole, can be adequately described by the multi-polar approximation method \cite{Dixon1974DynamicsOE, Dixon1973TheDO, PhysRevD.81.044019, Dixon1973TheDO}. It asserts that a set of multipole moments encode the effect of the internal structure of the secondary on its motion along a reference worldline $z^{\mu}$. Since the secondary object's size is much smaller than the curvature radius of the primary object, only a finite number of terms are required to describe the motion. Here, we consider terms up to quadrupolar order, which describes the secondary object as an extended spinning object subjected to quadrupolar deformation. Under this approximation, the energy-momentum tensor of the object can be written as follows \cite{PhysRevD.81.044019, PhysRevD.86.044033}, 
	\begin{equation}\label{SET}
		\begin{aligned}
			T^{\alpha \beta }&=\int \text{d$\tau $}\bigg[\frac{\delta ^4(x-z(\tau))}{\sqrt{-g}}p^{(\alpha }v^{\beta )}\bigg]\\
			&-\int \text{d$\tau $}\nabla _{\gamma }\left[S^{\gamma (\alpha }v^{\beta )}\frac{\delta ^4(x-z(\tau))}{\sqrt{-g}}\right]\\
			&-\frac{1}{3}\int \text{d$\tau $}\bigg[\left(J^{\gamma \delta \epsilon (\alpha }R^{\beta )}{}_{\epsilon \gamma \delta }\frac{\delta ^4(x-z(\tau))}{\sqrt{-g}}\right)\\&+2\nabla _{\gamma }\nabla _{\delta }\left(J^{\delta (\alpha \beta )\gamma }\frac{\delta ^4(x-z(\tau))}{\sqrt{-g}}\right)\bigg]+\mathcal{O}(\epsilon^3)
		\end{aligned}
	\end{equation}
	where, $v^{\mu}=dz^{\mu}/d\tau$ is the tangent to the object's worldline, $p^{\mu}$ is the momentum of the object, $S^{\mu\nu}$ is spin tensor, and $J^{\alpha \beta \gamma \delta }$ is the quadrupole tensor. Here, we choose the proper time $\tau$ as the affine parameter so that the following normalization condition is satisfied $v^{\mu}v_{\mu}=-1$. Note that the quadrupole tensor exhibits all the algebraic symmetries of the Riemann tensor $R_{\alpha \beta \gamma \delta }$. Following Ref.~\cite{PhysRevD.88.084005}, we introduce a small parameter $\epsilon$ to keep track of the terms with different multipole moment orders. The first bracketed term on the right-hand side of \autoref{SET} is the monopole term ($\mathcal{O}(\epsilon^0)$), which describes the energy-momentum tensor of a point particle. The effect of spin and quadrupolar deformation is specified through the inclusion of the second and third bracketed terms, which are $\mathcal{O}(\epsilon^1)$ and $\mathcal{O}(\epsilon^2)$ respectively.\par
	The equation of motion of the object is given by the Mathisson-Papapetrou-Dixon (MPD) equation, which can be written as follows \cite{Dixon1974DynamicsOE, PhysRevD.81.044019, PhysRevD.86.044033, PhysRevD.88.084005, Dixon1970DynamicsOE}
	\begin{equation}\label{MPD}
		\begin{aligned}
			\frac{D p^{\mu }}{d\tau }&=-\frac{1}{2} S^{\rho \sigma } v^{\nu } R^{\mu }{}_{\nu \rho \sigma }-\frac{1}{6}J^{\alpha \beta \gamma \delta }\nabla ^{\mu }R_{\alpha \beta \gamma \delta }+\mathcal{O}(\epsilon^3) \\
			\frac{DS^{\mu \nu }}{d \tau}&=2p^{[\mu }v^{\nu ]}-\frac{4}{3}R_{\alpha \beta \gamma }{}^{[\mu }J^{\nu ]\gamma \alpha \beta }+\mathcal{O}(\epsilon^3).
		\end{aligned}
	\end{equation}
	Here, $D/d\tau\equiv v^{\mu}\nabla_{\mu }$. However, the system of equations consisting of the MPD equation along with the tangent equation  $v^{\mu}=dz^{\mu}/d\tau$ is under-determined as the number of variables ($z^{\mu},~v^{\mu},~p^{\mu},~S^{\mu\nu}$) exceeds the number of equations. Thus, we need to impose some supplementary conditions. Here, we choose the Tulczyjew spin supplementary condition \cite{PhysRevD.86.044033, PhysRevD.88.084005, Dixon1970DynamicsOE}
	\begin{equation}\label{SSC}
		p_{\mu}S^{\mu\nu}=0~.
	\end{equation}
	The above condition fixes the centre of mass of the object. Moreover, it gives a relation between the 4-velocity $v^\mu$ and the momentum $p^\mu$ which can be written as follows \cite{PhysRevD.86.044033}
	\begin{equation}\label{rel_v_p}
		v^\mu=\hat{p}^{\mu}+\frac{2R_{\nu\gamma\alpha\beta}S^{\alpha\beta}S^{\mu\nu}}{4m_d^2+R_{\zeta\nu\alpha\beta}S^{\alpha\beta}S^{\zeta\nu}}\hat{p}^{\gamma}~,
	\end{equation}
	where,
	\begin{equation}\label{hat_p}
		\hat{p}^{\mu}=u^{\mu}+\frac{4}{3m_d^2}R_{\alpha\beta\gamma}{}^{[\mu}J^{\nu]\gamma\alpha\beta}p_\nu
	\end{equation}
	and $p^{\mu}=m_d u^{\mu}$. The parameter $m_d$ represents the dynamic mass of the object, which can be defined as follows $m_d^2=(-p_{\mu}p^{\mu})$. For convenience, we also introduce the monopole rest mass $m_0$ of the object, which can be defined in the following way, $m_0=-p_{\mu}v^{\mu}$.\par
	The quadrupole tensor $J^{\mu\nu\alpha\beta}$ contains information about the deformation due to spin and tidal forces. In this paper, we focus on the distortion caused by spin effects. Thus, we choose the following form of the quadrupole tensor \cite{PhysRevD.86.044033, PhysRevD.88.084005, Bini:2015zya}
	\begin{equation}\label{quadrupole}
		J^{\alpha \beta \gamma \delta }=-\frac{3}{m_d ^2}p^{[\alpha }Q^{\beta ][\gamma }p^{\delta ]}
	\end{equation}
	where, $Q^{\alpha\beta}=C_{Q} S^{\alpha}_{\mu}S^{\beta\mu}/m_d$ is the mass quadrupole tensor. Here, $C_Q$ is the spin-induced quadrupole moment (SIQM) parameter. For rotating Kerr black holes, $C_Q=1$ by black hole no-hair theorem \cite{PhysRevD.57.5287}. For rotating neutron stars, the value of $C_Q$ varies between $\sim 2-20$ depending on the equation-of-state \cite{Harry:2018hke, Pappas:2012ns, Pappas:2012qg}. Interestingly, for certain exotic compact objects, like boson stars, the $C_Q$ can be quite large, ranging from $\sim 10-150$ \cite{PhysRevD.55.6081}. 
For thin-shell gravastar, the parameter can even take negative values for small values of compactness parameter $m_s/R$, where $m_s$ and $R$ is the mass and radius of gravastar \cite{Uchikata:2015yma, PhysRevD.94.064015}. For instance, the SIQM parameter takes the value of  $C_Q\sim -0.5$ for a polytropic thin shell gravastar with polytropic index $n=1$ and $R=5\,m_s$ \cite{PhysRevD.94.064015}.
	
	\subsection{Conserved quantities}
	
	If the spacetime admits a Killing vector $\xi_{\mu}$, then the following quantity \cite{PhysRevD.86.044033, PhysRevD.88.084005, Bini:2015zya}
	\begin{equation}\label{Conserved}
		H_{\xi}=p^{\mu}\xi_{\mu}-\frac{1}{2}S^{\mu\nu}\nabla_{\nu }\xi_{\mu}
	\end{equation}
	is conserved along the trajectory of the object. Since we are interested in stationary, axisymmetric spacetime, the associated conserved quantities are the energy $E$ and angular momentum $J_z$ corresponding to Killing vector $(\partial_t)^{\mu}$ and $(\partial_{\phi})^{\mu}$ respectively. The conservation of the spin length $S^2=S_{\mu\nu}S^{\mu\nu}/2$ depends on the Tulczyjew spin supplementary condition, which can be seen from the following expression. 
	\begin{equation}\label{spin_length}
		\begin{aligned}
			S\frac{dS}{d\tau}&=\frac{1}{2}S_{\mu\nu}\frac{DS^{\mu\nu}}{d\tau}\\&=S_{\mu\nu}(p^{\mu }v^{\nu}-\frac{2}{3}R_{\alpha \beta \gamma }{}^{\mu }J^{\nu \gamma \alpha \beta })\\&=0~.
		\end{aligned}
	\end{equation}
	Here, we obtain the above result by substituting \autoref{quadrupole} in the second line and then using the Tulczyjew spin supplementary condition. 
The dynamical mass term $m_d$ is not conserved. To see this, let us consider the following term $\frac{Dp_\mu}{d\tau}p_\nu\frac{DS^{\mu\nu}}{d\tau}$. Note that, 
$p^\mu = m_d v^\mu + \mathcal O(\epsilon^2)$ (see \autoref{rel_v_p}), 
$S^{\mu\nu}=\mathcal{O}(\epsilon^1)$, and $J^{\mu\nu\alpha\beta}=\mathcal{O}(\epsilon^2)$. From \autoref{MPD}, we can check that $\frac{Dp_\mu}{d\tau}=\mathcal{O}(\epsilon^1)$.  By taking total derivative of \autoref{SSC}, we can show that $p_\nu\frac{DS^{\mu\nu}}{d\tau}=-S^{\mu\nu}\frac{Dp_\nu}{d\tau}=\mathcal{O}(\epsilon^2)$. Hence, the term $\frac{Dp_\mu}{d\tau}p_\nu\frac{DS^{\mu\nu}}{d\tau}=\mathcal{O}(\epsilon^3)$. Furthermore, using \autoref{MPD}, we obtain the following relation
\begin{equation}
\begin{aligned}
   \frac{Dp_\mu}{d\tau}p_\nu\frac{DS^{\mu\nu}}{d\tau}&= \frac{Dp_\mu}{d\tau}\left(-m_0 p^{\mu}+m_d^2v^\mu-\frac{4}{3}R_{\alpha \beta \gamma }{}^{[\mu }J^{\nu ]\gamma \alpha \beta }p_{\nu}\right)\\&=m_0m_d\left(\frac{dm_d}{d\tau}-\frac{m_d}{6m_0}\frac{D R_{\rho\alpha\beta\gamma}}{d\tau}J^{{\rho\alpha\beta\gamma}}+\mathcal{O}(\epsilon^{3})\right)\\&=\mathcal{O}(\epsilon^{3})
\end{aligned}
\end{equation}
where, we used the fact $p^{\mu}(Dp_\mu/d\tau)=-m_d (dm_d/d\tau)$ and $\frac{Dp_\mu}{d\tau}J^{\nu\gamma\alpha\beta}= \mathcal{O}(\epsilon^3)$. The above equation leads to the following relation
\begin{equation}\label{dyn_mass_not_conserved}
    \frac{dm_d}{d\tau}=\frac{m_d}{6m_0}\frac{D R_{\rho\alpha\beta\gamma}}{d\tau}J^{{\rho\alpha\beta\gamma}}+\mathcal{O}(\epsilon^{3})~.
\end{equation}
As can be seen, the dynamical mass term $m_d$ is not conserved.
	However, we can define the mass term $m_s$ given by the following expression \cite{PhysRevD.86.044033, Bini:2015zya}
	\begin{equation}\label{con_mass}
		m_s\equiv m_d-\frac{m_d}{6m_0}R_{\rho\alpha\beta\gamma}J^{{\rho\alpha\beta\gamma}}
	\end{equation}
	which is approximately conserved along the trajectory: using \autoref{dyn_mass_not_conserved} and the fact that $(D J^{\rho\alpha\beta\gamma} /d\tau)=\mathcal{O}(\epsilon^3)$,  we can easily show that  $dm_s/d\tau=\mathcal{O}(\epsilon^3)$.

	
	\section{Orbital motion of the extended object in Kerr background}\label{Sec_3_Orbital_motion}
	
	We start with  Kerr black holes in Boyer-Lindquist coordinate $(t,r,\theta,\phi)$ whose line element can be expressed as follows \cite{Chandrasekhar:1985kt}
	\begin{equation}\label{KDS_metric}
		\begin{aligned}
			ds^{2}=&-\frac{\Delta}{\Sigma}\left[dt-a \sin^{2}\theta~d\phi\right]^{2}+\Sigma\left[\frac{dr^{2}}{\Delta}+d\theta^{2}\right]\\&+\frac{\sin^{2}\theta}{\Sigma}\left[a dt-(r^{2}+a^{2})~d\phi\right]^{2}
		\end{aligned}
	\end{equation}
	where, ${\Delta}=(r^{2}+a^{2})-2M r$ and ${\Sigma}=r^{2}+a^{2}\cos^{2}\theta$ and $a$ is the rotation parameter of the black hole. The solutions of $\Delta=0$ give the position of the horizons as $r_\pm=M\pm\sqrt{M^2-a^2}$, where the upper (lower) sign corresponds to the event (Cauchy) horizon. The spacetime is stationary and axisymmetric; thus admits two Killing vectors $\xi_t^{\mu}=(\partial_{t})^{\mu}$ and $\xi_\phi^{\mu}=(\partial_{\phi})^{\mu}$. 
	For convenience, we introduce an orthonormal tetrad frame to describe the orbital motion \cite{PhysRevD.88.084005}
	\begin{equation}\label{tetrad_frame}
		\begin{aligned}
			{e^{(0)}_{\mu}}&=\frac{\sqrt{\Delta}}{\sqrt{\Sigma}}(1,0,0,-a\sin^{2}\theta),\quad{e^{(1)}_{\mu}}=\frac{\sqrt{\Sigma}}{\sqrt{\Delta}}(0,1,0,0),\\
			{e^{(2)}_{\mu}}&=(0,0,\sqrt{\Sigma},0),\quad{e^{(3)}_{\mu}}=\frac{\sin\theta}{\sqrt{\Sigma}}(-a,0,0,r^2+a^2)~.
		\end{aligned}
	\end{equation}
	We define the spin vector through the following relation \cite{Pani}
	\begin{equation}\label{spin_vector}
		\begin{aligned}
			S^{\ta}&=-\frac{1}{2}\varepsilon^{\ta\tb\tc\td}u_{\tb}S_{\tc\td}\\
			S^{\ta\tb}&=\varepsilon^{\ta\tb\tc\td}u_{\tc}S_{\td}
		\end{aligned}
	\end{equation}
	where,  $\varepsilon^{\ta\tb\tc\td}$ is the Levi-Civita tensor.
	\subsection{Equations of motion on the equatorial plane}
	
	In the following, we consider the secondary object is orbiting around the supermassive black hole in an equatorial plane ($\theta=\pi/2$). Moreover, we choose the spin vector of the secondary $S^{\mu}$ is parallel to $z$ axis i.e., $S^{\ta}=(0,0,-S,0)$. The negative sign implies that the secondary is moving in a spin-aligned configuration \cite{PhysRevD.88.084005}. Basically, for $\theta=\frac{\pi}{2},$ the $\partial_{\theta}$ and $\partial_{z}$ are anti-aligned. Hence this negative sign is making the spin of the secondary to align to $\partial_z$ and hence with that of the primary \cite{Pani}. Using \autoref{SSC} and \autoref{spin_vector}, we find that $p^{(2)}=0$, $S^{(2)\ta}=0$, $S^{(0)(1)}=-S u^{(3)}$, $S^{(0)(3)}=S u^{(1)}$, $S^{(1)(3)}=S u^{(0)}$. It is useful to define dimensionless variables,
	\begin{equation}
		\begin{aligned}\label{dimensionless}
			\hr&=\frac{r}{M},\quad{\ha}=\frac{a}{M},\quad{\hE}=\frac{E}{m_s},\\
			\hJ&=\frac{J_z}{Mm_s},\quad{\sigma}=\frac{S}{Mm_s}=q\chi~.
		\end{aligned}
	\end{equation}
	where, $q=m_s/M$ is the mass ratio. Using \autoref{Conserved}, we can write the energy and angular momentum of the object as follows
	\begin{equation}\label{energy_momentum}
		\begin{aligned}
			\hE&=\frac{\sqrt{\Delta }}{\hat{r}}w^{(0)}+\frac{ \left(\hat{a} \hat{r}+\sigma \right)}{\hat{r}^2}w^{(3)}\\
			\hJ&=\frac{\sqrt{\Delta }  \left(\hat{a}+\sigma \right)}{\hat{r}}w^{(0)}+\frac{\left(\hat{a} \left(\hat{r}+1\right) \sigma +\hat{a}^2 \hat{r}+\hat{r}^3\right)}{\hat{r}^2}w^{(3)}
		\end{aligned}
	\end{equation}
	where, we introduce a parameter $w^{\mu}=p^{\mu}/m_s$ for convenience. We invert the above expression to write $w^{(0)}$ and $w^{(3)}$ in terms of $\hE$ and $\hJ$, which is given as follows.
	\begin{equation}\label{w0_w3}
		\begin{aligned}
			w^{(0)}&=\frac{\left(\hat{r}^2+\hat{a}^2 \right)\hat{E}-\hat{a} \hat{J}_z}{\hat{r} \sqrt{\Delta}}\left(1+\frac{\sigma^2}{\hat{r}^3}\right)+\frac{\sigma  \left(\hat{a} \hat{E}\left(\hat{r}+1 \right)-\hat{J}_z\right)}{\hat{r}^2 \sqrt{\Delta}}\\
			w^{(3)}&=\frac{\hat{J}_z-\hat{a} \hat{E}}{\hat{r}}\left(1+\frac{\sigma^2}{\hat{r}^3}\right)-\frac{\hat{E} \sigma }{\hat{r}}.
		\end{aligned}
	\end{equation}
	Replacing the above expression in \autoref{con_mass}, we find conserved mass as follows \cite{PhysRevD.86.044033} 
	\begin{equation}\label{conmass_pi/2}
		\begin{aligned}
			m_s=m_d \left[1+\frac{\sigma ^2 C_Q }{2 \hat{r}^3}\left(1+3\left(\frac{\hJ-\hat{a}\hE}{\hr^2}\right)^2\right)\right]+\mathcal{O}(\epsilon^3).
		\end{aligned}
	\end{equation}
	We can obtain the expression for $w^{(1)}$ from the following relation $\left(w^{(0)}\right)^2-\left(w^{(1)}\right)^2-\left(w^{(3)}\right)^{2}=m_d^2/m_s^2$. The relationship between normalized momenta $u^{\ta}$ and the 4-velocity  $v^{\ta}$ turns out to be 
	\begin{equation}\label{v_and_u_pi/2}
		\begin{aligned}
			v^{(0)}&=\left(1+\frac{3  \left(1-8\, C_Q\right)\sigma ^2 (u^{(3)})^2}{\hat{r}^3}\right)u^{(0)}+\mathcal{O}(\epsilon^3)\\
			v^{(1)}&=\left(1+\frac{3  \left(1-8\, C_Q\right)\sigma ^2 (u^{(3)})^2}{\hat{r}^3}\right)u^{(1)}+\mathcal{O}(\epsilon^3)\\
			v^{(3)}&=\left(1+\frac{3  \left(1-8\, C_Q\right)\sigma ^2 \left(1+(u^{(3)})^2\right)}{\hat{r}^3}\right)u^{(3)}+\mathcal{O}(\epsilon^3)~,
		\end{aligned}
	\end{equation}
	where, $u^{\ta}$ follows the relation $u^{\ta}=m_sw^{\ta}/m_d$. The component of 4-velocity in \BL\ coordinate can be obtained with the following $v^{\mu}=e^{\mu}_{\ta}v^{\ta}$ which gives the equation of motion as follows \cite{PhysRevD.86.044033, PhysRevD.88.084005}
	\begin{equation}\label{eqn_of_motn}
		\begin{aligned}
			\Sigma_s\Lambda_s\left(\frac{m_d}{m_s}\right)\left(\frac{d\hht}{d\tau}\right)&=\hat{a} \left(\hJ-(\hat{a}+\sigma)\hE\right)Q_s+\frac{\left(\hat{r}^2+\hat{a}^2 \right)}{\Delta}P_s,\\
			\left(\frac{d\hr}{d\tau}\right)^2=V_{\sigma}(\hr)&\equiv\frac{\hr^4}{\Sigma_s^2}\left[\alpha\hE^2-2\beta\frac{\hJ}{\hr}\hE+\gamma\frac{\hJ^2}{\hr^2}-\delta\frac{m_d^2}{m_s^2}\right],\\
			\Sigma_s\Lambda_s\left(\frac{m_d}{m_s}\right)\left(\frac{d\phi}{d\tau}\right)&= \left(\hJ-(\hat{a}+\sigma)\hE\right)Q_s+\frac{\hat{a}}{\Delta}P_s~,
		\end{aligned}
	\end{equation}
	where,
	\begin{equation}\label{motn_variables}
		\begin{aligned}
			\Sigma_s&=\hr^2\left(1-\frac{\sigma^2}{\hr^3}\right),\quad{\Lambda_s}=1-\frac{3(1-8\,C_Q)(\hJ-\hE\ha)^2\sigma^2}{\hr^5},\\
			Q_s&=1-\frac{3(1-8\,C_Q)\sigma^2}{\hr^3},\\{P_s}&=\hE\left[\left(\hat{r}^2+\hat{a}^2 \right)+\frac{\ha\sigma}{\hr}(\hr+1)\right]-\hJ(1+\frac{\sigma}{\hr})\\
			\alpha&=\left(1+\frac{\hat{a}^2}{\hat{r}^2}+\frac{\hat{a}  \sigma }{\hat{r}^2}\left(1+\frac{1}{\hat{r}}\right)\right)^2-\frac{\Delta  }{\hat{r}^2}\left(\frac{\hat{a}}{\hat{r}}+\frac{\sigma }{\hat{r}}\right)^2\\
			\beta&=\left(1+\frac{\hat{a}^2}{\hat{r}^2}+\frac{\hat{a}  \sigma }{\hat{r}^2}\left(1+\frac{1}{\hat{r}}\right)\right) \left(\frac{\hat{a}}{\hat{r}}+\frac{\sigma }{\hat{r}^2}\right)-\frac{\Delta}{\hat{r}^2}\left(\frac{\hat{a}}{\hat{r}}+\frac{\sigma }{\hat{r}}\right)\\
			\gamma&=\left(\frac{\hat{a}}{\hat{r}}+\frac{\sigma }{\hat{r}^2}\right)^2-\frac{\Delta }{\hat{r}^2},\qquad{\delta}=\frac{\Delta  \Sigma_s^2}{\hr^6}.
		\end{aligned}
	\end{equation}
	\subsection{Circular orbit, ISCO and Orbital frequency}
	
	In this paper, we focus on circular orbits. For an object moving in a circular orbit, the radial velocity and acceleration vanish simultaneously, leading to the condition $V_{\sigma}=0$ and $dV_{\sigma}/d\hr=0$. The stability of such orbits against radial perturbation is dictated by the condition $d^2V_{\sigma}/d\hr^2<0$. It is more convenient to use an effective potential term $V_{\textrm{eff}}$ for the calculation, which can be written as follows \cite{PhysRevD.86.044033}
	\begin{equation}\label{eff_potential}
		V_{\textrm{eff}}(\hr)=\left[\alpha\hE^2-2\beta\frac{\hJ}{\hr}\hE+\gamma\frac{\hJ^2}{\hr^2}-\delta\frac{m_d^2}{m_s^2}\right]
	\end{equation}
	where, the $\alpha,~\beta,~\gamma,~\delta$ is given in \autoref{motn_variables}. Moreover, we adopt the variables $y= 1/\hr$ and $x=\hJ-\ha\hE$ in place of $\hr$ and $\hJ$. The condition for circular orbit then transformed as $V_{\textrm{eff}}=0$ and $dV_{\textrm{eff}}/dy=0$. Noting that the parameter $\sigma\ll1$, we can expand the equations mentioned above into a series of $\sigma$. Here, we seek a solution to the equations in the following form.
	\begin{equation}\label{circular}
		\begin{aligned}
			\hE=\hE_0+\sigma \hE_1+\sigma^2 \hE_2,\quad{\hx}=\hx_0+\sigma \hx_1+\sigma^2 \hx_2~,
		\end{aligned}
	\end{equation}
	where, $\{\hE_0,\hx_0\}$ corresponds to the value of $\{\hE,\hx\}$ for a spinless object, whereas $\{\hE_1,\hx_1\}$ and $\{\hE_2,\hx_2\}$ represent the linear and quadratic corrections due to spin respectively. For stable circular orbit, $\{\hE_0,\hx_0\}$ attains the value \cite{Jefremov:2015gza}
	\begin{equation}\label{0th order}
		\begin{aligned}
			\hE_0=\frac{1-2y\mp\hat{a} \sqrt{y^3}}{\sqrt{1-3 y\mp 2 \hat{a} \sqrt{y^3}}},\quad{\hx_0}=\frac{1\mp \hat{a} \sqrt{y}}{\sqrt{y \left(1-3 y\mp 2 \hat{a} \sqrt{y^3}\right)}}.
		\end{aligned}
	\end{equation}
	The upper sign represents a retrograde (counter-rotating) orbit, whereas the lower sign corresponds to a prograde (co-rotating) orbit. The equations for  $\{\hE_i,\hx_i\}$ ($i=1,2$) is presented in the Appendix \ref{App:Circular}. We solve these equations numerically and replace them in \autoref{circular} along with \autoref{0th order} to obtain the value of $\{\hE,\hx\}$ as a function of $y$. By replacing $y=1/\hr$ and $\hJ=\hx+\ha\hE$, we obtain the value of energy and angular momentum of the object hovering in a circular orbit with radius $\hr$.\par
	Determination of the parameters of the innermost stable circular orbit (ISCO) needs an additional condition $d^2V_{\textrm{eff}}/dy^2=0$ ($d^2V_{\textrm{eff}}/d\hr^2=0$). Series expansion of this condition into the series of $\sigma$ is presented in Appendix \ref{App:Circular}. Similar to \autoref{circular}, we seek a solution in the following form. 
	\begin{equation}\label{ISCO}
		\begin{aligned}
			y=y_0+\sigma y_1+\sigma^2 y_2~.
		\end{aligned}
	\end{equation}
	Solving the equations $V_{\textrm{eff}}=dV_{\textrm{eff}}/dy=d^2V_{\textrm{eff}}/dy^2=0$ simultaneously, we obtain the parameters $\{\hE,\hx,y\}$ in the form given by \autoref{circular} and \autoref{ISCO}, which in turn gives us the energy, angular momentum and position of the ISCO $\{\hE^{\textrm{isco}},\hJ^{\textrm{isco}},\hr^{\textrm{isco}}\}$ (see Appendix \ref{App:Circular} for more details).\par
	The angular frequency of the circular orbits is given by
	\begin{equation}\label{frequency}
		\begin{aligned}
			\hO\equiv \frac{d\phi/d\tau}{d\hht/d\tau}=\frac{\Delta \left(\hJ-(\hat{a}+\sigma)\hE\right)Q_s+\ha P_s}{\Delta\hat{a} \left(\hJ-(\hat{a}+\sigma)\hE\right)Q_s+\left(\hat{r}^2+\hat{a}^2 \right) P_s}~,
		\end{aligned}
	\end{equation}
	where in the second step, we use \autoref{eqn_of_motn}. Replacing \autoref{circular} in \autoref{frequency} and expanding the expression as 
	\begin{equation}\label{orb_freq}
		\begin{aligned}
			\hO(\hr)=\hO_0(\hr)+\sigma \hO_1(\hr)+\sigma^2 \hO_2(\hr)~,
		\end{aligned}
	\end{equation}
	we obtain the angular frequency of the circular orbit. Here, $\hO_0$ corresponds to the angular frequency of a non-spinning object, whereas $\hO_1$ and $\hO_2$ represent linear and quadrupolar correction due to spin, respectively.
	\section{Gravitational wave fluxes}\label{Sec_4_GW_flux}
	
In this section, we describe the gravitational wave radiation from the EMRI system, where the secondary object inspirals into the primary following quasi-circular, equatorial orbits. As a result of the system's tiny mass ratio, we can study the evolution of the secondary object through perturbation methods. The system loses energy and angular momentum due to gravitational radiation, the back-reaction of which (self-force) shrinks the binary separation, and the system goes through an inspiral phase. In this paper, we use \textit{adiabatic approximation} to study the gravitational back-reaction effects \cite{PhysRevD.78.064028, Hughes:2021exa}.
 The motivation behind this formalism is that the orbital time scale $T_o$ ($\sim M$) is much shorter than the dissipative $T_i$ ($\sim M^2/m_s\sim T_o/q\gg T_o$); thus allowing us to treat the orbital dynamics as geodesics over a short time scale. Furthermore, the rate of change of the orbit's energy 
is dictated by the time-averaged, dissipative part of the self-force \cite{ Hughes:2021exa} i.e., 
	\begin{equation}\label{adiabatic}
		\begin{aligned}
			\left(\frac{d\hE}{dt}\right)^{\textrm{orbit}}=-\Big\langle\frac{d\hE}{dt}\Big\rangle_{\textrm{GW}}
		\end{aligned}
	\end{equation}
	where $\langle\boldsymbol{\cdot}\rangle$ denotes the averaging over a time period much larger than $T_o$ but smaller than $T_i$. We calculate the back-reaction effect on the orbit by solving the Teukolsky equation. This gives the adiabatic evaluation of the object from orbit to orbit. Note that the adiabatic approximation breaks down as the object crosses the ISCO and transits onto a geodesic plunge orbit \cite{PhysRevD.78.064028}. Our study focuses only on the adiabatic part of the motion. Much of our discussions presented in this section and the two subsequent subsections (\ref{sec 4A}) and (\ref{sec 4B}) follow closely from \cite{Pani}. Interested readers are referred to \cite{Pani} for further details.
	\subsection{Teukolsky equation}\label{sec 4A}
	
 As discussed earlier, we consider that the secondary object perturbs the background Kerr spacetime. Here, we adopt Teukolsky formalism to obtain the perturbation equation and the gravitational wave flux. The information about the gravitational radiation is encoded in the perturbed Weyl tensor $C_{\mu\nu\alpha\beta}$ \cite{1973ApJ...185..635T}. In particular, the Weyl scalar $\Psi_4=-C_{\mu\nu\alpha\beta} n^{\mu}\bar{m}^{\nu}n^{\alpha}\bar{m}^{\beta}$ contains information about the outgoing part of the radiation. Here, $n^{\mu}$ and $\bar{m}^{\mu}$ are part of the orthonormal null tetrad, the expression of which is given in \autoref{extrarev}.
 Teukolsky showed that $\Psi_4$ could be decomposed as functions of Boyer-Lindquist coordinates in the following manner\cite{1973ApJ...185..635T}
	\begin{equation}\label{Psi_4}
		\Psi_4=\rho^4\sum_{\ell=2}^{\infty}\sum_{m=-\ell}^{\ell}\int_{-\infty}^{\infty}d\ho~R_{\mode}(\hr)~{}_{-2}\Ang e^{i(m\phi-\ho\hht)}
	\end{equation}
	where $\rho=[\hr-i\ha\cos\theta]^{-1}$. ${}_{-2}\Ang$ is the spin-weighted spheroidal harmonics with weight $-2$, which satisfies the angular Teukolsky equation
	\begin{equation}\label{ang_Teuk}
		\begin{aligned}
			\Bigg[&\frac{1}{\sin\theta}\frac{d}{d\theta}\left(\sin\theta\frac{d}{d\theta}\right)-\ha^2\ho^2\sin^2\theta-\left(\frac{m-2\cos\theta}{\sin\theta}\right)^2\\&+4\ha\ho\cos\theta-2+2\ha m\ho+\lambda_{\mode}\Bigg]{}_{-2}\Ang=0~.
		\end{aligned}
	\end{equation}
	where, $\lambda_{\mode}=E_{\mode}-2m\ha\ho+\ha^2\ho^2-2$. Here, $E_{\mode}$ is the separation constant. Hereafter, we denote ${}_{-2}\Ang$ by $\Ang$ for brevity. The eigenfunction of the angular Teukolsky equation $\Ang$ satisfies the following normalization condition. 
	\begin{equation}\label{ang_norm}
		\int \sin\theta d\theta d\phi |\Ang e^{im\phi}|^2=1~.
	\end{equation}
	\\
The radial function $R_{\mode}(\hr)$ satisfies the inhomogenious Teukolsky equation
	\begin{equation}\label{rad_Teuk}
		\begin{aligned}
			\Delta^2\frac{d}{d\hr}\left(\frac{1}{\Delta}\frac{dR_{\mode}}{d\hr}\right)-V(\hr)R_{\mode}=\mathcal{J}_{\mode}
		\end{aligned}
	\end{equation}
	where 
	\begin{equation}\label{Potential}
		\begin{aligned}
			V(\hr)&=-\frac{K^2+4i(\hr-1)K}{\Delta}+8i\ho\hr+\lambda_{\mode}\\
			K&=(\hr^2+\ha^2)\ho-\ha m~.
		\end{aligned}
	\end{equation}
	The source term $\mathcal{J}_{\mode}$ depends on the energy-momentum tensor of the secondary object (see  \autoref{source_1}). 
	The details of the calculation for the source term are presented in Appendix  \ref{App:source_term}.\par
	We employ the Green function method to obtain $R_{\mode}$. 
	In terms of the linearly independent solutions of homogeneous radial Teukolsky equation $R^{\In}_{\mode}(\hr)$ and $R^{\up}_{\mode}(\hr)$ following purely incoming boundary conditions at the horizon and purely outgoing boundary condition at the infinity respectively, the solution of \autoref{rad_Teuk} can be written as 
	\begin{equation}\label{sol_Teuk}
		\begin{aligned}
			R_{\mode}(\hr)=&\frac{1}{\mathcal W}\Biggl\{R^{\up}_{\mode}\int_{\hr_+}^{\hr}d\hr\frac{R^{\In}_{\mode}\mathcal{J}_{\mode}}{\Delta^2}\\&+R^{\In}_{\mode}\int_{\hr}^{\infty}d\hr\frac{R_{\mode}^{\up}\mathcal{J}_{\mode}}{\Delta^2}\Biggr\}
		\end{aligned}
	\end{equation}
	where $\mathcal W=\left(R^{\In}_{\mode}\partial_{\hr} R^{\up}_{\mode}-R^{\up}_{\mode}\partial_{\hr} R^{\In}_{\mode}\right)/\Delta$ is the constant Wronskian. 
	The asymptotic behaviour of the radial function $R_{\mode}$ is given as follows. 
	\begin{equation}\label{asymp_behave}
		\begin{aligned}
			R_{\mode}=
			\begin{cases}
				\mathcal{Z}_{\mode}^{H}\hr^3~e^{i\ho\hr_{*}}~, & \hr\to \infty\\
				\mathcal{Z}_{\mode}^{\infty}\Delta^2~e^{-i(\ho-m\hO_+)\hr_{*}}~, & \hr\to \hr_+
			\end{cases}
		\end{aligned}
	\end{equation}
	where 
	\begin{equation}\label{tortoise}
		\begin{aligned}
			\hr_{*}=\hr+\frac{2\hr_{+}}{\hr_{+}-\hr_{-}}\ln\left[\frac{\hr-\hr_{+}}{2}\right]-\frac{2\hr_{+}}{\hr_{+}-\hr_{-}}\ln\left[\frac{\hr-\hr_{-}}{2}\right]
		\end{aligned}
	\end{equation}
	is the tortoise coordinate, and $\hO_+=\ha/(2\hr_+)$ is the dimensionless angular frequency of the black hole. The amplitudes $	\mathcal{Z}_{\mode}^{H,\infty}$ are given by the following relation. 
	\begin{equation}\label{amp_def}
		\begin{aligned}
			\mathcal{Z}_{\mode}^{H,\infty}=\mathcal{C}_{\mode}^{H,\infty}\int_{\hr_+}^{\infty
			}d\hr\frac{R_{\mode}^{\In,\up}\mathcal{J}_{\mode}}{\Delta^2}\,,
		\end{aligned}
	\end{equation}
	where $\mathcal{C}_{\mode}^{H,\infty}$ are constants; (see \autoref{constant_term}). For the energy-momentum tensor presented in \autoref{SET}, the amplitudes takes the following form (see \autoref{app_amplitude})
	\begin{equation}\label{amp_form}
		\begin{aligned}
			\mathcal{Z}^{H,\infty}_{l m \hat{\omega}}
			=\mathcal C^{H,\infty}_{l m \hat{\omega}}\int_{-\infty}^{\infty} d\hat{t}\, e^{i(\hat \omega \hat t-m\,\phi(t))}I^{H,\infty}_{l m \hat{\omega}}[r(t),\theta(t)],
		\end{aligned}
	\end{equation}
	where
	\begin{equation}\label{amp_integrant}
		\begin{aligned}
			I^{H,\infty}_{l m \hat{\omega}}[r(t),\theta(t)]
			=&\Big[A_0-(A_1+B_0)\frac{d}{d\hat r}+(A_2+B_1+C_0)\frac{d^2}{d\hat r^2}\\&-(B_2+C_1)\frac{d^3}{d\hat r^3}+C_2\frac{d^4}{d\hat r^4}\Big]R^{\textrm{in},\textrm{up}}_{l m \hat{\omega}}(\hat r)\Big|_{r(t),\theta(t)}~.
		\end{aligned}
	\end{equation}
	\begin{figure*}[t]
		\centering
		\minipage{0.33\textwidth}
		\includegraphics[width=\linewidth]{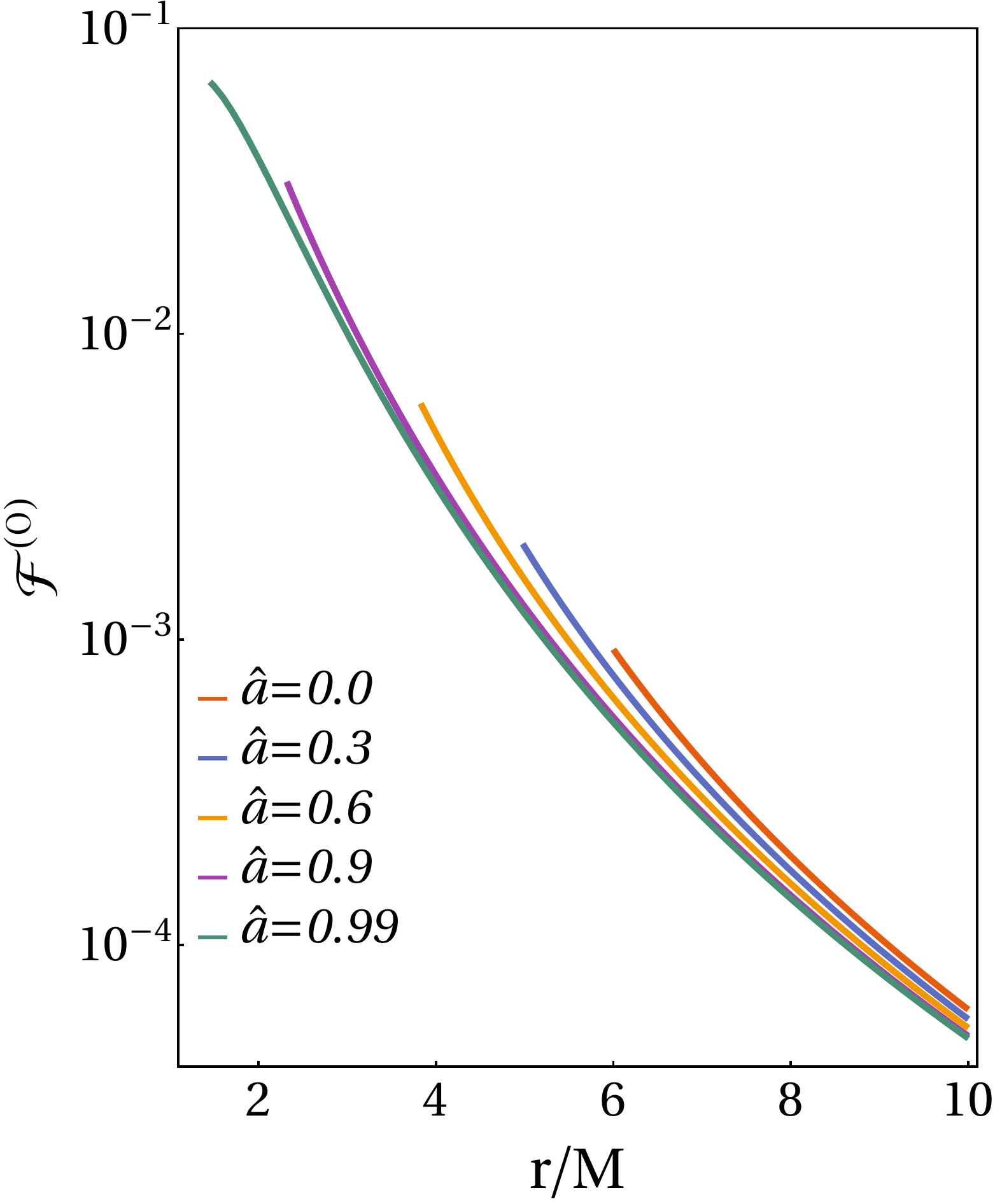}
		\endminipage\hfill
		\minipage{0.33\textwidth}
		\includegraphics[width=\linewidth]{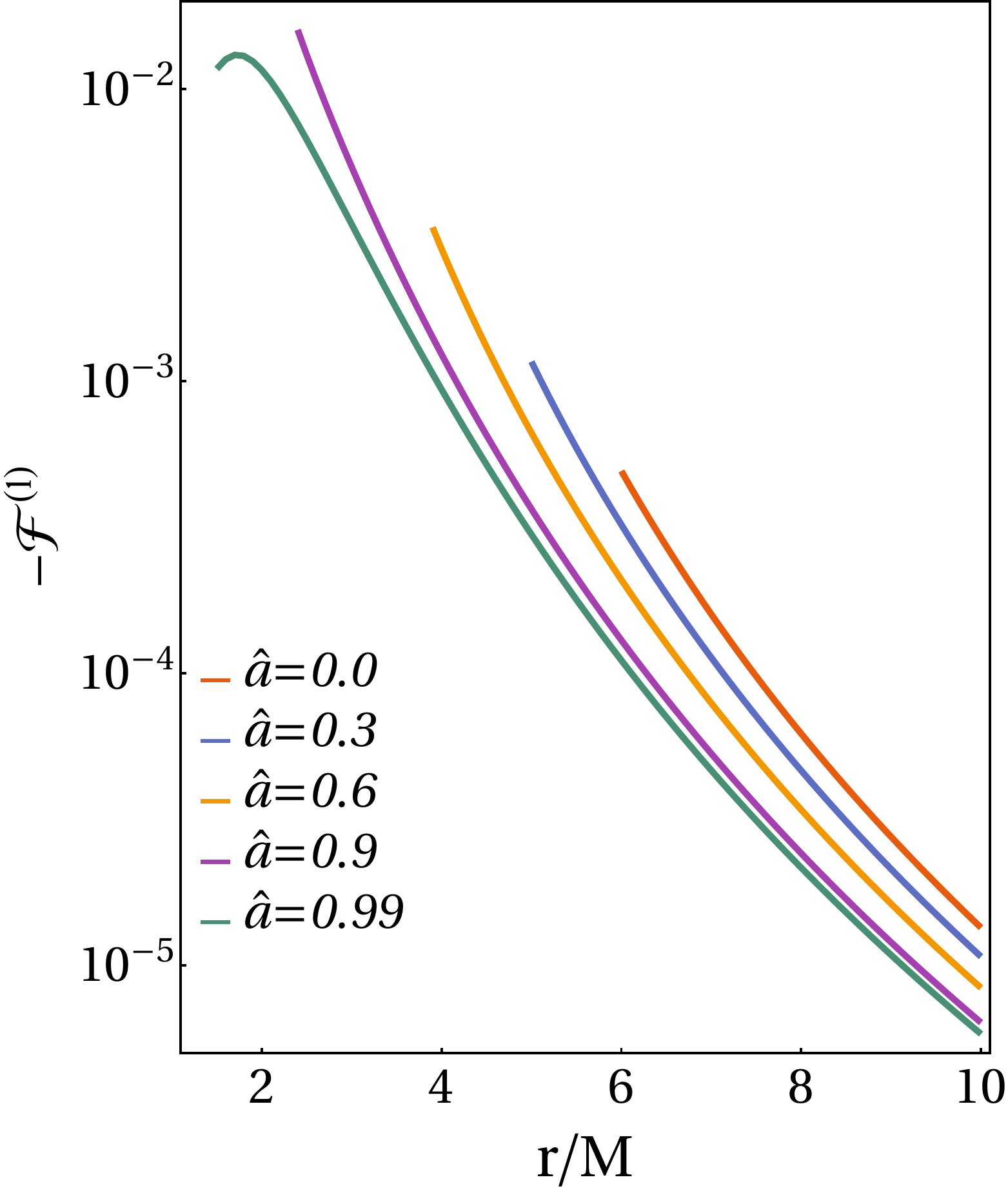}
		\endminipage\hfill
		\minipage{0.33\textwidth}%
		\includegraphics[width=\linewidth]{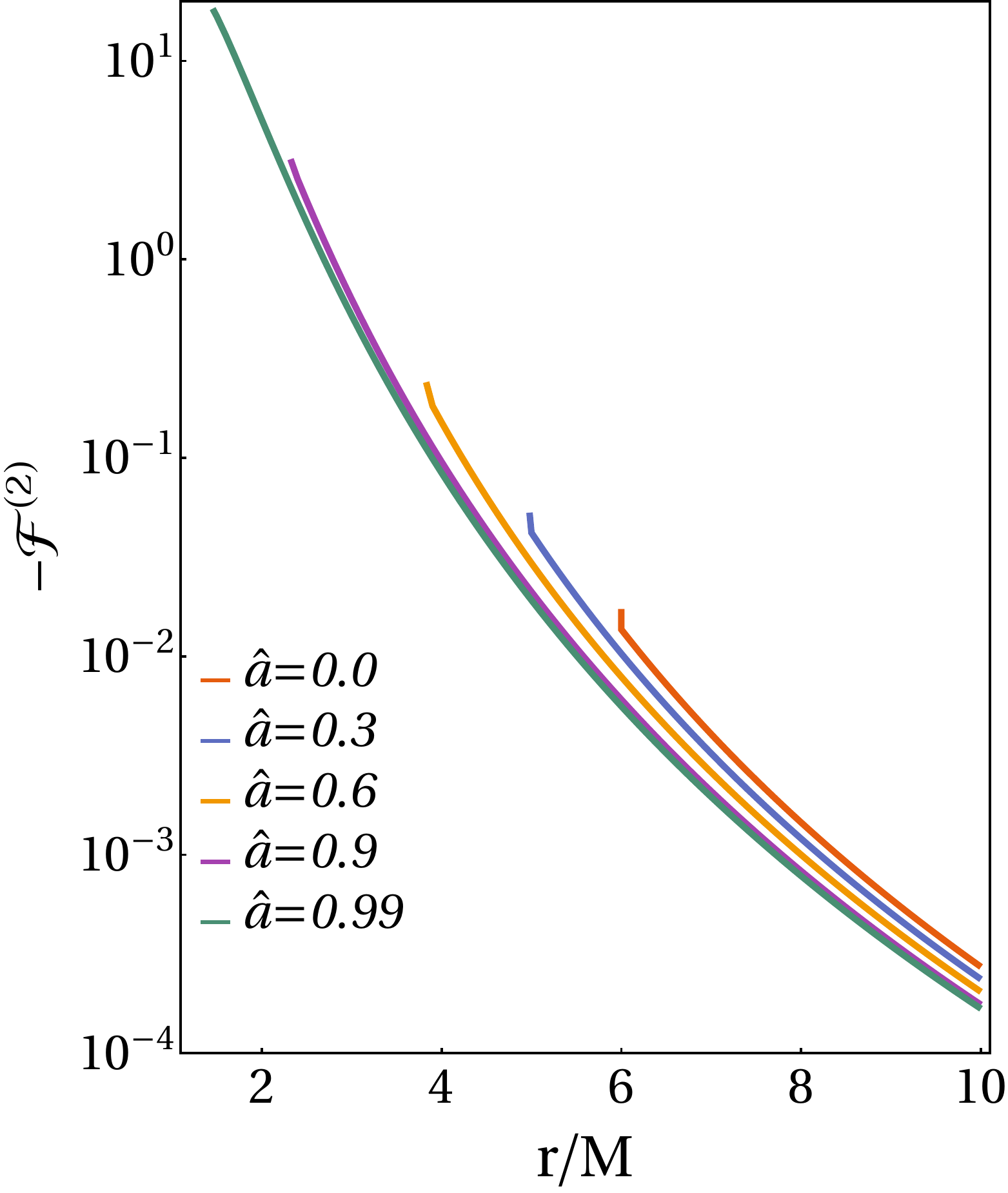}
		\endminipage
		\caption{The plot of total energy flux for stable, prograde orbits as a function of orbital radius $\hr$ for different values of spin of the primary object. In the left panel, the energy flux of a non-spinning secondary object  $\mathcal{F}^{(0)}$ is presented. The middle and right panels show the linear and quadratic correction coefficients in energy flux due to spin effects. Here, we take $q=10^{-4},~C_Q=10$.}\label{fig_flux}
	\end{figure*}
	We have presented the details to calculate these quantities as well as the explicit form of the coefficients $(A_i,~B_i,~C_i)$ ($i=0,1,2$) in Appendix \ref{App:source_term}.   \par
		In what follows, we turn our attention to equatorial, circular orbits. This hugely  simplifies the expressions of the coefficients $(A_i,~B_i,~C_i)$ (see \autoref{app13} in  Appendix \ref{App:source_term}). Moreover, for circular orbits, we have $\phi(\hht)=\hO \hht$ which simplifies the expression for the amplitude in \autoref{amp_form} as $\mathcal{Z}^{H,\infty}_{l m \hat{\omega}}
	=\mathcal A^{H,\infty}_{l m \hat{\omega}}\delta(\ho-m\hO)$ at some specific radius $r_0$, where  $\mathcal A^{H,\infty}_{l m \hat{\omega}}=2\pi~\mathcal C^{H,\infty}_{l m \hat{\omega}} I^{H,\infty}_{l m \hat{\omega}}[r_0,\pi/2]$.
At infinity, the Weyl scalar is related to gravitational wave strain $h$ ($	h\equiv h_+-ih_{\times}$) in the following manner $\Psi_4=\ddot{h}/2=(\ddot{h}_{+}-i\ddot{h}_{\times})/2$ , where overdot sign implies derivative with respect to $\hht$.	Using \autoref{Psi_4} and \autoref{amp_def}, we find that the gravitational wave strain for circular, equatorial orbits can be written in the following form	

	\begin{equation}\label{signal_equatorial}
		\begin{aligned}
			h=-\frac{2}{r}\sum_{\ell=2}^{\infty}\sum_{m=-\ell}^{\ell}\frac{\mathcal A^{H}_{\mode}}{(m\hO)^2}~S_{\mode}^{\ha\ho}(\vartheta) e^{im(\varphi-\hO(\hht-\hr_*))}~.
		\end{aligned}
	\end{equation}
	\par
	The above expression can be used to obtain the time-averaged energy flux at infinity which can be written as follows \cite{Pani}
	\begin{equation}\label{flux_inf}
		\begin{aligned}
			\Big\langle\frac{d\hE}{dt}\Big\rangle^{\infty}_{\textrm{GW}}&=\sum_{\ell=2}^{\infty}\sum_{m=1}^{\ell}\frac{|\mathcal A^{H}_{\mode}|^2}{2\pi( m\hO)^2}~,
		\end{aligned}
	\end{equation}
	Here, we use the property of the amplitude  $\mathcal{Z}_{\ell -m-\ho}^{H,\infty}=(-1)^{\ell}\bar{\mathcal{Z}}_{\mode}^{H,\infty}$ to restrict the sum over $m$ in \autoref{signal_equatorial} to positive $m$ values. 
Similarly, we can write energy flux at the horizon as \cite{Pani}
	\begin{equation}\label{flux_hor}
		\begin{aligned}
			\Big\langle\frac{d\hE}{dt}\Big\rangle^{H}_{\textrm{GW}}&=\sum_{\ell=2}^{\infty}\sum_{m=1}^{\ell}\alpha_{\ell m}\frac{|\mathcal A^{\infty}_{\mode}|^2}{2\pi( m\hO)^2}~,
		\end{aligned}
	\end{equation}
	where,
	\begin{equation}
	  \alpha_{\ell m}=\frac{256(2\hr_+)^5\kappa(\kappa^2+4\varepsilon^2)(\kappa^2+16\varepsilon^2)(m\hO)^3}{|C_{\ell m}|^2}  
	\end{equation}
	 with $\kappa=\ho-m\hO_+$, $\varepsilon=\sqrt{1-\ha^2}/4\hr_+$ and 
	\begin{equation}
		\begin{aligned}
			|C_{\ell m}|^2=&\Big[(\lambda_{\mOde}+2)^2+4\ha(m\hO)-4\ha^2(m\hO)^2\Big]\times\\&\Big[\lambda_{\mOde}^2+36m\ha(m\hO)-36\ha^2(m\hO)^2\Big]\\
			&+(2\lambda_{\mOde}+3)\Big[96\ha^2(m\hO)^2-48m\ha(m\hO)\Big]\\&+144(m\hO)^2(1-\ha^2)\nonumber~.
		\end{aligned}
	\end{equation}
	\begin{figure*}[th]
		\minipage{0.50\textwidth}
		\includegraphics[width=\linewidth]{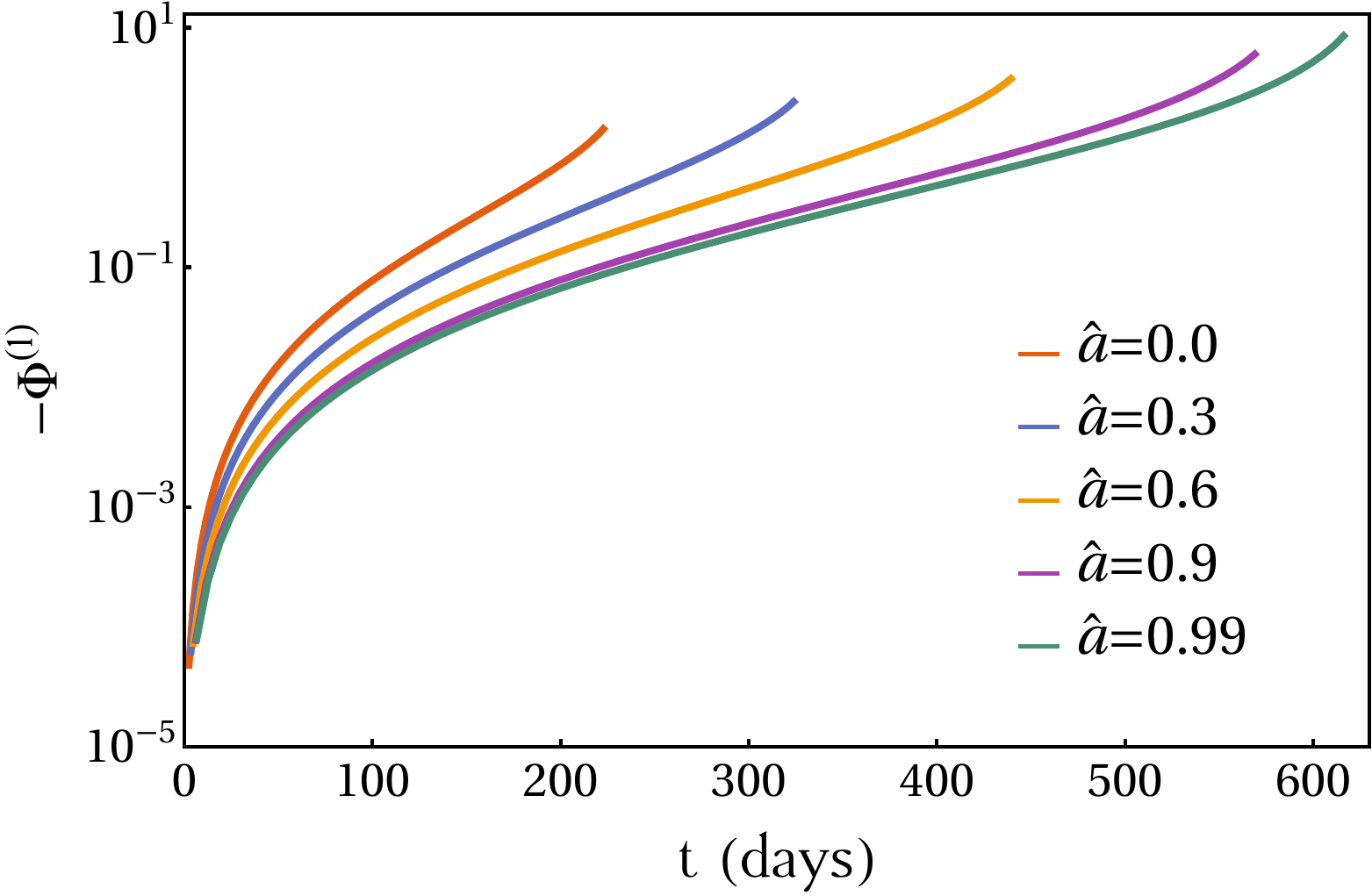}
		\endminipage\hfill
		\minipage{0.50\textwidth}
		\includegraphics[width=\linewidth]{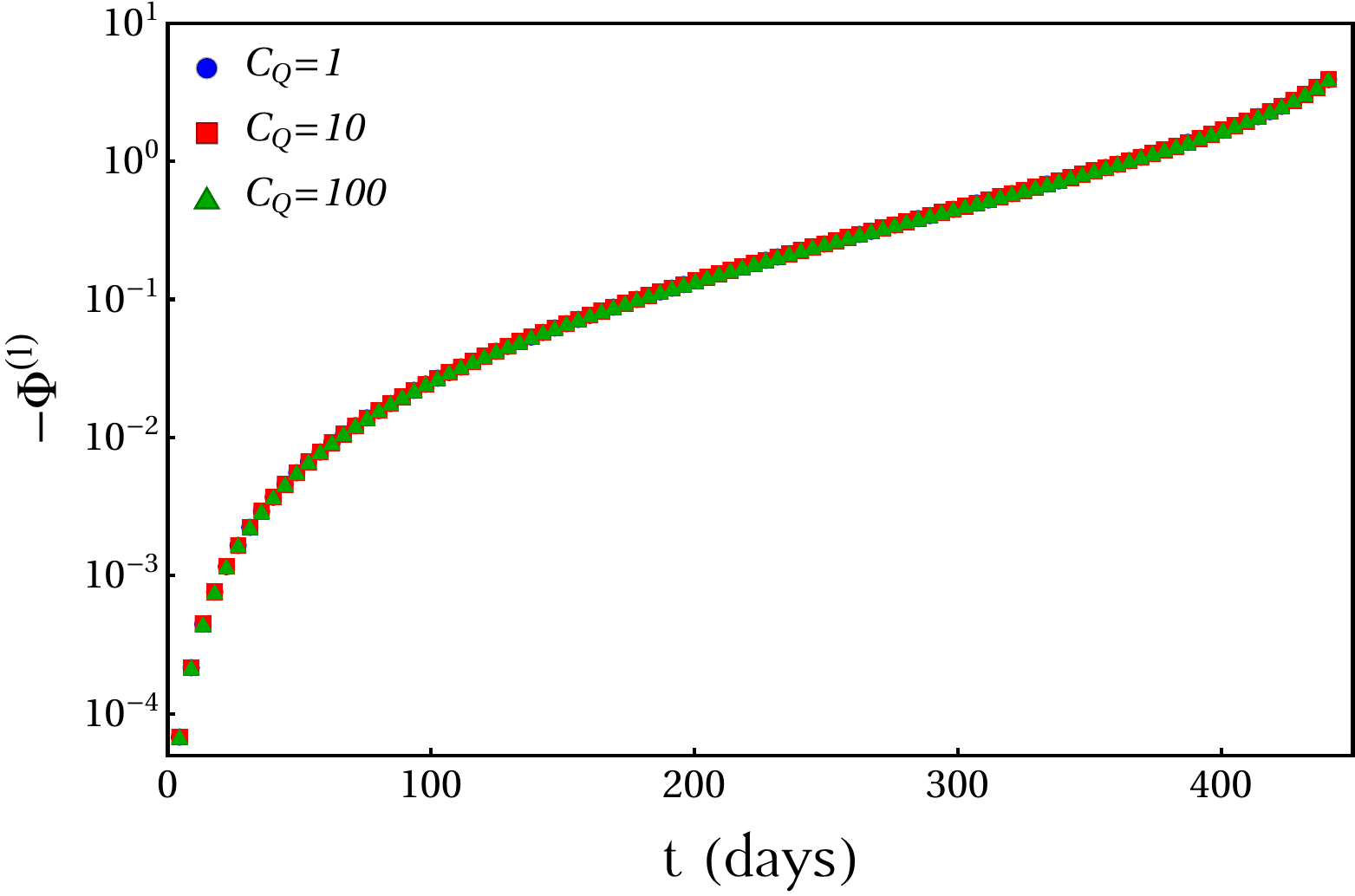}
		\endminipage
		\caption{The plot of linear order spin correction coefficients $\Phi^{(1)}$ as a function of $t$ for stable, prograde orbits. The left panel shows the time evolution of $\Phi^{(1)}$ for $C_Q=10$ and different values of $\ha$. The right panel shows the same for $\ha=0.6$ and different values of $C_Q$. Here, we consider the inspiral of a $1~M_\odot$ compact object into a supermassive black hole of mass $M=10^4~M_\odot$. The $\Phi^{(1)}$ depends only on $\ha$.}\label{fig_phi1}
	\end{figure*}
	\subsection{Adiabatic evolution of the orbit and gravitational wave phase} \label{sec 4B}
	With the expression of energy radiation at the horizon and infinity in \autoref{flux_hor} and \autoref{flux_inf}, respectively, we can calculate the total energy flux from the following equation. 
	\begin{equation}\label{tot_flux}
		\begin{aligned}
			\mathcal{F}
			=\frac{1}{q}\left[\Big\langle\frac{d\hE}{dt}\Big\rangle^{H}_{\textrm{GW}}+\Big\langle\frac{d\hE}{dt}\Big\rangle^{\infty}_{\textrm{GW}}\right]=\sum_{\ell=2}^{\infty}\sum_{m=1}^{\ell}\mathcal{F}_{\ell m}.
		\end{aligned}
	\end{equation}
	where, $\mathcal{F}_{\ell m}=(|\mathcal{A}^{\infty}_{\mOde}|^2+\alpha_{\ell m}|\mathcal{A}^{H}_{\mOde}|^2)/2\pi q (m\hO)^2$.
	The energy and angular momentum of the orbit evolves adiabatically due to the gravitational back-reaction effect over timescales $\sim T_i$ (see \autoref{adiabatic}). Here, we assume that the secondary object's mass, spin, and internal structure remain unaltered during evolution. The evolution of the orbital radius and phase as a result of the back-reaction effect is dictated by the following expression \cite{Pani}
	\begin{align}
		\frac{d\hr}{d\hht}&=-q\,\mathcal{F}(\hr)\left(\frac{d\hE}{d\hr}\right)^{-1},\label{r_eqn}\\
		\frac{d\phi}{d\hht}&=\hO\left(\hr(\hht)\right)~,\label{phi_eqn}
	\end{align}
	where, the expression of $\hE$ and $\hO$ as a function of $\hr$ is given in \autoref{circular} and \autoref{orb_freq} respectively. The solution of \autoref{phi_eqn} gives the expression for instantaneous orbital phase, which is related to the dominant mode gravitational-wave phase by $\Phi_\textrm{GW}(\hht)=2\phi(\hht)$. As discussed earlier, the adiabatic approximation breaks down as the object crosses the ISCO radius. Since we focus on the adiabatic evolution of the orbit, we consider the evolution in the domain $\hr\in (\hr^{\textrm{ini}}, \hr^{\textrm{isco}})$. Here, $\hr^{\textrm{ini}}$ is the starting point of the inspiral.
	\begin{figure*}[th]
		\minipage{0.46\textwidth}
		\includegraphics[width=\linewidth]{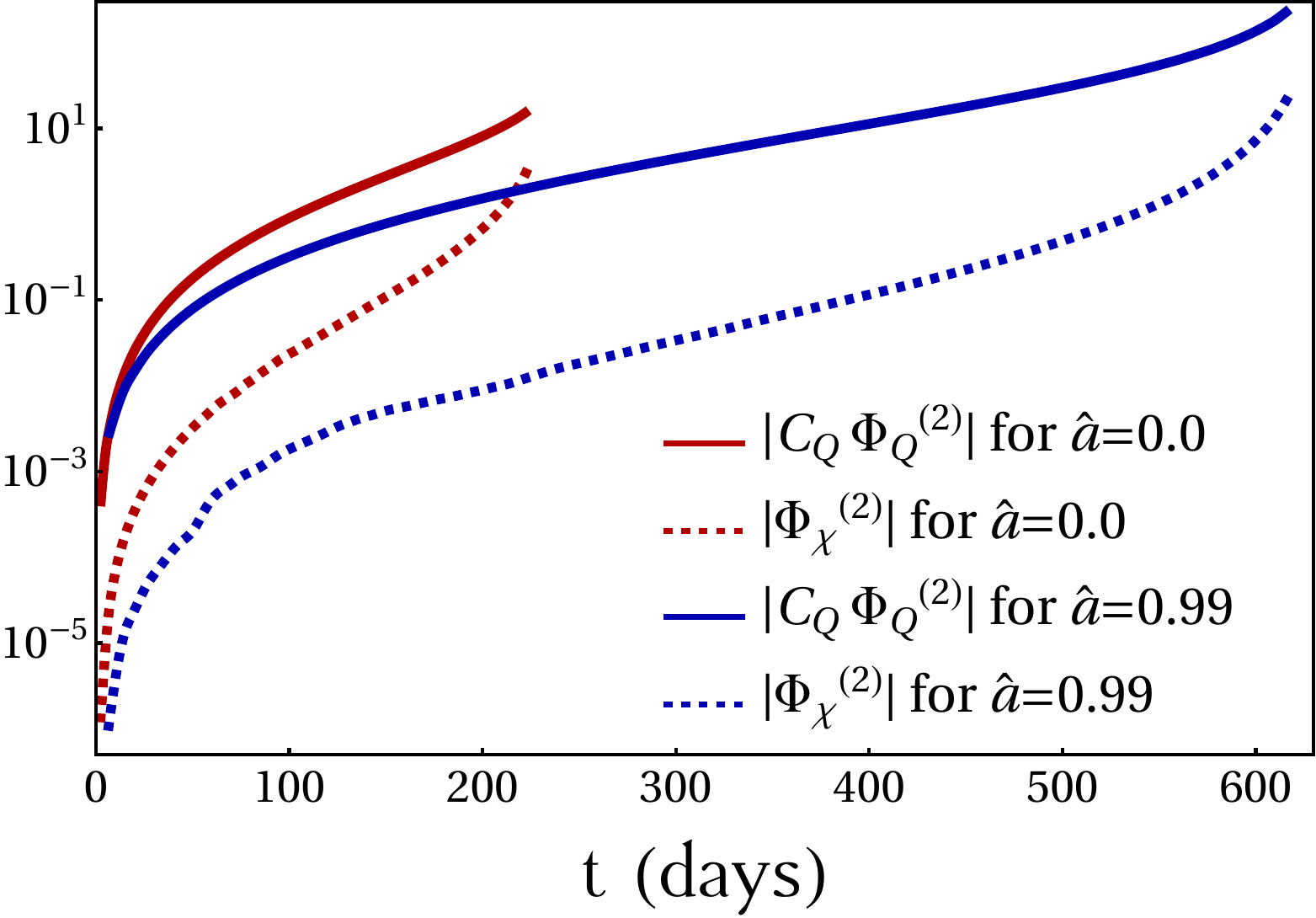}
		\endminipage\hfill
		\minipage{0.50\textwidth}
		\includegraphics[width=\linewidth]{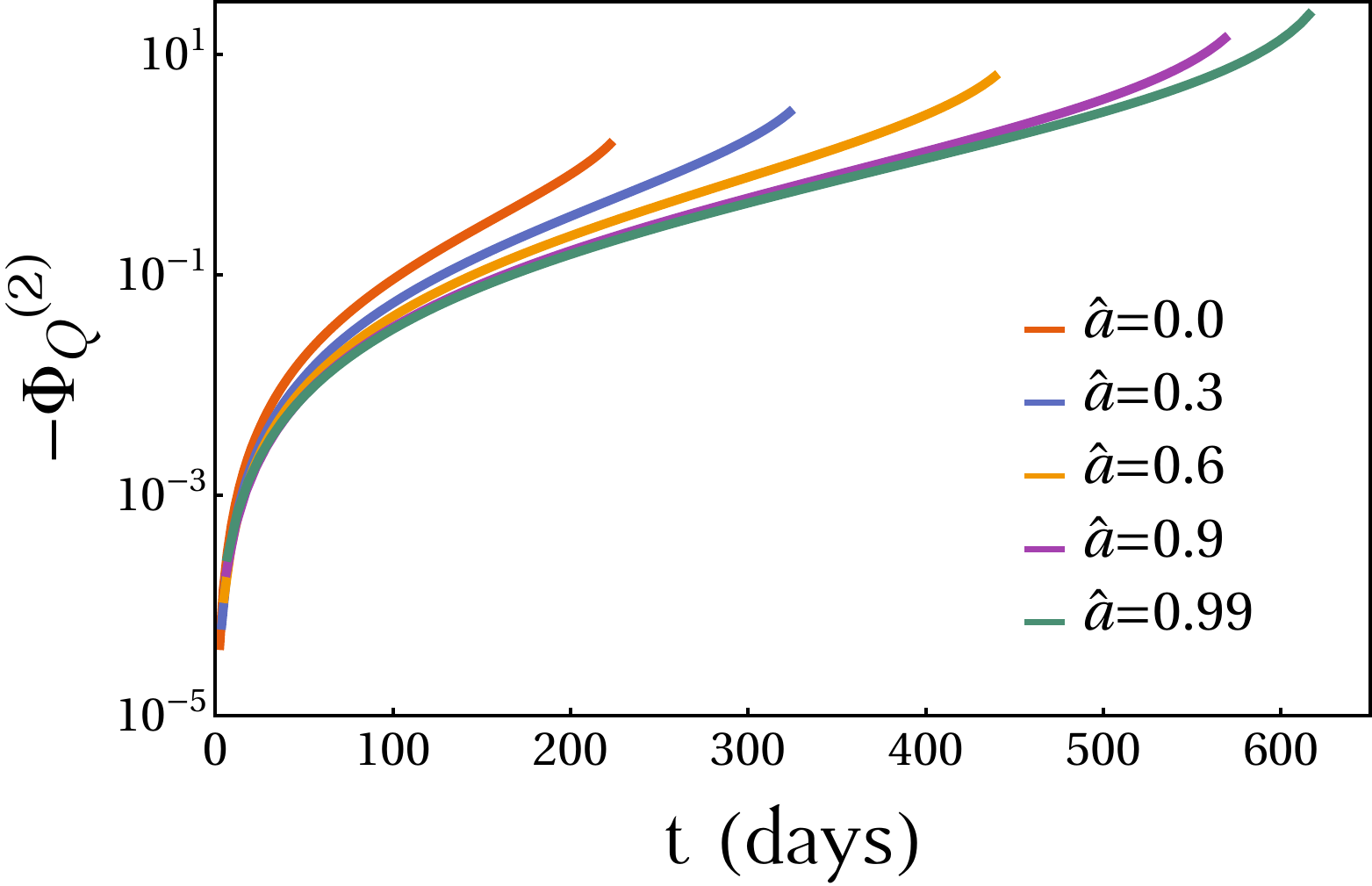}
		\endminipage
		\caption{Left: A comparison plot between spin-induced quadrupolar deformation term $C_Q\Phi_Q^{(2)}$ (represented by solid lines) and second order spin correction term $\Phi_\chi^{(2)}$ (represented by dashed lines) as a function of $t$ for stable, prograde orbit and $\ha=0.0$ and $\ha=0.99$. Right: The plot of $\Phi_Q^{(2)}$ as a function of $t$ for different values of $\ha$. In both of these plots, we consider the inspiral of a $1~M_\odot$ compact object with $C_Q=10$ into a supermassive black hole of mass $M=10^4~M_\odot$.
		}\label{fig_phi2_a}
	\end{figure*}
	\section{Numerical method and results}\label{Sec_5_Result}
	In this section, we briefly describe the numerical methods implemented to calculate the energy flux $\mathcal{F}$. 
	One of the main tasks to do so is to find the solutions of homogeneous Teukolsky equation, $R^{\In}_{\mode}$ and $R^{\up}_{\mode}$. Here, we have considered two different methods to calculate these functions: (i) Mano-Suzuki-Takasugi (MST) method \cite{10.1143/PTP.95.1079, 10.1143/PTP.112.415, 10.1143/PTP.121.843} as implemented in \texttt{Mathematica} package \texttt{Black hole Perturbation Toolkit} \cite{BHPT}, (ii) Sasaki-Nakamura (SN) method as described in \cite{10.1143/PTP.67.1788, Pani}. This is because the MST method, albeit faster, fails to deliver results with significant numerical precision for large values of $\ell$ (see \cite{Pani} for further discussion). We consider the SN method to calculate the energy flux in this scenario. In our calculation, we set the numerical precision to 22 significant digits. Furthermore, we have calculated the eigenvalue $\lambda_{\mode}$ and the eigenfunction $\Ang$ of the angular Teukolsky equation \autoref{ang_Teuk} using \texttt{Black hole Perturbation Toolkit} package. \par
	With $R^{\In,\up}_{\mode}$ in our hand, we can calculate the total flux using \autoref{tot_flux}. However, we need to truncate the infinite sum in those equations. Here, we set $\ell_{\textrm{max}}=22$ since the contribution of terms beyond $\ell>\ell_{\textrm{max}}$ to the total flux is negligible. In \autoref{table1}, we have presented the fractional truncation error in energy flux $\Delta\mathcal{F}_{\textrm{trun}}=|\mathcal{F}^{\ell=23}-\mathcal{F}^{\ell=22}|/\mathcal{F}^{\ell=23}$ for $q=10^{-4}$, $C_Q=10$ and $\chi=2$ and different values of primary spin $a$. Here, $\mathcal{F}^{\ell=23}$ and $\mathcal{F}^{\ell=22}$ is the flux considering $\ell_{\textrm{max}}=23$ and $\ell_{\textrm{max}}=22$ respectively. The fractional truncation error is computed at $\hr^{\textrm{isco}}$. The errors are even smaller for $\hr>\hr^{\textrm{isco}}$.  We find that $\Delta\mathcal{F}_{\textrm{trun}}$ is practically independent of $C_Q$ and $\chi$. However, as seen from \autoref{table1}, it depends on $\ha$.
	For each value of $\ell$, $m$ varies from 1 to $\ell$ starting with $m=\ell$ mode, which is the leading contributor to the flux. However, to speed up the computation, we compare $m=\ell$ mode with $m=\ell-i$ ($i=1,2,...,\ell-1$) mode at $\hr^{\textrm{isco}}$ and neglect the contributions of the terms for which $\mathcal{F}_{\ell \ell}\gg\mathcal{F}_{\ell \ell-i}$, or equivalently $|\mathcal{F}_{\ell \ell-i}/(\mathcal{F}_{\ell \ell}-\mathcal{F}_{\ell \ell-i})|\ll 1$. Following \cite{Pani}, we truncate the $m$ series whenever $|\mathcal{F}_{\ell \ell-i}/(\mathcal{F}_{\ell \ell}-\mathcal{F}_{\ell \ell-i})|<10^{-6}$. This gives us the energy flux $\mathcal{F}$ as a function of $\hr,~\ha,~q,~\chi$, and $C_Q$.\par
\begin{table}[b]
	\begin{center}
	\def\arraystretch{1.0}      	
	\setlength{\tabcolsep}{2em}
	\begin{tabular}{c c c} 
	\hline
		\hline
		$\ha$ & $\Delta\mathcal{F}_{\textrm{trun}}$ & $\Delta \Phi_{\textrm{trun}}$ \\
		\hline
		$0.0$ & $1.5\times 10^{-9}$ & $1.4\times 10^{-5} $\\
	
		$0.3$ & $1.28\times 10^{-8}$ & $1.2\times 10^{-5} $ \\

		$0.6$ & $2.05\times 10^{-7}$& $9.9\times 10^{-6} $\\

		$0.9$ & $2.05\times 10^{-5}$& $3.2\times 10^{-6} $ \\
		
		$0.99$ & $1.02\times 10^{-4}$& $1.1\times 10^{-6} $\\
		\hline
		\hline
	\end{tabular}
	\end{center}
	\caption{Fractional truncation error in flux  $\Delta\mathcal{F}_{\textrm{trun}}$ and gravitational wave phase $\Delta \Phi_{\textrm{trun}}$ at $\hr^{\textrm{isco}}$ for different values of $\ha$. Here, we take $q=10^{-4}$, $C_Q=10$ and $\chi=2$. }\label{table1}
\end{table} 
Since we are interested in the adiabatic evaluation of the orbit, we calculate the flux $\mathcal{F}(\hr)$ in the range $\hr\in (\hr^{\textrm{ini}}, \hr^{\textrm{isco}})$ for different values of $\chi\in [-2,2]$ and fixed values of $\ha,~q$, and $C_Q$. Following \cite{Pani}, we choose the starting point of the inspiral $\hr^{\textrm{ini}}$ such that all the spinning objects have the same orbital frequency as a non-spinning ($\chi=0$) secondary object at $\hr=10$. Note that $\mathcal{F}$ for $\chi=0$ corresponds to the flux for a non-spinning secondary object which we represent by $\mathcal{F}^{(0)}$. We have fitted the difference $\mathcal{F}-\mathcal{F}^{(0)}$ with a quadratic polynomial of $\sigma$, $b_0+b_1 \sigma+b_2 \sigma^2$. We use the \texttt{Fit} function in \texttt{Mathematica} to fit the numerical data with the quadratic polynomial. We have checked that value of $b_0\sim 10^{-18} (10^{-10})$ at $\hr=\hr^{\textrm{ini}} (\hr^{\textrm{isco}})$, irrespective of the value of primary spin. Consideration of higher order polynomial does not change the order of magnitude of $b_0$. Thus the total flux can be written as
	\begin{equation}\label{flux_expand}
		\begin{aligned}
			\mathcal{F}(\hr,\sigma)=\mathcal{F}^{(0)}(\hr)+\sigma  \mathcal{F}^{(1)}(\hr)+\sigma^2 \mathcal{F}^{(2)}(\hr)+\mathcal{O}(\sigma^3)~,
		\end{aligned}
	\end{equation}
	where  $\mathcal{F}^{(1)}(\hr)$ and $\mathcal{F}^{(2)}(\hr)$ describe the linear and quadratic corrections to flux due to spin effect. In \autoref{fig_flux}, we show the $\mathcal{F}^{(0)},~\mathcal{F}^{(1)}$ and $\mathcal{F}^{(2)}$ for stable prograde orbits as a function of orbital radius $\hr$ for different values of $\ha$. Here, we take $q=10^{-4},~C_Q=10$.\par
	We can calculate the adiabatic evaluation of the orbit by integrating \autoref{r_eqn}. The integration starts at $\hr^{\textrm{ini}}$, which marks the beginning of the inspiral phase. The integration stops when the object reaches $\hr^{\textrm{end}}=\hr^{\textrm{isco}}+\varepsilon$. Here, we choose $\varepsilon=10^{-6}$. We obtain the instantaneous orbital phase $\phi(\hht)$ by replacing the solution of \autoref{r_eqn} in \autoref{phi_eqn} and solving it using Euler's method \cite{holmes2006introduction}.
	The instantaneous gravitational wave phase can be obtained through the relation $\Phi_{\textrm{GW}}(\hht)=2\phi(\hht)$. In \autoref{table1}, we present the truncation error in the gravitational wave phase  $\Delta\Phi_{\textrm{trun}}=|\Phi_{\textrm{GW}}^{\ell=23}-\Phi_{\textrm{GW}}^{\ell=22}|/\Phi_{\textrm{GW}}^{\ell=23}$ at $\hr^{\textrm{isco}}$ for the cut off value $\ell_{\textrm{max}}=22$ and for $q=10^{-4}$, $C_Q=10$ and $\chi=2$ and different values of primary spin $\ha$. As evident, the contribution of the terms beyond $\ell=\ell_{\textrm{max}}$ is negligible.
The gravitational wave phase can be expressed in the following form \cite{Pani}
	\begin{equation}\label{phi_expand}
		\begin{aligned}
			\Phi_{\textrm{GW}}(\hht)
			=\Phi^{(0)}(\hht)+\chi \Phi^{(1)}(\hht)+q\chi^2 \Phi^{(2)}(\hht)+\mathcal{O}(\sigma^3)~
		\end{aligned}
	\end{equation}
	where $\Phi^{(0)}(\hht)$ denotes the phase of a non-spinning secondary object whereas $\Phi^{(1)}(\hht)$ and $\Phi^{(2)}(\hht)$ represents a shift in phase due to secondary's spin respectively.
	We have calculated the value of $\Phi_{\textrm{GW}}(\hht)$ for different values of $\chi$, including $\chi=0$ which corresponds to $\Phi^{(0)}(\hht)$ . The information about the dependence of gravitational wave phase on secondary spin is encoded in $\Phi_{\textrm{GW}}(\hht)-\Phi^{(0)}(\hht)$. By fitting the $\Phi_{\textrm{GW}}(\hht)-\Phi^{(0)}(\hht)$ with a quadratic polynomial $a_0+a_1 \chi+a_2 \chi^2$, we obtain the value of $\Phi^{(1)}(\hht)$ and $\Phi^{(2)}(\hht)$.
	Again, we use the \texttt{Fit} function in \texttt{Mathematica} to fit the numerical data with the quadratic polynomial in \autoref{phi_expand}. 
  Note that we have checked that $a_0\sim 10^{-10} (10^{-6})$ at $\hr=\hr^{\textrm{ini}} (\hr^{\textrm{isco}})$, irrespective of the spin of the primary object. Consideration of higher order polynomial does not alter the values of $\Phi^{(0)}$, $\Phi^{(1)}$, and $\Phi^{(2)}$ significantly. The coefficient $\Phi^{(0)}$ only depends on the spin of the primary object and has no dependence on the SIQM parameter $C_Q$.
	In \autoref{fig_phi1}, we present the linear correction in phase as a function of time for different values of primary spin $\ha$ and secondary's SIQM parameter $C_Q$ for prograde orbits. As seen from the left panel of the figure, the time of adiabatic evolution up to the ISCO increases with the increase of $\ha$. Furthermore, the right panel of the figure shows $\Phi^{(1)}$ does not depend on the $C_Q$. Although we have shown here the result for $\ha=0.6$, we have explicitly checked that this behavior persists for other values of $\ha$. This confirms that the body's internal structure does not affect the gravitational phase up to linear order in the spin. \par
\begin{figure*}[t]
		\centering
		\minipage{0.33\textwidth}
		\includegraphics[width=\linewidth]{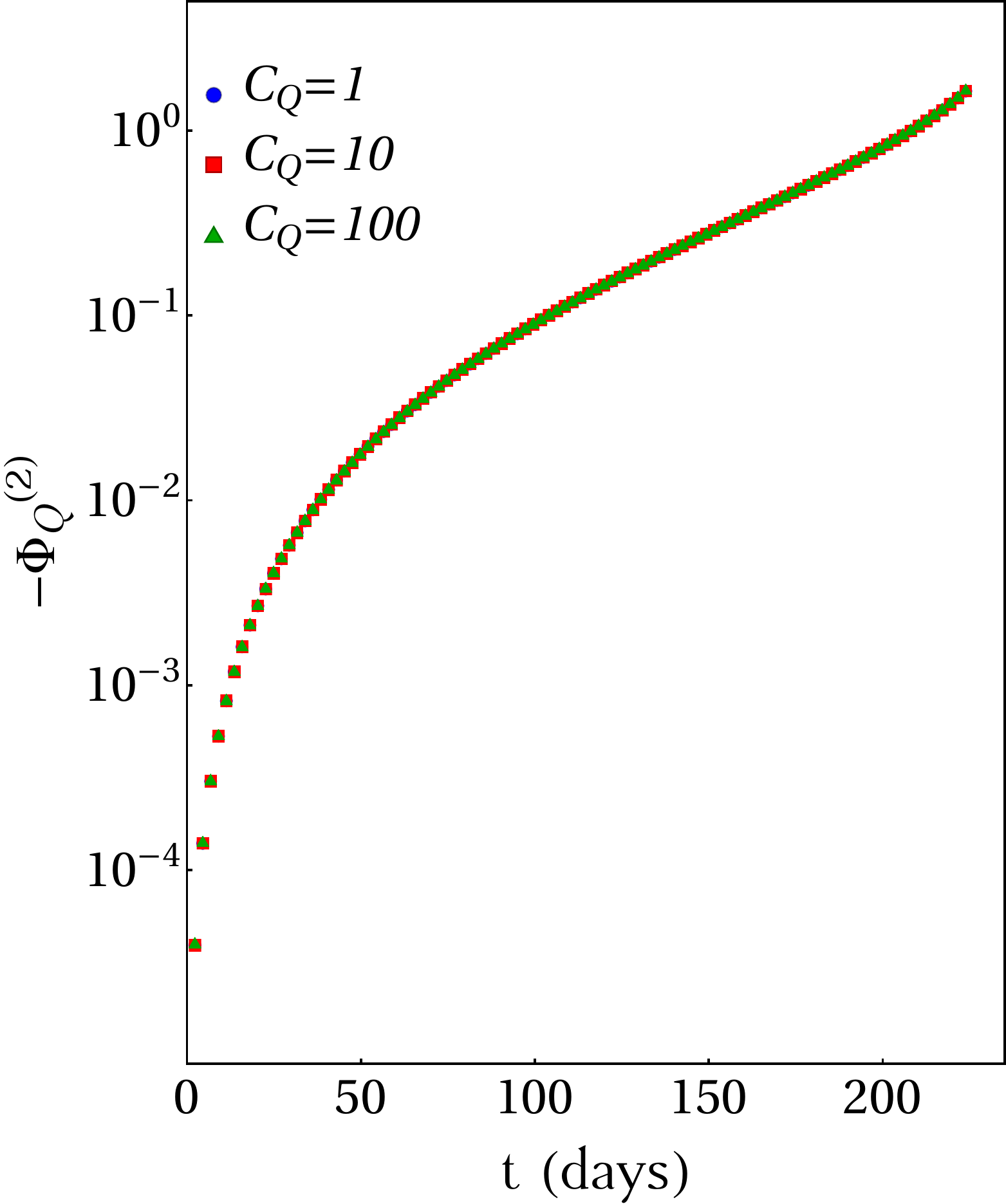}
		\endminipage\hfill
		\minipage{0.33\textwidth}
		\includegraphics[width=\linewidth]{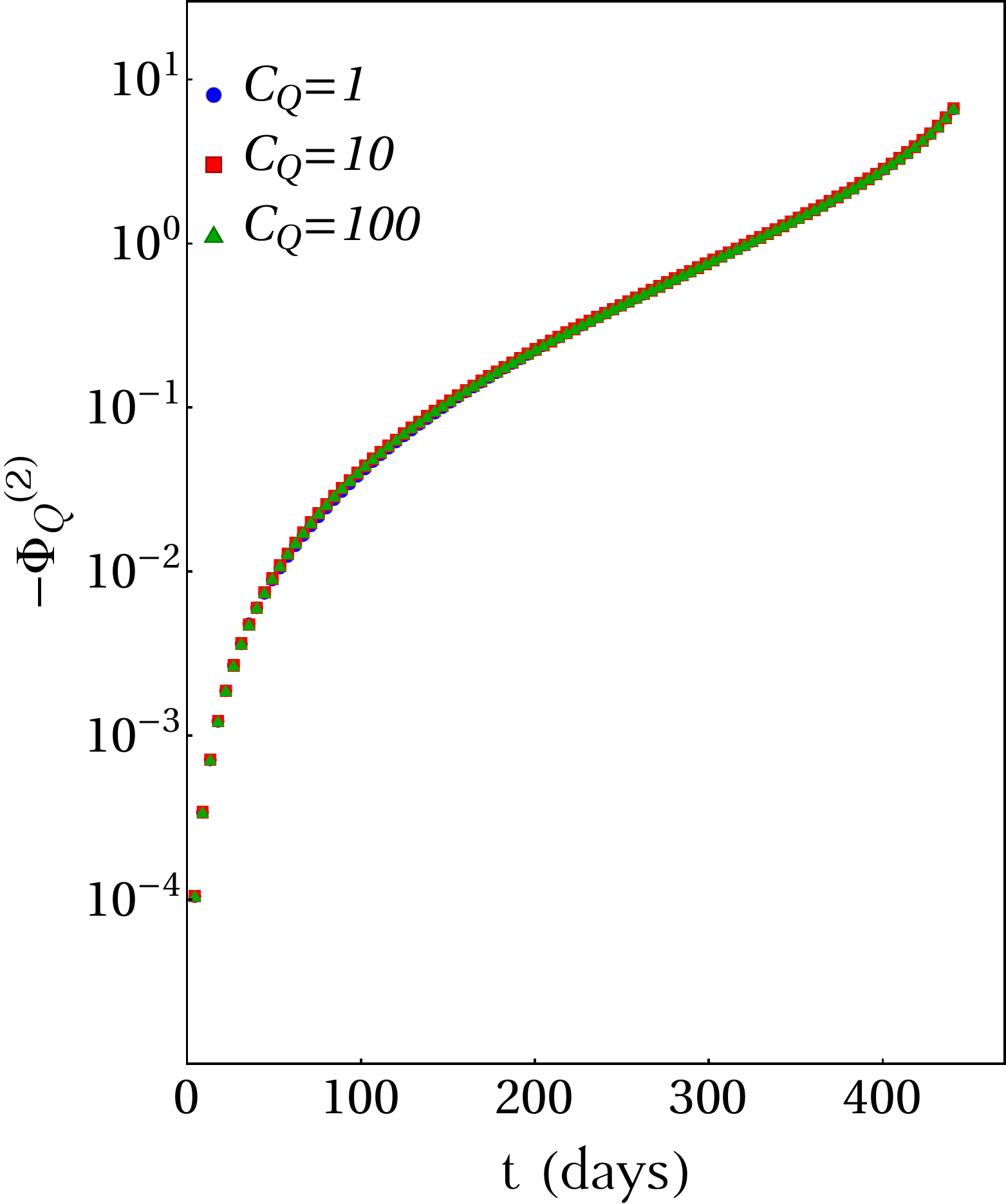}
		\endminipage\hfill
		\minipage{0.33\textwidth}
		\includegraphics[width=\linewidth]{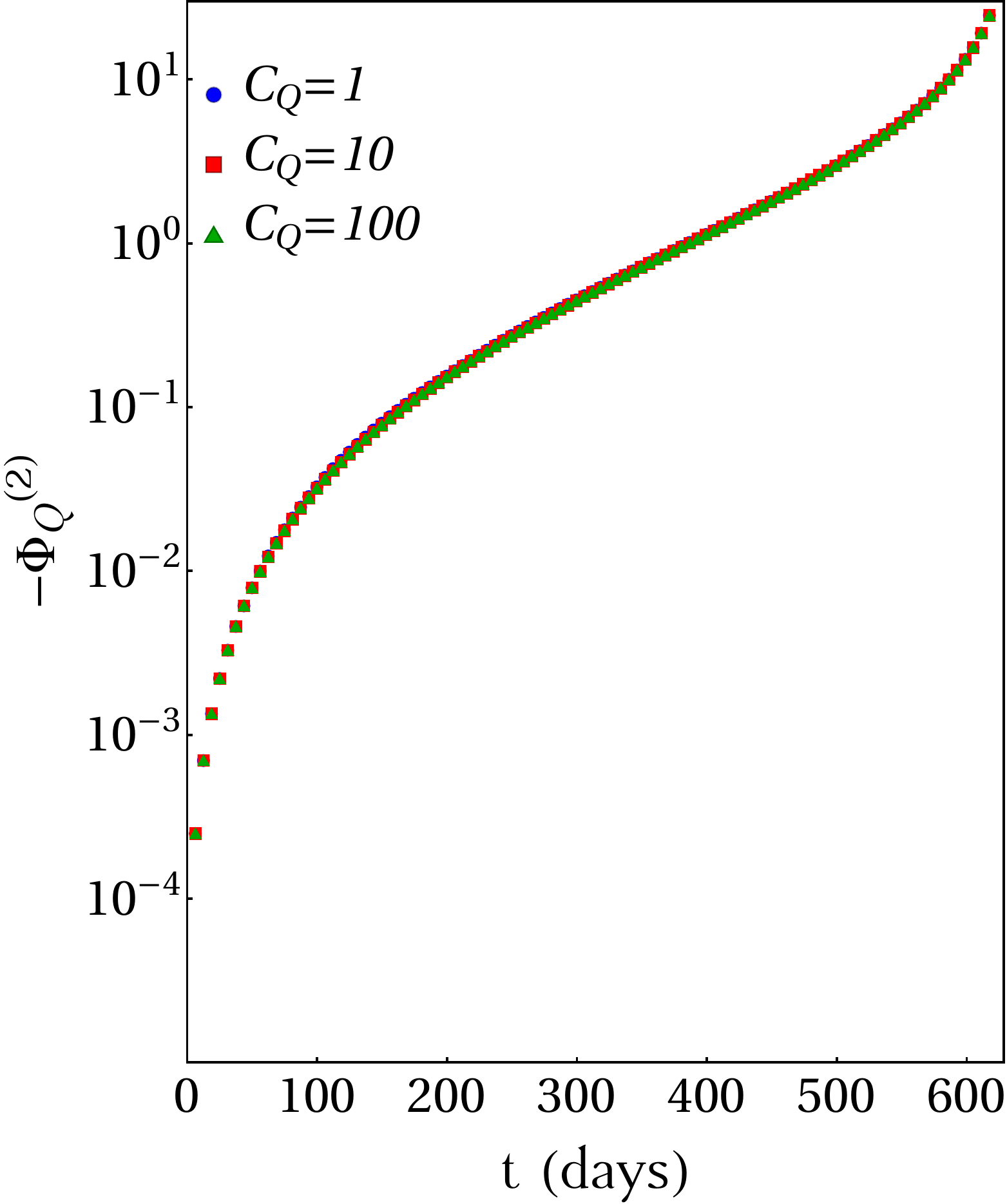}
		\endminipage
		\caption{The plot of $\Phi_Q^{(2)}$ as a function of $t$  for stable, prograde orbits for $\ha=0.0$ (left panel), $\ha=0.6$ (middle panel), and $\ha=0.99$ (right panel). 
		Here, we consider the inspiral of a $1~M_\odot$ compact object into a supermassive black hole of mass $M=10^4~M_\odot$.  As evident, $\Phi_Q^{(2)}$ is independent of $C_Q$. }\label{fig_phi2}
	\end{figure*}

	To see the behaviour of quadratic corrections in $\Phi_{\textrm{GW}}(t)$, we start with the following ansatz,
	\begin{equation}\label{phi_2}
		\begin{aligned}
			\Phi^{(2)}(\hht)=\Phi_\chi^{(2)}(\hht)+C_Q \Phi_Q^{(2)}(\hht)~.
		\end{aligned}
	\end{equation}
	Here, $\Phi_\chi^{(2)}(\hht)$ is the quadratic correction to GW phase for a spinning undeformed object, whereas $C_Q \Phi_Q^{(2)}(\hht)$ is the correction for the same due to spin-induced quadrupolar deformation. In the left panel of \autoref{fig_phi2_a}, we show the magnitude of these terms for $\ha=0$ and $\ha=0.99$.  $C_Q \Phi_Q^{(2)}(\hht)$ is represented by solid curves whereas $\Phi_\chi^{(2)}(\hht)$ is represented by dashed lines. Here, we consider that a $1~M_\odot$ compact object is inspiraling into a $10^4~M_\odot$ supermassive black hole. The right panel of \autoref{fig_phi2_a} shows the dependence of $\Phi_Q^{(2)}(t)$ on $\ha$.
	In \autoref{fig_phi2}, we show  the dependence of $\Phi_Q^{(2)}$ on the SIQM parameter $C_Q$ for $\ha=0.0$ (left panel), $\ha=0.6$ (middle panel), and $\ha=0.99$ (right panel). As can be seen from these plots, $\Phi_Q^{(2)}$ depends only on the spin of the primary object. Moreover, the quadratic correction to the accumulated phase at the end of the inspiral period $\Phi_Q^{(2)}(t_{\textrm{end}})$ increases with the increase of $\ha$. The values of $\Phi_Q^{(2)}(t_{\textrm{end}})$ for different values of $\ha$ is presented in \autoref{table2}.

	\subsection{ Measurability of the effects of quadrupolar deformation}
	
	In this section, we discuss whether the effects of quadrupolar deformation are strong enough for detection. Recently, Bonga et al. gave a rough estimate of the phase resolution for the EMRI measurement \cite{PhysRevLett.123.101103}. Considering the average signal-to-noise ratio (SNR) for LISA observation as $\sim 30$, they showed that distinction of two model waveforms is possible through LISA observation when the phase difference between these waveforms $\Delta \Phi$ is greater than $0.1$ radian. From \autoref{phi_expand} and \autoref{phi_2}, we find that the effect of quadrupolar deformation would be significant for LISA observations if the following condition is satisfied   
	\begin{equation}\label{final_detection}
		\Delta\Phi=q C_Q \chi^2\Phi_Q^{(2)}(t_{\textrm{end}})>0.1 ~\textrm{rad}~.
	\end{equation}
	where, $\Phi_Q^{(2)}(t_{\textrm{end}})$ is the quadratic correction to the accumulated phase at the end inspiral period. As discussed above, the value of $\Phi_Q^{(2)}(t_{\textrm{end}})$ depends only on the spin of the primary object. For convenience, we denote $\Phi_Q^{(2)}(t_{\textrm{end}})$ as $\Phi_{\textrm{Q,end}}^{(2)}(\ha)$ from now on to show its explicit dependence on $\ha$. 
	In \autoref{table2}, we present the value of $\Phi_{\textrm{Q,end}}^{(2)}(\ha)$ for different values of $\ha$. We use \autoref{final_detection} to check whether LISA can distinguish black holes from neutron stars and exotic objects like boson stars or gravastars.\par
\begin{figure*}[t!]
		\centering
		\minipage{0.33\textwidth}
		\includegraphics[width=\linewidth]{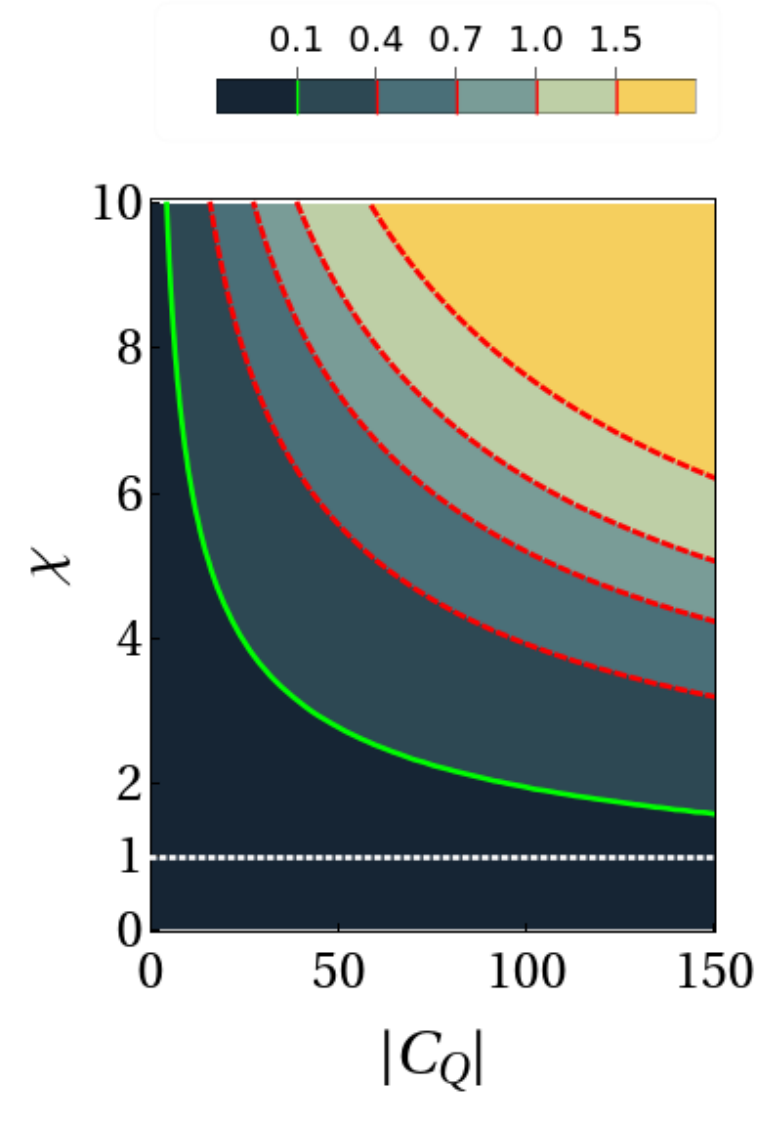}
		\endminipage\hfill
		\minipage{0.33\textwidth}
		\includegraphics[width=\linewidth]{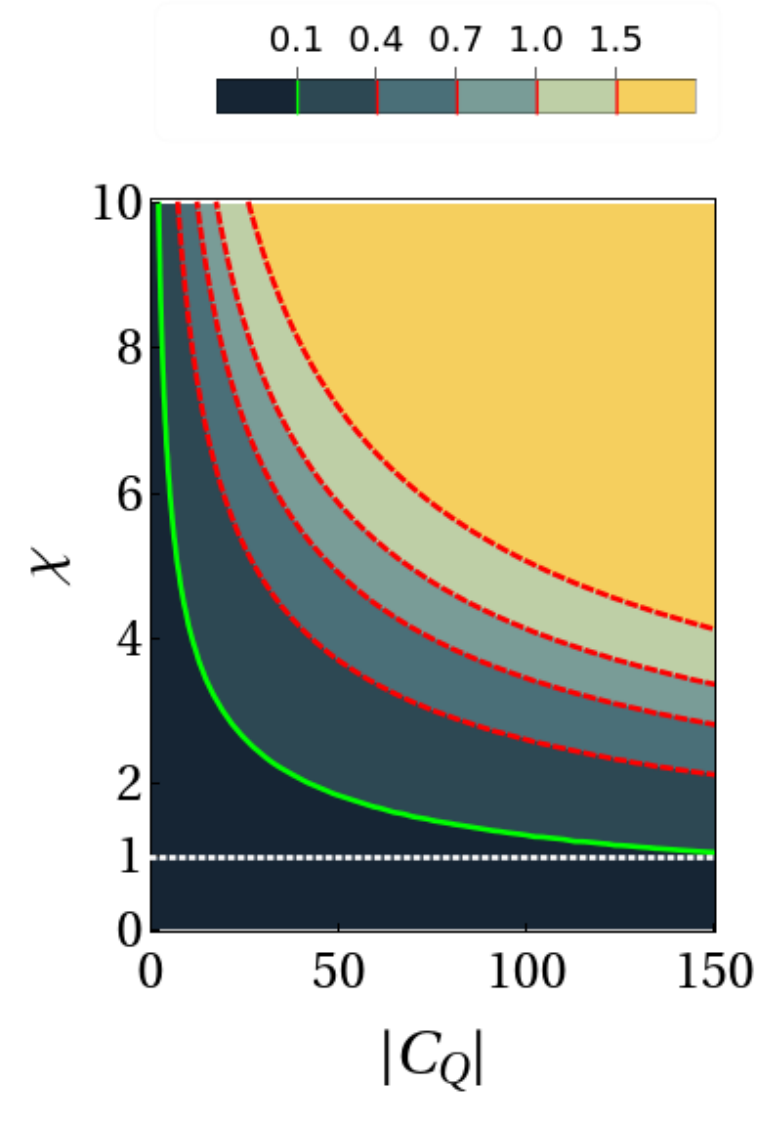}
		\endminipage\hfill
		\minipage{0.33\textwidth}
		\includegraphics[width=\linewidth]{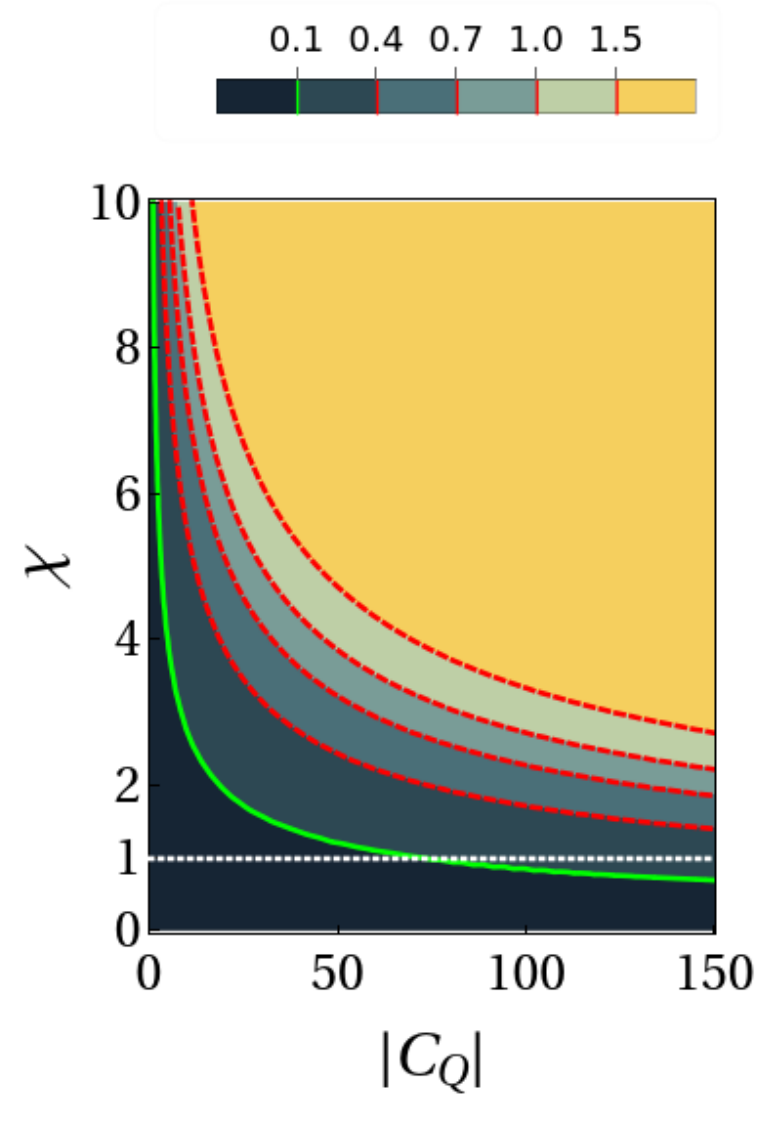}
		\endminipage
		\caption{The contour plot of quadratic corrections due to spin $\Delta \Phi$ in $(C_Q,\chi)$ space for $\ha=0.3$ (left panel), $\ha=0.6$ (middle panel), $\ha=0.9$ (right panel). Here, the green contour represents the threshold $\Delta\Phi=0.1$ rad for detecting the corrections due to quadrupolar deformation. The white dotted line corresponds to $\chi=1$. As evident, the parameter space for which the LISA can probe the effect of quadrupolar deformation is considerable, and it increases with the increase of $\ha$.}\label{fig_detect}
	\end{figure*}

\begin{table}[b]
	\begin{center}
	\def\arraystretch{1.3}      	
	\setlength{\tabcolsep}{4em}
	\begin{tabular}{c c} 
	\hline
		\hline
		$\ha$ & $\Phi_{\textrm{Q,end}}^{(2)}(\ha)$ \\
		\hline
$0$   & $1.63344$ \\
$0.3$ & $3.17133$ \\
$0.6$ & $6.65017$ \\
$0.9 $& $14.8259 $\\
$0.99$  &$ 24.5308$\\
		\hline
		\hline
	\end{tabular}
	\end{center}
	\caption{The quadratic  correction to accumulated phase at the end inspiral period due to quadrupolar deformation  $\Phi_Q^{(2)}(t_{\textrm{end}})$ for different values of $\ha$. }\label{table2}
\end{table} 
	In \autoref{fig_detect}, we present our main result. Here, we show the contour plot of $\Delta\Phi$ in the $(\chi, C_Q)$ plane for $\ha=0.3$ (left panel), $\ha=0.6$ (middle panel) and $\ha=0.9$ (right panel). Here, we fix the mass ratio as $q=10^{-4}$. As discussed earlier, the SIQM parameter $C_Q$ for Kerr black holes is 1, whereas it can take values $\sim 2-20$ for neutron stars and $\sim 10-150$ for boson stars. It can also take negative values for gravastar \cite{Uchikata:2015yma}. Thus, we vary the SIQM parameter in the range $C_Q\in(0,150)$. \par

The maximum value of the secondary's spin parameter $\chi_{\textrm{max}}$ depends on the nature of the object. For Kerr black holes, the spin parameter is restricted by the Kerr bound $\chi\leq \chi_{\textrm{max, Kerr}}\equiv 1$. However, for objects like neutron stars and white dwarfs, $\chi_{\textrm{max}}$ depends on the mass-shedding limit. Beyond this limit, the centripetal force on the particles at the star's surface surpasses the gravitational pull, and the star becomes unstable. This puts a limit on the angular velocity ($\Omega_{\textrm{max}}=\beta \sqrt{m_s/R^3}$) of the rotating star and consequently on its spin parameter \cite{PhysRevD.67.024005}
\begin{equation}\label{chi_maxval}
\chi_{\textrm{max}}\equiv \frac{I \Omega_{\textrm{max}}}{m_s^2}=\alpha \beta\sqrt{\frac{R}{m_s}}~,
\end{equation}
where $m_s$ is the mass and $I=\alpha m_s R^2$ is the moment of inertia of the object, and  $ R$ is its radius. The parameter $\alpha$ and $\beta$ depends on the stellar model. For instance, we get $\alpha=0.2045 ~(0.11804)$ and $\beta=0.5365~(0.46111)$ for isolated, self-gravitating fluid model with polytropic index $n=1.5~(2.5)$ \cite{Lai:1993ve, 1994ApJ...420..811L}. As can be seen from the above equation, $\chi_{\textrm{max}}$ depends strongly on the mass-radius relation of the object. Most of the neutron star mass-radius relation models estimates $\chi_{\textrm{max}}\approx 0.7$ \cite{mass_shedding, Lo_2011}. The fastest-spinning millisecond pulsar J1748+2446ad has a rotational frequency of  716 Hz \cite{Hessels:2006ze}. It has been noted that the spin parameter of this pulsar can be as large as $\chi\sim 0.5$, depending on its mass and equation of state \cite{Stein:2013ofa}. By taking its mass and radius as $m_s=2M_\odot$ and $R=16$ km \cite{Hessels:2006ze}, we find that value of the spin parameter for this object is $\chi=0.43707$. Here, we use  $\alpha=0.237+0.674 ~(m_s/R)+4.48~ (m_s/R)^4$ following \cite{Lattimer:2004nj}.  
\par
\begin{figure*}[t!]
		\minipage{0.50\textwidth}
		\includegraphics[width=\linewidth]{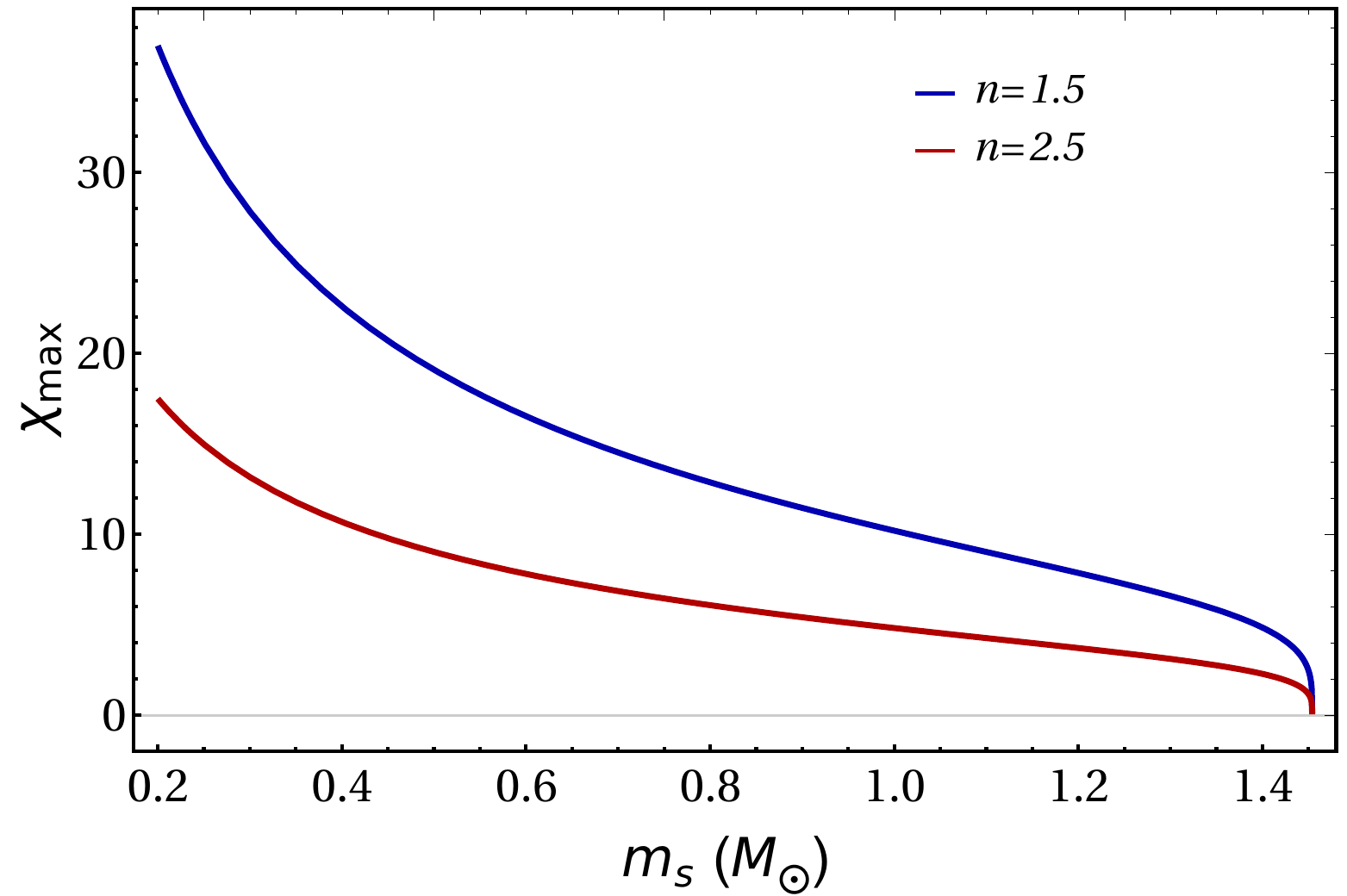}
		\endminipage\hfill
		\minipage{0.50\textwidth}
		\includegraphics[width=\linewidth]{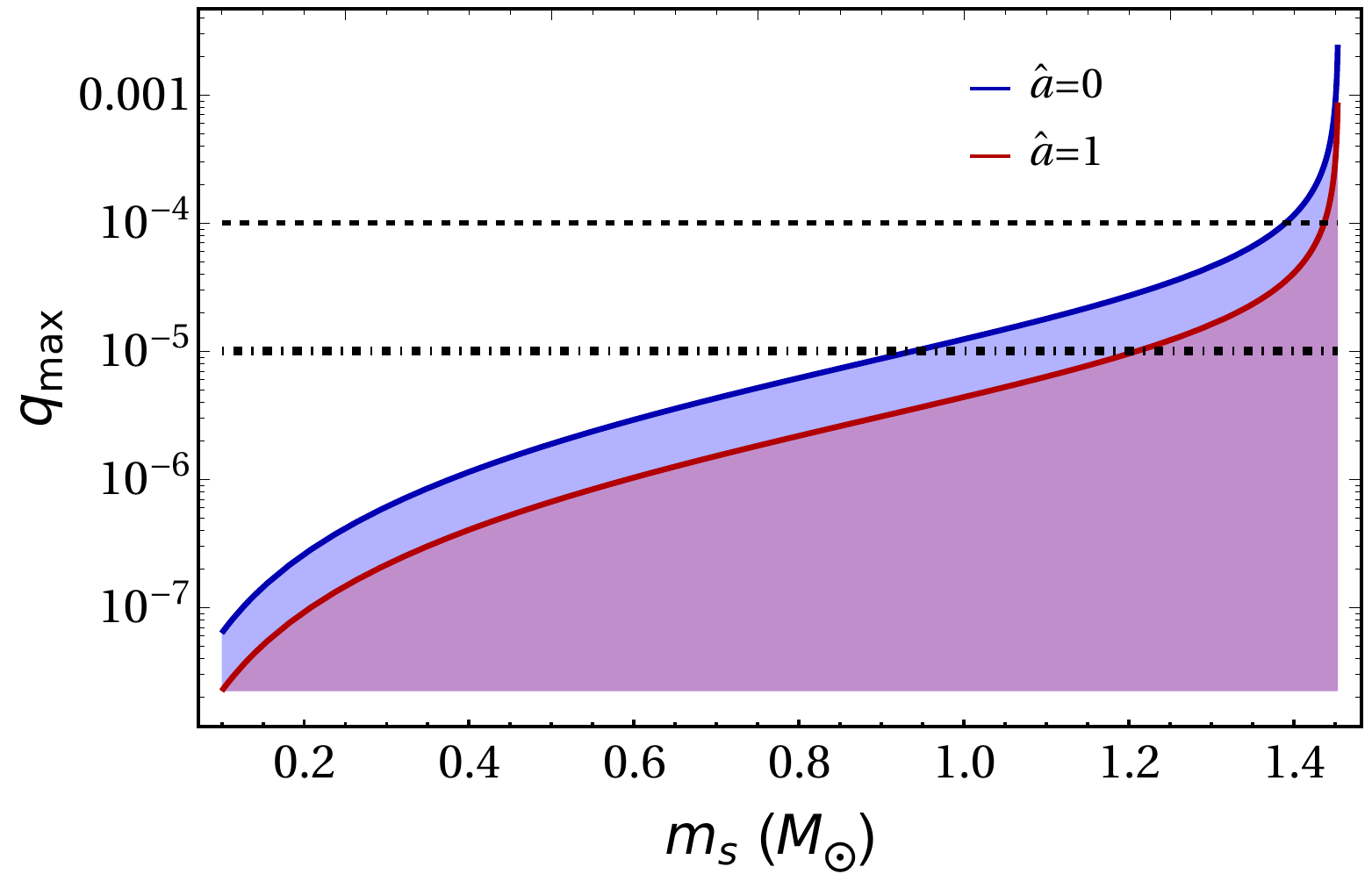}
		\endminipage
		\caption{Left: The plot of $\chi_{\textrm{max}}$ as a function of the mass of the white dwarf for different values of polytropic index $n$. Right: A simplistic order-of-magnitude estimate of the maximum value of mass ratio $q_{\textrm{max}}$ of a white dwarf-EMRI system below which there is no tidal disruption (represented by the shaded region) as a function of white dwarf mass for $\ha=0$ and $\ha=1$ (considering co-rotating orbits). Here, the black dashed line represents $q=10^{-4}$, whereas the black dot-dashed line represents the $q=10^{-5}$. As can be seen from the plot, white dwarfs with mass  $\sim M_\textrm{Ch}$ can withstand the tidal disruption of the primary object for $q=10^{-4}$. However, with smaller mass ratio values, white dwarfs with smaller masses can withstand tidal disruption.}\label{chi_max}
	\end{figure*}

However, for white dwarfs, $R$ strongly depends on the mass of the object. Following \cite{Lai:1993ve}, we rewrite \autoref{chi_maxval} in the following manner $\chi_{\textrm{max}}=\gamma \sqrt{R_0/m_s}~,$ where $R_0$ is the radius of a non-rotating polytrope with equal mass $m_s$ and the parameter $\gamma$ is dependent on stellar model. For a self-gravitating fluid with polytropic index $n=1.5~(2.5)$, we get $\gamma=0.1660~(0.0785)$ \cite{Lai:1993ve}. To get an estimate of $\chi_{\textrm{max}}$, we consider the following mass-radius relation of a non-rotating white dwarf \cite{1972ApJ...175..417N}
\begin{equation}\label{M_R_WD}
\frac{R_0}{R_\odot}=0.01125\left(\frac{m_s}{M_\odot}\right)^{-1/3}f(m_s)^{1/4}~,
\end{equation}
where $f(m_s)=1-(m_s/1.454 M_\odot)^{4/3}$. Using the above relation, we can find the maximum value of the spin parameter of a white dwarf as \cite{PhysRevD.67.024005}
\begin{equation}
\chi_{\textrm{max}}=77.68 \gamma \left(\frac{m_s}{M_\odot}\right)^{-2/3}f(m_s)^{1/4}~,
\end{equation}
\In \autoref{chi_max}, we plot $\chi_{\textrm{max}}$ as a function of mass of the white dwarf for $n=1.5$  and $n=2.5$. As evident, white dwarfs can have $\chi>1$. As illustrated in   \cite{PhysRevD.67.024005}, $\chi_{\textrm{max}}$ value for more realistic white dwarf models (e.g. see \cite{2000ApJ...534..359G}) are expected to lie between these curves. Ref.~\cite{Otoniel:2020gls}  obtained a lower bound on the mass of the rapidly rotating white dwarf CTCV J2056–3014 as $m_s=0.56 M_\odot$ by considering its rotational period 29.6 s is close to the mass-shedding limit. The radius is found to be $R=10965$ km. This leads to the value of the spin parameter as $\chi=20.95$. Here, we use $\alpha=0.2045$ corresponding to a self-gravitating isolated fluid model with polytropic index $n=1.5$.
Like white dwarfs, quark stars can also have dimensionless spin $\chi$ (slightly) larger than the unity\cite{ Lo_2011}. Furthermore, Chirenti and Rezzolla constructed stable gravastar models that can have large spin values $\chi\sim 1.2$ \cite{PhysRevD.78.084011}. Boson stars can also have $\chi>1$ \cite{Siemonsen:2020hcg, PhysRevD.55.6081, Vaglio:2022flq}. In \cite{PhysRevD.55.6081}, Ryan presented stationary, axisymmetric stable solutions of Einstein-Klein-Gordon equation for complex, self-interacting scalar fields with mass $m_{\textrm{SF}}$ and self-interaction parameter $\lambda_{\textrm{SF}}$. Considering strong self-interaction limit (i.e, $\lambda_{\textrm{SF}}\gg m_{\textrm{SF}}^{2}$), Ryan showed that the maximum allowed mass $M_{\textrm{Max}}^{\textrm{BS}}$ of a boson star (beyond which the configuration collapses to form a black hole) increases with the increase of its spin $\chi$. For instance, to obtain a boson star with mass $M_{\textrm{BS}}=0.15 \sqrt{\lambda_{\textrm{SF}}}/m_{\textrm{SF}}^{2}$, the object should be spun faster than $\chi=1.4$ \footnote{We like to emphasize that the rotating boson stars are only stable in restricted regions of parameter space. For a more detailed analysis of this (in) stability issue, readers are referred to \cite{Siemonsen:2020hcg}. There the authors made some studies of the stability of rotating boson stars by considering a few models. Allthough in the non-relativistic regime, i.e. $ \frac{\omega}{\mu} \lesssim 1, $  where $\omega$ and $\mu$ are the angular frequency and the mass parameter of the complex scalar field, these stars are stable, but they develop linear non-axisymmetric instability (NAI) in the relativistic regime \cite{Sanchis-Gual:2019ljs,Siemonsen:2020hcg}. For $m=1$ ($m$ being the azimuthal number) rotating mini boson star (described by a  scalar field with quadratic potential), authors of \cite{Siemonsen:2020hcg} found NAI for all values of the parameter space. On the other hand, when there are nonlinear interactions, apart from the usual mass terms, in the potential, NAI can be found when $\frac{\omega}{\mu}$ is greater than some critical values (which in turn related to the compactness parameter of the star) in the relativistic regime. One may refer to equation (15) and (16) of \cite{Siemonsen:2020hcg} for these critical values (as well as the Table (1) of \cite{Siemonsen:2020hcg} for the stable solutions along with the corresponding values of compactness parameters) for various interacting rotating boson star models.}.\par
	
 Moreover,  string theory predicts the existence of exotic compact objects, superspinars, that can breach the Kerr bound \cite{Gimon:2007ur}. It is still debatable whether these objects are stable against linear perturbation. This is because the nature of these objects is still unknown. Thus, we are still unsure about the boundary conditions needed to solve the perturbation equations. A study by Pani et al. shows that these objects are unstable if perfectly reflecting or perfectly absorbing boundary conditions are imposed \cite{PhysRevD.82.044009}. However, a more recent study shows that there exists an infinite set of boundary conditions for which these objects are stable against linear perturbation \cite{NAKAO2018410}. They concluded that we need more information about the physical nature of these objects to confirm their stability. For this study, we assume that these objects can exist in nature and are potential candidates for the secondary in the EMRI system. For our study, we set the parameter range for the secondary spin as $\chi\in (0,10)$.\par
  In \autoref{fig_detect}, the green contour line represents the threshold $\Delta\Phi=0.1$ rad for detecting the quadrupolar deformation. The plot shows that the parameter space that allows distinction between black holes and exotic compact objects like boson stars and superspinars is quite significant. The size of the parameter space which allows this distinction increases with the increase of primary spin $\ha$. \par
 To have a more qualitative understanding of the parameter space that allows us to distinguish the exotic compact objects, we can write \autoref{final_detection} as follows,
 \begin{equation}\label{ECO}
    \Delta\Phi^{\textrm{ECO}}=0.4411\left(\frac{q}{10^{-4}}\right) \left(\frac{C_Q}{50}\right)\left(\frac{\chi}{2}\right)^2\left(\frac{\Phi_{\textrm{Q,end}}^{(2)}(\ha)}{24.5308}\right)
\end{equation}
The reference parameters   $(q_\textrm{ref},~C_Q^\textrm{ref},~\chi_\textrm{ref},~\ha_\textrm{ref})^{\textrm{ECO}}=(10^{-4},50,2,0.99)$ are chosen to represent the values of these parameters for a typical EMRI system with an exotic compact object as its secondary. Choosing the value of parameters $(q,C_Q,\chi,\ha)$ at the reference point, we can easily check that it is possible to distinguish exotic compact objects like boson stars even for $\ha\sim 0.6$ (see \autoref{table2}).
However, for a smaller mass ratio, $q\sim 10^{-5}$, we can only distinguish black holes from very fast-spinning exotic compact objects for larger values of $\ha$.\par
It is improbable to distinguish between a black hole and a neutron star through EMRI observations. To see this, we write \autoref{final_detection} as follows, 
 \begin{equation}\label{NS}
    \Delta\Phi^{\textrm{NS}}=0.0216\left(\frac{q}{10^{-4}}\right) \left(\frac{C_Q}{20}\right)\left(\frac{\chi}{0.7}\right)^2\left(\frac{\Phi_{\textrm{Q,end}}^{(2)}(\ha)}{24.5308}\right)~.
\end{equation}

Here,  we choose the reference parameters as $(q_\textrm{ref},~C_Q^\textrm{ref},~\chi_\textrm{ref},~\ha_\textrm{ref})^{\textrm{NS}}$ as $(10^{-4},20,0.7,0.99)$. The above equation reflects the fact that even when we choose large values of SIQM parameter, primary spin ($\ha\sim 1$) and secondary spin (close to the mass-shedding limit $\chi\sim \chi_{\textrm{max}}$), the small mass ratio of the system does not allow us to identify the effect of spin-induced quadrupolar deformation.
\par 

When we consider objects like white dwarfs as possible candidates for the secondary \cite{Vazquez-Aceves:2021xwl}, we have to ensure that these objects can withstand the tidal disruption caused by the supermassive black hole \cite{Maguire:2020lad, Gourgoulhon:2019iyu}. This leads to the concept of the tidal-disruption radius \cite{Maguire:2020lad, PhysRevD.99.123025} 
\begin{equation}\label{TDR}
    R_t\approx R\left(\frac{M}{m_s}\right)^{1/3}~,
\end{equation}
 defined as a critical radius inside which the object is torn apart by the tidal forces of the primary. The expression for white dwarf radius $R$ is presented in \autoref{M_R_WD}. We provide a simplistic order-of-magnitude estimate on the tidal disruption radius. Furthermore, we consider only the objects that can reach the ISCO without tidal disruption, which requires that the tidal-disruption radius should lie inside the ISCO (consideration of dephasing up to tidal disruption radius would be quite interesting. However, we are not considering such a scenario here).
 Equating \autoref{TDR} with the ISCO radius, we can obtain the minimum value of the black hole mass $M_{\textrm{min}}$ beyond which there is no tidal disruption \cite{Maguire:2020lad, Gourgoulhon:2019iyu, PhysRevD.99.123025}. We can also define the maximum mass ratio as $q_{\textrm{max}}=m_s/M_{\textrm{min}}$ below which there is no tidal disruption. In the right panel of  \autoref{chi_max}, we plot $q_{\textrm{max}}$ as a function of white dwarf mass for two different primary configurations: (a) the primary is non-rotating, and (b) the primary is an extreme Kerr black hole, and the secondary object is co-rotating with the black hole. The shaded region in the plot signifies the parameter space for which there is no tidal disruption. The black dashed line represents $q=10^{-4}$. As evident from the plot, for $q=10^{-4}$, the white dwarf can withstand the tidal forces of the primary when its mass is very close to  Chandrasekhar mass $M_\textrm{Ch}=1.454 M_\odot$. However, the parameter space for the same is much larger for a smaller mass ratio (say for $q=10^{-5}$, represented by a black dot-dashed line in the plot).
 The SIQM parameter for the white dwarfs can take very large values $C_Q\sim 10^3-10^5$ \cite{Taylor:2019hle}. 
 Similar to \autoref{ECO} and \autoref{NS}, we can write \autoref{final_detection} as 
 \begin{equation}\label{WD}
    \Delta\Phi^{\textrm{WD}}=0.882\left(\frac{q}{10^{-6}}\right) \left(\frac{C_Q}{10^4}\right)\left(\frac{\chi}{2}\right)^2\left(\frac{\Phi_{\textrm{Q,end}}^{(2)}(\ha)}{24.5308}\right)
\end{equation}
to get a quantitative idea about the dephasing of gravitational waves due to the quadrupolar deformation of white dwarfs. Here,  we choose the reference parameters as $(q_\textrm{ref}, C_Q^\textrm{ref},\chi_\textrm{ref},\ha_\textrm{ref})^{\textrm{WD}}$ as $(10^{-6},10^4,2,0.99)$.
 \autoref{WD} suggests that LISA can probe the effect of quadrupolar deformation of the white dwarfs even for moderate values of primary and secondary spin (see \autoref{table2}).  \par
Another interesting candidate for the secondary could be the brown dwarfs. The value of $q_{\textrm{max}}$ for these objects is extremely small $\lesssim 10^{-8}$ \cite{Gourgoulhon:2019iyu, PhysRevD.99.123025}. However, the value of spin and SIQM parameters can be very large. For instance, consider three brown dwarfs 2MASS J0348-6022, 2MASS J1219+3128, and 2MASS J0407+1546. The physical parameters of these stars are presented in \autoref{tableBD} (see Table 5 of \cite{Tannock_2021}). Using the relation $\chi=I\omega_{\textrm{BD}}/m_s^2=2\pi \alpha R^2/(m_s P_{\textrm{BD}})$, we obtain spin parameter of these objects as $102.23$, $97.6$ and $66.4$ respectively. Here, $\omega_{\textrm{BD}}$ is the rotational frequency and $P_{\textrm{BD}}$ is the rotational period. Following \cite{Ni,ubjak_2020}, we consider $\alpha=0.275$. In order to calculate the SIQM parameter, we make use of the fact that spin and tidal deformation parameters are the same for a Newtonian star for any equation of state, i.e., $\bar{\lambda}^{(\textrm{rot})}=\bar{\lambda}^{(\textrm{tid})}$. Here, $\bar{\lambda}^{(\textrm{rot})}=(I/m_s^3)^2 C_Q$ is the dimensionless rotational Love number and $\bar{\lambda}^{(\textrm{tid})}=2k_2^{\textrm{(tid)}}R^5/(3 m_s^5)$ is the dimensionless tidal Love number with $k_2^{\textrm{(tid)}}$ as the tidal apsidal constant \cite{PhysRevD.88.023009}. A simple manipulation gives the expression for the SIQM parameter $C_Q=2k_2^{\textrm{(tid)}} R/(3m_s \alpha^2)$. The above relation gives the SIQM parameter for 2MASS J0348-6022, 2MASS J1219+3128 and 2MASS J0407+1546 as $2.6\times 10^{6}$, $2.5\times 10^{6}$ and $1.85\times 10^{6}$ respectively. Here, we consider polytropic equation of state with $n=1.5$, for which 
$k_2^{\textrm{(tid)}}=0.286$ \cite{BrookerOlle,2010A&A...514A..22H}. Similar to boson stars, neutron stars, and white dwarf cases, we write \autoref{final_detection} as 
 \begin{equation}\label{BD}
    \Delta\Phi^{\textrm{BD}}=45\left(\frac{q}{10^{-10}}\right) \left(\frac{C_Q}{2\times10^6}\right)\left(\frac{\chi}{80}\right)^2\left(\frac{\Phi_{\textrm{Q,end}}^{(2)}(\ha)}{3.17}\right)
\end{equation}
to get a quantitative idea about the dephasing due to brown dwarf deformation. Here,  we choose the reference parameters as $(q_\textrm{ref}, C_Q^\textrm{ref},\chi_\textrm{ref},\ha_\textrm{ref})^{\textrm{BD}}$ as $(10^{-10},2\times10^6,80,0.3)$. The above equation suggests that LISA can probe the spin-induced quadrupolar deformation effect of brown dwarfs.

\begin{table}[htb]
\def\arraystretch{1.3}      	
	\setlength{\tabcolsep}{1em}
\begin{adjustwidth}{}{}
\begin{equation*}
\begin{array}{c c c c} 
\hline
\hline
& \textrm{2Mass} & \textrm{2Mass} & \textrm{2Mass}\\
		\textrm{Object} & \textrm{J0348-6022}&  \textrm{J1219+3128} & \textrm{J0407+1546}\\
		\hline 
	\textrm{Mass} (M_{\odot})   & 0.041 & 0.047 & 0.067 \\
        \textrm{Radius} (R_{\odot}) & 0.093 & 0.100 & 0.100 \\
       \textrm{ Period (hr)} & 1.080 & 1.14 & 1.23\\
\hline
\hline
\bottomrule
\end{array}
\end{equation*}
\end{adjustwidth}
\caption{Physical parameters of brown dwarfs 2MASS J0348-6022, 2MASS J1219+3128 and 2MASS J0407+1546 \cite{Tannock_2021}.}
\label{tableBD}
\end{table}
\section{Conclusion and Discussion}\label{Sec_Conclusion}
	
	Detection of gravitational waves by LIGO-VIRGO detectors taught us a valuable lesson: accurately modelling the coalescence process is as vital as extracting accurate data to maximize the scientific return from the observation. Realistic modelling of the binary system is of utmost importance because LISA will observe hundreds of EMRI events each year. In this paper, we have considered a system where a spinning stellar-mass object orbits around a supermassive Kerr black hole in the equatorial plane and studied the system's orbital dynamics and the emitted gravitational radiation. Moreover, we considered the effect of spin-induced quadrupolar deformation of the secondary on the gravitational wave phase. The effect of quadrupolar deformation is often ignored from the expectation that the information about the effect gets suppressed by the tiny mass ratio of the system. In this paper, we have shown that the impact of quadrupolar deformation on the gravitational wave phase can be pretty significant for certain astrophysical objects; thus, ignoring the contribution of such effects can create considerable estimation biases. In the paper, we have provided an order of magnitude estimation of the possible identification of different astrophysical objects by LISA  through GW phase measurement by considering spin-induced quadrupolar deformation effects.\par
Moreover, our analysis shows that the gravitational signals from the EMRI system can distinguish different astrophysical objects. We show that the quadrupolar deformation adds a correction term $\Delta \Phi=qC_Q\chi^2\Phi_Q^{(2)}(t_\textrm{end})$ to total accumulated phase, where $\Phi_Q^{(2)}(t_\textrm{end})$ is the numerical parameter which depends only on the dimensionless spin $\ha$ of the central black hole. The no-hair theorem sets the value of the SIQM parameter to unity ($C_Q=1$) for a Kerr black hole. However, for other astrophysical objects, the parameter's value depends on their internal structure, ranging between $2-20$ for neutron stars and $10-150$ for boson stars, and can even take negative values for gravastars. Although the spin parameter of the black holes and neutron stars is restricted by Kerr bound and mass-shedding limit, respectively, it can take large values ($\chi>1$) for objects like boson stars, gravastars, superspinars, white dwarfs, and brown dwarfs. The effect of quadrupolar deformation would be significant for LISA observation for $\Delta\Phi>0.1$ rad \cite{PhysRevLett.123.101103}. In \autoref{fig_detect}, we have shown that the condition is satisfied for a large parameter space (in $\chi-C_Q$ plane). Moreover, the parameter space increases with the increase of $\ha$. This allows us to distinguish black holes from a large variety of astrophysical objects, including boson stars, superspinars, white dwarfs, and brown dwarfs. However, \autoref{NS} dictates that it is impossible to distinguish between a black hole and from neutron star from EMRI observations due to the small mass ratio of the system. However, 
the perturbation analysis as presented in this paper may remain valid for an intermediate-mass ratio inspiral (IMRI) system, a binary system with $q\approx 10^{-4}-10^{-2}$.
Furthermore, \cite{Wardell:2021fyy} considered second-order self-force effects to generate the gravitational waveforms. They found a good agreement between these perturbative and numerical relativity waveforms, even for comparable-mass binary systems. From \autoref{NS}, we can easily check that it is possible to distinguish black holes from neutron stars in such a scenario. Moreover, from \autoref{WD} and \autoref{BD}, we can see that the effect of spin-induced quadrupolar deformation on the gravitational wave phase for white dwarfs and brown dwarfs can be pretty significant even for smaller values of mass-ratio ($q\lesssim 10^{-6}$) and moderate values of primary and secondary spin. \par

However, to get a more accurate estimate, we need to perform a complete Fisher-matrix error analysis following \cite{Piovano:2021iwv}. This analysis will also show whether the measurement of the SIQM parameter correlates with other parameters.

\par
	A possible extension of this work is to study the effect of tidally-induced quadrupolar deformation due to gravito-electric and gravito-magnetic tidal forces on gravitational wave production \cite{PhysRevD.86.044033}. Like spin-induced quadrupolar deformation, tidal deformation contains information about the object's internal structure and thus can potentially distinguish different astrophysical objects \cite{PhysRevD.94.064015, Narikawa:2021pak, Saleem:2021vph}.  Moreover, for objects like white dwarfs and brown dwarfs, the tidal love number can be pretty large \cite{10.1093/mnras/stw2614}. Thus, we can hope to probe the effect of tidal deformation through LISA observations for these objects even when the object is not rapidly rotating. 
 Other possible extensions include the relaxation of this paper's assumptions, like equatorial circular orbit and aligned spin.
	Furthermore, as we discuss in the introduction, the second-order force terms can considerably affect the dynamics of the EMRI system over the long inspiral period. Thus, an exciting extension of the work includes the contribution of the second-order self-force effects \cite{PhysRevD.103.064048}. Several authors recently studied such effects for EMRI systems consisting of a massive Schwarzschild black hole and a point particle \cite{PhysRevLett.127.151102, PhysRevLett.124.021101, Wardell:2021fyy}. Interestingly, the waveforms constructed considering these effects have good agreement even with the numerical relativity waveforms for comparable mass binaries. So, it is important to consider this effect in the presence of spinning secondary, which we left for the future.  

	\section*{Acknowledgements}
	We thank Anand Sengupta and speakers of for useful discussion. The authors like to thank the speakers of the online conference funded by Shastri Indo-Canadian Institute's Shastri Conference \& Lecture Series Grant (SCLSG) ``Testing Aspects of General Relativity," held between 11-14th March, 2022, for helpful discussion.  M.R is supported by the postdoctoral fellowship (MIS/IITGN/PD-SCH/201415-006) by IIT-Gandhinagar and the National Post Doctoral Fellowship grant (Reg. No. PDF/2021/001234) by  SERB, Government of India. A.B. is supported by  Mathematical Research Impact Centric Support Grant (MTR/2021/000490), Start-Up Research Grant (SRG/2020/001380) by the Department of Science and Technology Science and Engineering Research Board (India) and Relevant Research Project grant (58/14/12/2021- BRNS) by the Board Of Research In Nuclear Sciences (BRNS), Department of Atomic Energy, India. Last but not the least, the authors also like to thank the anonymous referee for his/her comments in improving this manuscript.
	\appendix
	
	\section{Circular orbit, ISCO and angular frquency}\label{App:Circular}
	The condition for circular orbit is given by $V_{\textrm{eff}}=0$ and $dV_{\textrm{eff}}/dy=0$. We expand the equations up to quadratic order of spin $\sigma$. We seek solutions in the form given by \autoref{circular}. The equations for a non-spinning object are given by 
	\begin{equation}
		\begin{aligned}\label{E0_x0}
			&-2 \hat{a} \hat{E}_0 \hat{x}_0 y^2-y^2 \left(\hat{a}^2+\hat{x}_0^2\right)+\hE_0^2+2 \hat{x}_0^2 y^3+2 y-1=0~,\\
			&-2 \left(2 \hat{a} \hE_0 \hat{x}_0 y+\hat{a}^2 y+\hat{x}_0^2 \left(y-3 y^2\right)-1\right)=0~.
		\end{aligned}
	\end{equation}
	The solutions of these equations gives the value  $\{\hE_0,\hx_0\}$. Linear order corrections $\{\hE_1,\hx_1\}$ can be found by solving the following equations
	\begin{equation}
		\begin{aligned}\label{E1_x1}
			&2 \Big(\hat{a} \hE_0^2 y^2+\hat{x}_0 y^2 \left(-\hat{a} \hE_1+\hat{a} \hat{x}_0 y^3+2 \hat{x}_1 y-\hat{x}_1\right)\\&+\hE_0 \left(y^2 \left(-\hat{a} \hat{x}_1-3 \hat{x}_0 y+\hat{x}_0\right)+\hE_1\right)\Big)=0~,\\
			&\hat{a} \left(2 \hE_0^2+\hat{x}_0 \left(5 \hat{x}_0 y^3-2 \hE_1\right)-2 \hE_0 \hat{x}_1\right)\\&+\hat{x}_0 \left(-9 \hE_0 y+2 \hE_0+6 \hat{x}_1 y-2 \hat{x}_1\right)=0~.
		\end{aligned}
	\end{equation}
	The equation for quadratic corrections $\{\hE_2,\hx_2\}$ is given by
	\begin{equation}\label{E2_x2}
		\begin{aligned}
			&y^2 \big(\hat{a}^2 y^3 \left(C_Q \left(3 \hat{x}_0^2 y^2+1\right)+2\right)-2 \hat{a} \hat{x}_0 \left(\hE_2-2 \hat{x}_1 y^3\right)\\&+\hat{x}_0^2 y^3 \left((3-6 y) C_Q+y\right)+(1-2 y) y C_Q\\&+2 \hat{x}_1^2 y+2 \hat{x}_0 \hat{x}_2 (2 y-1)-\hat{x}_1^2+2 (1-2 y) y\big)\\&+2 \hE_1 y^2 \left(\hat{a} \left(2 \hE_0-\hat{x}_1\right)+\hat{x}_0 (1-3 y)\right)\\&+2 \hE_0 \left(\hE_2-y^2 \left(\hat{a} \left(\hat{x}_0 y^3+\hat{x}_2\right)+\hat{x}_1 (3 y-1)\right)\right)\\&+\hE_0^2 (2 y-1) y^2+\hE_1^2=0~,\\
			&\hat{a}^2 y^3 \left(C_Q \left(21 \hat{x}_0^2 y^2+5\right)+10\right)+2 \hat{a} \Big(-2 \hat{x}_1 \left(\hE_1-5 \hat{x}_0 y^3\right)\\&+\hE_0 \left(4 \hE_1-5 \hat{x}_0 y^3-2 \hat{x}_2\right)-2 \hE_2 \hat{x}_0\Big)\\&+y C_Q \left(3 \hat{x}_0^2 y^2 (5-12 y)-8 y+3\right)\\&+2 \Big(\hE_0 \hat{x}_1 (2-9 y)+\hE_1 \hat{x}_0 (2-9 y)+\hE_0^2 (3 y-1)\\&+3 y \left(\hat{x}_0^2 y^3+\hat{x}_1^2\right)+2 \hat{x}_0 \hat{x}_2 (3 y-1)\\&-\hat{x}_1^2+y (3-8 y)\Big)=0~.
		\end{aligned}
	\end{equation}
	The ISCO is requires additional condition, $d^2V_{\textrm{eff}}/dy^2=0$. For an non-spinning object (zero-th order in spin), this condition translates as
	\begin{equation}\label{y_0}
		\begin{aligned}
			-2 \hat{a} \hE_0 \hat{x}_0-\hat{a}^2+\hat{x}_0^2 \left(6 y_0-1\right)=0
		\end{aligned}
	\end{equation}
	Linear order corrections in spin is dictated by  
	\begin{equation}\label{y_1}
		\begin{aligned}
			&	2 \Big(\hat{a} \left(\hat{x}_0 \left(10 \hat{x}_0 y_0^3-\hE_1\right)-\hE_0 \hat{x}_1+\hE_0^2\right)\\&+\hat{x}_0 \left(\hE_0 \left(1-9 y_0\right)+3 \hat{x}_0 y_1+\hat{x}_1 \left(6 y_0-1\right)\right)\Big)=0
		\end{aligned}
	\end{equation}
	Quadratic correction follows the equation given below
	\begin{equation}\label{y_2}
		\begin{aligned}
			&10 y_0^3 \left(\hat{a}^2 \left(C_Q+2\right)-2 \hat{a} \hat{x}_0 \left(\hE_0-2 \hat{x}_1\right)+3 \hat{x}_0^2 C_Q\right)\\&+63\, \hat{a}^2 \hat{x}_0^2 y_0^5 C_Q-12 y_0^2 \left(-5 \hat{a} \hat{x}_0^2 y_1+C_Q+2\right)\\&+2 \hE_0 \left(2 \hat{a} \hE_1-\hat{a} \hat{x}_2-9 \hat{x}_0 y_1+\hat{x}_1\right)-2 \hat{a} \hE_2 \hat{x}_0-2 \hat{a} \hE_1 \hat{x}_1\\&+3 y_0 \Big(C_Q+2 \Big(-3 \hE_0 \hat{x}_1+\hat{x}_0 \Big(2 \hat{x}_2-3 \hE_1\Big)\\&+\hE_0^2+\hat{x}_1^2+1\Big)\Big)+15 \hat{x}_0^2 y_0^4 \left(1-6 C_Q\right)+2 \hE_1 \hat{x}_0\\&-\hE_0^2+6 \hat{x}_0^2 y_2+12 \hat{x}_0 \hat{x}_1 y_1-\hat{x}_1^2-2 \hat{x}_0 \hat{x}_2=0
		\end{aligned}
	\end{equation}
	Expanding the parameter $y$ in \autoref{E0_x0}, \autoref{E1_x1}, \autoref{E2_x2} as \autoref{ISCO} and solving them together with the conditions \autoref{y_0}, \autoref{y_1} and \autoref{y_2}, we obtain the values of $\{\hE_i,\hx_i,y_i\}$ ($i=0,1,2$). Replacing these values in \autoref{circular} and \autoref{ISCO}, we obtain the values of $\{\hE,\hx,y\}$.
	\section{The Teukolsky source term}\label{App:source_term}
	
In \autoref{Sec_4_GW_flux}, we have calculated the flux due to the gravitational wave. We now provide some more details in this appendix.
 \autoref{amp_def} gives the amplitude at the horizon and at infinity
	\begin{equation}
		\begin{aligned}
			\mathcal{Z}_{\mode}^{H,\infty}=\mathcal{C}_{\mode}^{H,\infty}\int_{\hr_+}^{\infty
			}d\hr\frac{R_{\mode}^{\In,\up}\mathcal{J}_{\mode}}{\Delta^2}\,,\nonumber
		\end{aligned}
	\end{equation}
	As discussed in the main text, $\mathcal{J}_{\mode}$ is the source term for the radial Teukolsky equation \autoref{rad_Teuk}, $R_{\mode}^{\In,\up}$ are the solution of homogeneous Teukolsky equation with the following boundary condition 
\begin{equation}\label{boundary_condition}
	\begin{aligned}
		R^{\In}_{\mode}\sim
		\begin{cases}
			B^{\textrm{out}}_{\mode}\hr^3~e^{i\ho\hr_{*}}+B^{\textrm{in}}_{\mode}\frac{1}{\hr}~e^{-i\ho\hr_{*}}~, & \hr\to \infty\\
			B^{\textrm{tran}}_{\mode}\Delta^2~e^{-i\kappa\hr_{*}}~, & \hr\to \hr_+
		\end{cases}\\
			R^{\up}_{\mode}\sim
	\begin{cases}
		D^{\textrm{tran}}_{\mode}\hr^3~e^{-i\kappa\hr_{*}}	~, & \hr\to \infty\\D^{\textrm{out}}_{\mode}\hr^3~e^{i\kappa\hr_{*}}+D^{\textrm{in}}_{\mode}\Delta^2~e^{-i\kappa\hr_{*}}
	~, & \hr\to \hr_+
	\end{cases}
	\end{aligned}
\end{equation}
where, $\kappa=(\ho-m\hO_+)$. The constant terms $\mathcal{C}_{\mode}^{H,\infty}$ are given by,
\begin{equation}\label{constant_term}
\mathcal{C}_{\mode}^{H}=\frac{1}{2i\ho B^{\textrm{in}}_{\mode}},\quad{\mathcal{C}_{\mode}^{\infty}}=\frac{B^{\textrm{tran}}_{\mode}}{2i\ho B^{\textrm{in}}_{\mode}D^{\textrm{tran}}_{\mode}},
\end{equation}
where following \cite{Pani}, we fix the value of the $B^{\textrm{tran}}_{\mode}$ and $D^{\textrm{tran}}_{\mode}$ as $B^{\textrm{tran}}_{\mode}=\frac{1}{d_{\mode}},\,{D}^{\textrm{tran}}_{\mode}=-\frac{4\ho^2}{c_0}.$
with
\begin{equation}\label{c_d}
	\begin{aligned}
c_0=&-12i\ho+\lambda_{\mode}(\lambda_{\mode}+2)-12\ha\ho(\ha\ho-m)\\
d_{\mode}=&2\sqrt{2\hr_{+}}\Big[(2-6i\ho-4\ho^2)\hr_{+}^2+(3i\ha m-4\\&+4\ha\ho m+6i\ho)\hr_{+}-\ha^2m^2-3i\ha m+2\Big]
	\end{aligned}
\end{equation}
and $B^{\textrm{in}}_{\mode}$ satisfies the following relation with the constant Wronskian $\mathcal{W}=2i\ho B^{\textrm{in}}_{\mode}D^{\textrm{tran}}_{\mode}$.\\
	In Teukolsky formalism, the source term is given by the following relation \cite{Pani}
		\begin{align}\label{source_1}
		\begin{split}
			\mathcal{J}_{l m \hat \omega}=\int d\hat t d\theta d\phi\, \Delta^2\, e^{i(\hat \omega \hat t-m\,\phi)}\Big(\mathcal{J}_{NN}+\mathcal{J}_{\bar M N}+\mathcal{J}_{\bar M\bar M}\Big),
		\end{split}
	\end{align}
	where,
\begin{align}
	\begin{split} \label{app004}
		\mathcal{J}_{NN}&=-\frac{2\sin(\theta)}{\Delta^2\rho^3\bar\rho}\Big[\Big(\mathcal{L}^{\dagger}_{1}-2\,i\,\hat{a}\,\rho\sin(\theta)\Big)\mathcal{L}^{\dagger}_{2}S^{\hat{a} \hat{\omega}}_{l m}\Big]T_{NN},\\
		\mathcal{J}_{\bar M N}=\partial_{\hat r}&\Big\{T_{\bar M N}\Big[\frac{4\,\sin(\theta)}{\sqrt{2}\rho^3\Delta}\Big(\mathcal{L}_2^{\dagger}S^{\hat{a}\hat{\omega}}_{l m}+i\,\hat{a}\,\sin(\theta)(\bar\rho-\rho)S^{\hat{a}\hat{\omega}}_{l m}\Big)\Big]\Big\}\\&+T_{\bar M N}\Big[\frac{4\,\sin(\theta)}{\sqrt{2}\rho^3\Delta}\Big\{\Big(i\,\frac{K}{\Delta}+\rho+\bar \rho\Big)\mathcal{L}_2^{\dagger}\,S^{\hat{a}\hat{\omega}}_{l m}\\&-\hat{a}\,\sin(\theta)\frac{K}{\Delta}(\bar\rho-\rho)S^{\hat{a}\hat{\omega}}_{l m}\Big\}\Big]\\
		\mathcal{J}_{\bar M \bar M}=\Big\{&\partial_{\hat r}^2\Big(T_{\bar M\bar M}\Big[-\frac{\bar \rho}{\rho^3}\Big]\Big)+\partial_{\bar r}\Big(T_{\bar M\bar M}\Big[-\Big(\frac{\bar\rho}{\rho^2}+i\,\frac{\bar\rho}{\rho^3}\frac{K}{\Delta}\Big)\Big]\Big)\\&+ T_{\bar M\bar M}\Big[\frac{\bar \rho}{\rho^3}\Big(\frac{\partial}{\partial \hat r}\Big(\frac{i\,K}{\Delta}\Big)-2\rho\frac{i\, K}{\Delta}+\frac{K^2}{\Delta^2}\Big)\Big]\Big\}\sin(\theta)S^{\hat{a}\hat{\omega}}_{l m},
	\end{split}
\end{align}
with
 \begin{align}
	\begin{split} \label{extrarev1}
&	K=((\hat r^2+\hat a^2)\,\hat \omega-\hat a \, m),\\&
		\rho=\frac{1}{\hat r-i\, \hat{a}\cos(\theta)},\quad \bar{\rho}=\frac{1}{\hat r+i\, \hat{a}\cos(\theta)},\\& \mathcal{L}^{\dagger}_s=\frac{\partial}{\partial\theta}-\frac{m}{\sin(\theta)}+\hat{a}\,\hat{\omega}\,\sin(\theta)+s\cot(\theta).\end{split}
\end{align}
and $	T_{ NN}$, $T_{ \bar M N}$ and $ T_{ \bar{M}\bar{M}}$ are the projection of energy-momentum tensor along the null tetrad i.e.,
	\begin{align}
	\begin{split}\label{SET_NP}
		&	T_{ NN}=n^{\mu}n^{\nu}e_{\mu(a)}e_{\nu(b)}T^{(a)(b)},\\& T_{ \bar M N}=\bar{m}^{\mu}n^{\nu}e_{\mu(a)}e_{\nu(b)}T^{(a)(b)}\\&\,\, T_{ \bar{M}\bar{M}}=\bar{m}^{\mu}\bar{m}^{\mu}e_{\mu(a)}e_{\nu(b)}T^{(a)(b)},
	\end{split}
\end{align}
where the null-Tetrads are are defined as,
\begin{align} \label{extrarev}
	\begin{split}
		& l^{\mu}=\sqrt{\frac{\Sigma}{\Delta}}(e^{\mu}_{(0)}+e^{\mu}_{(1)})=\frac{\sqrt{\Sigma}}{\Delta} \tilde{l}^{\mu},\\&
		n^{\mu}=\sqrt{\frac{\Sigma}{\Delta}}(e^{\mu}_{(0)}-e^{\mu}_{(1)})=\frac{\sqrt{\Sigma}}{\Delta} \tilde{n}^{\mu},\\&
		m^{\mu}=\bar{\rho}\sqrt{\frac{\Sigma}{2}}(e^{\mu}_{(2)}+i\,e^{\mu}_{(3)})=\bar{\rho}\sqrt{\Sigma} \tilde{m}^{\mu},\\&
		\bar{m}^{\mu}=\rho\sqrt{\frac{\Sigma}{2}}(e^{\mu}_{(2)}-i\,e^{\mu}_{(3)})=\rho\sqrt{\Sigma} \tilde{k}^{\mu}~.
	\end{split}
\end{align}	
\begin{widetext} 
The energy-momentum tensor as defined in \autoref{SET} takes the following form in the Tetrad frame,
\begin{align}
	\begin{split}
	T^{(a) (b) }&=\int \text{d$\tau $}\frac{\delta ^4(x-z(\tau))}{\sqrt{-g}}\left(p^{((a) }v^{(b))}-\frac{1}{3}J^{(c) (d) (e)  ((a) }R^{(b) )}{}_{(e)(c)(d) }\right)\\&-\int d\tau \frac{\delta ^4(x-z(\tau))}{\sqrt{-g}}\Big(e^{((a)}_{\alpha}e^{(b))}_{\beta}\nabla_{\gamma}[e^{\alpha}_{(a_1)}e^{\beta}_{(b_1)}]\,S^{\gamma (a_1)}v^{(b_1)}\Big)\\&-\frac{2}{3}\int d\tau\frac{\delta ^4(x-z(\tau))}{\sqrt{-g}}\Big(e^{((a)}_{\alpha}e^{(b))}_{\beta}\nabla_{\gamma}\nabla_{\delta}[e^{\alpha}_{(a_1)}e^{\beta}_{(b_1)}]\,J^{\gamma (a_1)(b_1)\delta}\Big)\\
		&-\int \text{d$\tau $}\nabla _{\gamma }\left(S^{\gamma ((a) }v^{(b))}\frac{\delta ^4(x-z(\tau))}{\sqrt{-g}}\right)\\&-\frac{2}{3}\int \text{d$\tau $}\nabla _{\gamma }\nabla _{\delta }\left(J^{\delta ((a) (b))\gamma }\frac{\delta ^4(x-z(\tau))}{\sqrt{-g}}\right).
	\end{split}
\end{align}
Then after simplifying we get,
\begin{align} \label{app1}
	\begin{split}
		T^{(a) (b) }=\int \text{d}\hat{t}\, &\Big(\frac{\delta ^4(x-z(\tau))}{\sqrt{-g}} \mathcal{P}^{(a)(b)}+\partial_{\gamma}\Big[\frac{\delta ^4(x-z(\tau))}{\sqrt{-g}}\mathcal{Q}^{\gamma (a)(b)}\Big]+\\& \partial_{\gamma}\partial_{\delta}\Big[\frac{\delta ^4(x-z(\tau))}{\sqrt{-g}}\mathcal{I}^{\delta (a)(b)\gamma}\Big]\Big),
	\end{split}
\end{align}
where,
\begin{align}
	\begin{split}
		\mathcal{P}^{(a)(b)}=\frac{1}{\dot{\hat{t}}}\Big[& v^{((a)}p^{(b))}-\frac{1}{3}J^{(c) (d) (e) ((a)}R^{(b))}{}_{(c) (d) (e)}+\omega_{(c)(a_1)}{}^{((a)}v^{(b))}S^{(c)(a_1)}-\omega_{(c)(b_1)}{}^{((a)}S^{(b))(c)}v^{b_1}\\&-\frac{4}{3}J^{(d)((a_1)(b_1) )(c)}\Big(-e^{\gamma}_{(c)}\nabla_{\gamma}\omega_{(d)(a_1)}{}^{((a)}\delta^{(b))}_{(b_1)}+\omega_{(c)(m)}{}^{((a)}\delta^{(b))}_{(b_1)}\omega_{(d)(a_1)}{}^{(m)}\\&-\omega_{(d_1)(a_1)}{}^{((a)}\delta^{(b))}_{(b_1)}\omega_{(c)(d)}{}^{(d_1)}+\omega_{(c)(a_1)}{}^{((a)}\omega_{(d)(b_1)}{}^{(b))}\Big)\Big],
	\end{split}
\end{align}
\begin{align}
	\begin{split}
		\mathcal{Q}^{\gamma(a)(b)}=-\frac{1}{\dot{\hat{t}}
		}S^{\gamma ((a)}v^{(b))},\quad \mathcal{I}^{\delta (a)(b)\gamma}=-\frac{2}{3\, \dot{\hat{t}}}J^{\delta (a)(b)\gamma},\quad \dot{\hat{t}}=\Big|\frac{d\hat{t}}{d\tau}\Big|.
	\end{split}
\end{align}
We have used the expression for the Ricci rotation coefficients in terms of the Tetrad is
\begin{align}
	\begin{split}
\omega_{(a)(m)}{}^{(n)}= \quad e^{\alpha}_{(a)}e^{\beta}_{(m)}\nabla_{\alpha}e_{\beta}^{(n)}
	\end{split}
\end{align}
and we have used the following identities, 
\begin{align}
	\begin{split}
	e^{\gamma}_{(c)}e^{(a)}_{\alpha}\nabla_{\gamma}e^{\alpha}_{(a_1)}=-\omega_{(c)(a_1)}{}^{(a)},\, e^{(a)}_{\alpha}e^{\delta}_{(d)}e^{\gamma}_{(c)}\nabla_{\gamma}\nabla_{\delta}e^{\alpha}_{(a_1)}=-e^{\gamma}_{(c)}\nabla_{\gamma}\omega_{(d)(a_1)}{}^{(a)}+\omega_{(c)(m)}{}^{(a)}\omega_{(d)(a_1)}{}^{(m)}-\omega_{(d_1)(a_1)}{}^{(a)}\omega_{(c)(d)}{}^{(d_1)}.
	\end{split}
\end{align}
Note that, $T^{(a)(b)}$ mentioned in the \autoref{app1} should be viewed as a \textit{tempered distribution} which acts on an arbitrary smooth function of the form $\tilde{h}(x)= h(r,\theta) e^{i\,(\hat\omega\, \hat t- m\, \phi)}.$ 	Then  after performing the integral over $\hat t,$
\begin{align} \label{app2}
	\begin{split}
		T^{(a)(b)}=&\frac{\delta^{(3)}}{\sqrt{-g}}\Big(\mathcal{P}^{(a)(b)}-\mathcal{Q}^{\hat{t} (a)(b)}\partial_{\hat{t}}+\mathcal{I}^{\hat{t}(a)(b)\hat{t}}\partial^2_{\hat{t}}\Big)\\&-\frac{2}{\sqrt{-g}}\partial_{\hat{i}}\Big(\mathcal{I}^{\hat{t} (a)(b)\hat{i}}\delta^{(3)}\Big)\partial_{\hat{t}}+\frac{1}{\sqrt{-g}}\partial_{\hat{i}}\Big(\mathcal{Q}^{\hat{i} (a)(b)}\delta^{(3)}\Big)\\&+\partial_{\hat{i}}\partial_{\hat{j}}\Big(\mathcal{I}^{\hat{i}(a)(b)\hat{j}}\frac{\delta^{(3)}}{\sqrt{-g}}\Big).
	\end{split}
\end{align}
$\hat{i}$ and $\hat{j}$ take the following values $\{\hat{r},\theta,\phi\},$  $\hat{r}=\frac{r}{M}$ and $\delta^{(3)}=\delta(r-\hat r(t))\delta(\theta-\theta(t))\delta(\phi-\phi(t))$ is the three-dimensional Dirac delta function.
\end{widetext} 
Also, we have used the following,
\begin{equation}
	\int f(y)\delta^{[n]}(y-x)\, dy= (-1)^n f^{[n]}(x), 
\end{equation}
where the superscript $[n]$ of the function denotes the number of derivatives. So if a smooth function $g$ vanishes at $\hat{t}=\pm \infty$, then 
\begin{align}
	\begin{split}
		& g \, \delta (\hat{t}-t_0) \Big|_{-\infty}^{\infty}=0 \rightarrow \int_{-\infty}^{\infty} \text{dt}\, \partial_t\,\Big(g\, \delta (\hat{t}-t_0)\Big)=0,\\&
		\int_{-\infty}^{\infty} \text{d}\hat{t}\, \partial_{\hat{t}}\, g\, \delta (\hat{t}-t_0)=-\int_{-\infty}^{\infty} \text{d}\hat{t}\, g\, \partial_{\hat{t}} \delta(\hat{t}-t_0).
	\end{split}
\end{align}
We have used the last equality to obtain \autoref{app2}.
\par
	Next we have to calculate the quantities in \autoref{SET_NP}. In order to do so, we consider the smooth function  $\tilde{h}(x).$ 
	Then,
	\begin{align}
		\begin{split}\label{app001}
			T_{NN} \tilde{h}(x)=\,&\mathcal{N}_{NN}\Big[\delta^{(3)}\Big(\mathcal{P}_{\tilde{N}\tilde{N}}-i\,\hat{\omega}\,\mathcal{Q}^{\hat{t}}{}_{\tilde{N}\tilde {N}}-\hat{\omega}^2\,\mathcal{I}^{\hat{t}}{}_{\tilde{N}\tilde{N}}{}^{\hat{t}}\Big)\tilde{h}(x)\Big]\\&+\mathcal{N}_{NN}\partial_{\hat{i}}\Big[\Big(\mathcal{Q}^{\hat{i}}_{\tilde{N}\tilde {N}}-2\,i\,\hat{\omega}\,\mathcal{I}^{\hat{t}}{}_{\tilde{N}\tilde{N}}{}^{\hat{i}}\Big)\delta^{(3)}\Big]\tilde{h}(x)\\&+\mathcal{N}_{NN}\partial_{\hat{i}}\partial_{\hat{j}}\Big[\mathcal{I}^{\hat{i}}{}_{\tilde N \tilde N}{}^{\hat{j}}\delta^{(3)}\Big]\tilde{h}(x),
		\end{split}
	\end{align}
	where, $\mathcal{N}_{NN}=\frac{\Sigma}{\Delta\sqrt{-g}}.$ Also, we have used the  notation for the tilde indices following the similar notation mentioned in (\ref{SET_NP}), e.g $\mathcal{P}_{\tilde N\tilde N}= \tilde n^{\mu}\tilde n^{\nu} e_{\mu (a)}e_{\nu (b)}\mathcal{P}^{(a)(b)},$ where $\tilde n^{\mu}$ is defined in (\ref{extrarev}).
	Finally we can write it in the following compact form,
	\begin{equation}
		T_{NN}\tilde{h}(x)=\delta^{(3)} D^{\Omega}_{\tilde N\tilde N}\Big[\mathcal{N}_{NN}\tilde{h}(x)\Big]+D^{r}_{\tilde N\tilde N}\Big[\mathcal{N}_{NN}\tilde{h}(x)\Big],
	\end{equation}
	where,
	\begin{align}
		\begin{split}\label{app02}
			D^{\Omega}_{\tilde N\tilde N}\Big[\mathcal{N}_{NN}\tilde{h}(x)\Big]=&\Big[ \mathcal{P}_{\tilde{N}\tilde{N}}-i\,\hat{\omega}\,\mathcal{Q}^{\hat{t}}{}_{\tilde{N}\tilde {N}}-\hat{\omega}^2\,\mathcal{I}^{\hat{t}}{}_{\tilde{N}\tilde{N}}{}^{\hat{t}}\\&+i\,m\,\Big(\mathcal{Q}^{\phi}{}_{\tilde{N}\tilde{N}}-2\,i\,\hat{\omega}\,\mathcal{I}^{\hat{t}}{}_{\tilde N\tilde N}{}^{\phi}\Big)\\&-m^2\,\mathcal{I}^{\phi}{}_{\tilde N\tilde N}{}^{\phi}\Big] \mathcal{N}_{NN}\tilde{h}(x)\\&-\Big(\mathcal{Q}^{\theta}{}_{\tilde N\tilde N}-2\,i\,\hat{\omega}\,\mathcal{I}^{\hat{t}}{}_{\tilde N\tilde N}{}^{\phi}\\&+2\,i\, m\,\mathcal{I}^{\theta}{}_{\tilde N\tilde N}{}^{\phi}\Big)\partial_{\theta}\Big(\mathcal{N}_{NN}\tilde{h}(x)\Big)\\&+\mathcal{I}^{\theta}{}_{\tilde N\tilde N}{}^{\theta}\partial_{\theta}^2\Big(\mathcal{N}_{NN}\tilde{h}(x)\Big),
		\end{split}
	\end{align}
	\begin{align}
		\begin{split} \label{app3}
			D^{r}_{\tilde N\tilde N}\Big[\mathcal{N}_{NN}\tilde{h}(x)\Big]=&\Big\{\partial_{\hat{r}}\Big[\Big(\mathcal{Q}^{\hat{r}}{}_{\tilde N\tilde N}-2\,i\,\hat{\omega}\,\mathcal{I}^{\hat{t}}{}_{\tilde N\tilde N}{}^{\hat{r}}\\&+2\,i\, m\,\mathcal{I}^{\hat{r}}{}_{\tilde N\tilde N}{}^{\phi}\Big)\delta^{(3)}\Big]+\\&\partial_{\hat{r}}^2\Big[\mathcal{I}^{\hat{r}}{}_{\tilde N\tilde N}{}^{\hat{r}}\delta^{(3)}\Big]\Big\}\mathcal{N}_{NN}\tilde{h}(x)\\&-2\,\partial_{\hat{r}}\Big[\mathcal{I}^{\hat{r}}{}_{\tilde N\tilde N}{}^{\theta}\delta^{(3)}\Big]\partial_{\theta}\Big(\mathcal{N}_{NN}\,\tilde{h}(x)\Big).
		\end{split}
	\end{align}
	Similarly,
	\begin{align}
		\begin{split} \label{app04}
			& T_{\bar{M}N}=\delta^{(3)}D^{\Omega}_{\tilde K\tilde N}\Big[\mathcal{N}_{\bar M N}\tilde h(x)\Big]+D^{\hat{r}}_{\tilde K \tilde N}\Big[\mathcal{N}_{\tilde M N}\tilde h(x)\Big],\\&
			T_{\bar M\bar N}=\delta^{(3)} D^{\Omega}_{\tilde K\tilde K}\Big[\mathcal{N}_{\bar \bar M}\tilde h(x)\Big]+D^{\hat{r}}_{\tilde K \tilde K}\Big[\mathcal{N}_{\bar M \bar M}\tilde h(x)\Big],
		\end{split}
	\end{align}
	with, $ \mathcal{N}_{\bar M M}=\frac{\sqrt{\Delta}\, \rho}{\sqrt{-g}}\,\, \&\,\, \mathcal{N}_{\bar M\bar M}=\frac{\Sigma\, \rho^2}{\sqrt{-g}}.$ Here, $D^{\Omega}_{\tilde K\tilde N},D^{\hat{r}}_{\tilde K \tilde N},D^{\Omega}_{\tilde K\tilde K},D^{\hat r}_{\tilde K \tilde K} $ satisfy similar equations as \autoref{app02} and \autoref{app3} with the appropriate indices. Also we have used the following: $\partial_{\hat{i}}\partial_{\hat{j}}\mathcal{I}^{\hat{i}}{}_{\tilde N\tilde N}{}^{\hat{j}}=\partial_{\hat{j}}\partial_{\hat{i}}\mathcal{I}^{\hat{i}}{}_{\tilde N\tilde N}{}^{\hat{j}}.$ \par
	With $ T_{NN}$, $ T_{\bar{M}N}$ and $\mathcal{N}_{\bar M\bar M}$ in our hand, we can calculated the quantities in \autoref{app004} which turns out to be
		\begin{align}
		\begin{split} \label{app4}
			\mathcal{J}_{NN}
			=\delta^{(3)} D^{\Omega}_{\tilde N\tilde N}\Big[f^{(0)}_{N N}\Big]+D^{\hat{r}}_{\tilde N\tilde N}\Big[f^{(0)}_{N N}\Big],
		\end{split}
	\end{align}
\begin{align}
	\begin{split} \label{app5}
		\mathcal{J}_{\bar M N}&=\partial_{\hat r}\Big\{\delta^3 D^{\Omega}_{\tilde M N}\Big[f^{(1)}_{\bar M N}\Big]+D^{\hat{r}}_{\tilde M N}\Big[f^{(1)}_{\bar M N}\Big]\Big\}\\&+\delta^3 D^{\Omega}_{\tilde M N}\Big[f^{(0)}_{\bar M N}\Big]+D^{\hat{r}}_{\tilde M N}\Big[f^{(0)}_{\bar M N}\Big],
	\end{split}
\end{align}
	\begin{align}
	\begin{split} \label{app6}
		\mathcal{J}_{\bar M \bar M}=&
		=\partial_{\hat r}^2\Big\{\delta^3 D^{\Omega}_{\tilde K \tilde K}\Big[f^{(2)}_{\bar M \bar M}\Big]+D^{\hat r}_{\tilde K \tilde K}\Big[f^{(2)}_{\bar M \bar M}\Big]\Big\}\\&+\partial_{\hat r}\Big\{\delta^3 D^{\Omega}_{\tilde K \tilde K}\Big[f^{(1)}_{\bar M \bar M}\Big]+D^{\hat r}_{\tilde K \tilde K}\Big[f^{(1)}_{\bar M \bar M}\Big]\Big\}\\&+\delta^3 D^{\Omega}_{\tilde K \tilde K}\Big[f^{(0)}_{\bar M \bar M}\Big]+D^{\hat r}_{\tilde K \tilde K}\Big[f^{(0)}_{\bar M \bar M}\Big],
	\end{split}
\end{align}
where,
\begin{align}
	\begin{split}
		f^{(0)}_{N N}&=-\frac{2\,\bar\rho}{\Delta\, \rho}\Big[\Big(\mathcal{L}^{\dagger}_{1}-2\,i\,\hat{a}\,\rho\sin(\theta)\Big)\mathcal{L}^{\dagger}_{2}S^{\hat{a}\hat{\omega}}_{l m}\Big],\\
		 f^{(0)}_{\bar M N}&=\frac{4\, \bar\rho}{\sqrt{2}\rho\sqrt{\Delta}}\Big\{\Big(i\,\frac{K}{\Delta}+\rho+\bar \rho
		\Big)\mathcal{L}_2^{\dagger}\, S^{\hat{a}\hat{\omega}}_{l m}\\&-\hat{a}\,\sin(\theta)\frac{K}{\Delta}(\bar\rho-\rho)S^{\hat{a}\hat{\omega}}_{l m}\Big\},\\
		f^{(1)}_{\bar M N}&=\frac{4\, \bar\rho}{\sqrt{2}\rho\sqrt{\Delta}}\Big(\mathcal{L}_2^{\dagger}\,S^{a \omega}_{l m}+i\,\hat{a}\,\sin(\theta)(\bar\rho-\rho)S^{a \omega}_{l m}\Big\},\\
		f^{(0)}_{\bar M\bar M}&=\frac{\bar \rho}{\rho}\Big[\frac{d}{d \hat{r}}\Big(\frac{i\, K}{\Delta}\Big)-2\rho\frac{i\, K}{\Delta}+\frac{K^2}{\Delta^2}\Big]S^{\hat{a}\hat{\omega}}_{l m},\\ f^{(1)}_{\bar M \bar M}&=-\Big(\bar \rho+\frac{\bar \rho}{\rho}\frac{i\, K}{\Delta}\Big)S^{\hat{a}\hat{\omega}}_{l m},\quad{f}^{(2)}_{\bar M\bar M}=-\frac{\bar \rho}{\rho}S^{\hat{a}\hat{\omega}}_{l m}.
	\end{split}
\end{align}
	Replacing \autoref{app4}, \autoref{app5} and \autoref{app6} in \autoref{source_1} and doing the integration over $\theta$ and $\phi$ we get the expression for source term, 
	\begin{align}
		\begin{split} \label{app7}
			\mathcal{J}_{l m \hat \omega}&=\int d\hat t\, \Delta^2\, e^{i(\hat \omega \hat t-m\,\phi)}\Big[\delta(\hat r-\hat r(\hat t))\,J_{D}^{(0)}\\&+\partial_{\hat r}\Big(J_{D}^{(1)}\,\delta(\hat r-\hat r(\hat t))\Big)+\partial_{\hat r}^2\Big(J_{D}^{(2)}\,\delta(\hat r-\hat r(\hat t))\Big)\\&+J_{\hat r}^{(0)}+\partial_{\hat r}\Big(J_{\hat r}^{(0)}\Big)+\partial_{\hat r}^2\Big(J_{\hat r}^{(2)}\Big)\Big]\Big |_{\theta=\theta(\hat t),\phi=\phi(\hat t)},
		\end{split}
	\end{align}

where,
\begin{align}
	\begin{split} \label{app8}
		 J_{D}^{(0)}&=D^{\Omega}_{\tilde N \tilde N} \Big(f^{(0)}_{NN}\Big)+D^{\Omega}_{\tilde K \tilde N} \Big(f^{(0)}_{\bar M N}\Big) +D^{\Omega}_{\tilde K \tilde K} \Big(f^{(0)}_{\bar M 
			\bar M}\Big),\\
		J_{D}^{(1)}&=D^{\Omega}_{\tilde K \tilde N} \Big(f^{(1)}_{\bar M N}\Big)+D^{\Omega}_{\tilde K \tilde K} \Big(f^{(1)}_{\bar M \bar M}\Big),\, J_{\infty}^{(2)}=D^{\Omega}_{\tilde K\tilde K}\Big(f^{(2)}_{\bar M\bar M}\Big),\\
		J_{\hat r}^{(0)}&=D^{\hat r}_{\tilde N \tilde N} \Big(f^{(0)}_{N N}\Big)\delta(\hat r-\hat r(\hat t))+D^{\hat r}_{}\Big(f^{(0)}_{\bar M N}\delta(\hat r-\hat r(\hat t))\Big)\\&+D^{\hat r}_{}\Big(f^{(0)}_{\bar M \bar M}\Big)\delta(\hat r-\hat r(\hat t)),\\
		J_{\hat r}^{(1)}&=D^{\hat r}_{\bar M N}\Big(f^{(1)}_{\bar M N}\delta(\hat r-\hat r(\hat t))\Big)+D^{r}_{\bar M \bar M}\Big(f^{(1)}_{\bar M\bar M}\Big)\delta(\hat r-\hat r(\hat t)),\\J_{\hat r}^{(2)}&=D^r_{\bar M\bar M}\Big(f^{(2)}_{\bar M\bar M}\delta(\hat r-\hat r(\hat t))\Big)
	\end{split}
\end{align}

With the source term for Teukolsky equation in our hand, we can calculate the amplitude given in \autoref{amp_def}, which turns out to be
\begin{align}
	\begin{split}
		\mathcal{Z}^{H,\infty}_{l m \hat{\omega}}
		= C^{H,\infty}_{l m \hat{\omega}}\int_{\hat{r}_+}^{\infty} & d\hat{r}'\int_{-\infty}^{\infty} d\hat{t}\, e^{i(\hat \omega \hat t-m\,\phi)}\\&\Big[\delta(\hat r-\hat r(\hat t))\,J_{D}^{(0)}+\partial_{\hat r}\Big(J_{D}^{(1)}\,\delta(\hat r-\hat r(\hat t))\Big)\\&+\partial_{\hat r}^2\Big(J_{D}^{(2)}\,\delta(\hat r-\hat r(\hat t))\Big) +J_{\hat r}^{(0)}+\partial_{\hat r}\Big(J_{\hat r}^{(0)}\Big)\\&+\partial_{\hat r}^2\Big(J_{\hat r}^{(2)}\Big)\Big]R^{\textrm{in},\textrm{up}}_{l m \hat{\omega}}(\hat r')\,
	\end{split}
\end{align}
\begin{widetext} 
Note that the integrand has to be evaluated at $$\theta=\theta(\hat t),\, \phi=\phi(\hat t).$$

	Next we first do the integration over $\hat{r}$ and utilizing the delta function we get after integrating by parts and throwing away the surface terms, 
\begin{align}
	\begin{split}
		\mathcal{Z}^{H,\infty}_{l m \hat{\omega}}
		=C^{H,\infty}_{l m \hat{\omega}}\int_{-\infty}^{\infty} &d\hat{t}\, e^{i(\hat \omega \hat t-m\,\phi)}\Big[J_{D}^{(0)}-J_{D}^{(1)}\,\partial_{\hat r}+J_{D}^{(2)}\,\partial_{\hat r}^2\\&+J_{\hat r}^{(0)}+\partial_{\hat r}(J_{\hat r}^{(0)})+\partial_{\hat r}^2(J_{\hat r}^{(2)})\Big]R^{\textrm{in},\textrm{up}}_{l m \hat{\omega}}(\hat r)\,
	\end{split}
\end{align}
Here $\hat r$ is evaluated at $\hat r(\hat t).$ Then using \autoref{app8} and doing integrating by parts we arrive at the following expression,
	\begin{align}
		\begin{split} \label{app9}
			\mathcal{Z}^{H,\infty}_{l m \hat{\omega}}
			=&C^{H,\infty}_{l m \hat{\omega}}\int_{-\infty}^{\infty} d\hat{t}\, e^{i(\hat \omega \hat t-m\,\phi)}\\&\Big[\Big\{O_{NN}f^{(0)
			}_{NN}+O_{\bar M N}f^{(0)
			}_{\bar M N}+O_{\bar M\bar M}f^{(0)
			}_{\bar M \bar M }\Big\}R^{\textrm{in},\textrm{up}}_{l m \hat{\omega}}(\hat r)\\&-\Big\{\Big(O_{\bar M N}f^{(1)
			}_{\bar M N}+O_{\bar M \bar M }f^{(1)
			}_{\bar M \bar M }\Big)-\Big(J_{NN}f^{(0)
			}_{NN}+J_{\bar M N}f^{(0)
			}_{\bar M N}+ J_{\bar M\bar M}f^{(0)}_{\bar M \bar M }\Big)\Big\}\partial_{\hat r} R^{\textrm{in},\textrm{up}}_{l m \hat{\omega}}(\hat r)\\&+\Big\{O_{\bar M \bar M}f^{(2)}_{\bar M \bar M}-\Big(J_{\bar M N}f^{(1)}_{\bar M N}+J_{\bar M \bar M }f^{(1)
			}_{\bar M \bar M }\Big)+\Big(J_{NN}f^{(0)
			}_{NN}+J_{\bar M N}f^{(0)
			}_{\bar M N}+J_{\bar M\bar M}f^{(0)
			}_{\bar M \bar M }\Big)\Big\}\partial_{\hat r}^2 R^{\textrm{in},\textrm{up}}_{l m \hat{\omega}}(\hat r)\\&- \Big\{-J_{\bar M \bar M}f^{(2)}_{\bar M \bar M}+\Big(K_{\bar M N}f^{(1)}_{\bar M N}+K_{\bar M \bar M }f^{(1)
			}_{\bar M \bar M }\Big)\Big\}\partial_{\hat r}^3 R^{\textrm{in},\textrm{up}}_{l m \hat{\omega}}(\hat r)+\Big\{K_{\bar M \bar M}f^{(2)}_{\bar M \bar M}\Big\}\partial_{\hat r}^4 R^{\textrm{in},\textrm{up}}_{l m \hat{\omega}}(\hat r)\Big],
		\end{split}
	\end{align}
where,
\begin{align}
	\begin{split}\label{app10}
		&	 K_{NN}= \mathcal{I}^{\hat r}{}_{\tilde N\tilde N}{}^{\hat r},\\& J_{NN}=I_{NN}+\mathcal{I}^{\hat r}{}_{\tilde N\tilde N}{}^{\hat r}\partial_{\hat r},
	\end{split}
\end{align}
\begin{align}
	\begin{split} \label{app10a}
		I_{NN}=&-\Big(\mathcal{Q}^{\hat r}_{\tilde N\tilde N}-2\,i\,\hat\omega\, \mathcal{I}^{\hat t}{}_{\tilde N\tilde N}{}^{\hat r}+ 2\,i\, m \,\mathcal{I}^{\hat t}{}_{\tilde N\tilde N}{}^{\phi}\Big)+\mathcal{I}^{\hat r}{}_{\tilde N\tilde N}{}^{\hat r}\partial_{\hat r}+2\,\mathcal{I}^{\hat r}{}_{\tilde N\tilde N}{}^{\theta}\partial_{\theta}
	\end{split}
\end{align}
and
\begin{align}
	\begin{split} \label{app11}
		O_{NN}= &\mathcal{P}_{\tilde N \tilde N}-i\,\hat \omega\, \mathcal{Q}^{\hat t}{}_{\tilde N \tilde N}-\hat \omega^2\,\mathcal{I}^{\hat t}_{\tilde N\tilde N}{}^{\hat t}+i\,m\Big(\mathcal{Q}^{\phi}{}_{\tilde{N}\tilde{N}}-2\,i\,\hat{\omega}\,\mathcal{I}^{\hat{t}}{}_{\tilde N\tilde N}{}^{\phi}\Big)
		-m^2\,\mathcal{I}^{\phi}{}_{\tilde N\tilde N}{}^{\phi}\\&-\Big(\mathcal{Q}^{\theta}{}_{\tilde N\tilde N}-2\,i\,\hat{\omega}\,\mathcal{I}^{\hat{t}}{}_{\tilde N\tilde N}{}^{\phi}+2\,i\, m\,\mathcal{I}^{\theta}{}_{\tilde N\tilde N}{}^{\phi}\Big)-\mathcal{I}^{\theta}_{\tilde N\tilde N}{}^{\theta}\partial_{\theta}^2 +I_{NN}\partial_{\hat r}.
	\end{split}
\end{align}
\end{widetext}
Other terms in \autoref{app9} can be obtained by replacing the indices appropriately in \autoref{app10}, \autoref{app10a} and \autoref{app11}.
Finally \autoref{app9} can be written in the following compact form, 
\begin{widetext}
	\begin{equation}\label{app_amplitude}
		\mathcal{Z}^{H,\infty}_{l m \hat{\omega}}
		=C^{H,\infty}_{l m \hat{\omega}}\int_{-\infty}^{\infty} d\hat{t}\, e^{i(\hat \omega \hat t-m\,\phi)}\Big[A_0-(A_1+B_0)\frac{d}{d\hat r}+(A_2+B_1+C_0)\frac{d^2}{d\hat r^2}-(B_2+C_1)\frac{d^3}{d\hat r^3}+C_2\frac{d^4}{d\hat r^4}\Big]R^{\textrm{in},\textrm{up}}_{l m \hat{\omega}}(\hat r),
	\end{equation}
	where,
	\begin{align}
		\begin{split}\label{app12}
			& A_0=O_{N N}f^{(0)}_{NN}+O_{\bar M N}f^{(0)}_{\bar M N}+O_{\bar M \bar M}f^{(0)}_{\bar M \bar M},\\&
			A_1=O_{\bar M N}f^{(1)}_{\bar M N}+O_{\bar M \bar M}f^{(1)}_{\bar M \bar M}\\& A_2=O_{\bar M\bar M}f^{(2)}_{\bar M\bar M},\\&
			B_0=-\Big(J_{N N}f^{(0)}_{NN}+J_{\bar M N}f^{(0)}_{\bar M N}+J_{\bar M \bar M}f^{(0)}_{\bar M \bar M}\Big),\\&
			B_1=-\Big(J_{\bar M N}f^{(1)}_{\bar M N}+J_{\bar M \bar M}f^{(1)}_{\bar M \bar M},\\& B_2=-J_{\bar M\bar M}f^{(2)}_{\bar M\bar M},\\&
			C_0=K_{N N}f^{(0)}_{NN}+K_{\bar M N}f^{(0)}_{\bar M N}+K_{\bar M \bar M}f^{(0)}_{\bar M \bar M},\\&
			C_1=K_{\bar M N}f^{(1)}_{\bar M N}+K_{\bar M \bar M}f^{(1)}_{\bar M \bar M},\\& C_2=k_{\bar M\bar M}f^{(2)}_{\bar M\bar M}.
		\end{split}
	\end{align}
In this paper we have set $\theta=\frac{\pi}{2}.$ This further simplifies certain term. On the equatorial plane whenever one of the indices set to $\theta$, the corresponding components of the tensor will be zero, i.e 
\begin{align}\label{app13}
	\begin{split}
		&\mathcal{Q}^{\theta}{}_{\tilde N\tilde N}=\mathcal{Q}^{\theta}{}_{\tilde K \tilde N}=\mathcal{Q}^{\theta}{}_{\tilde K\tilde K}=0,\\&
		\mathcal{I}^{\alpha}{}_{\tilde N\tilde N}{}^{\theta}=\mathcal{I}^{\alpha}{}_{\tilde K\tilde N}{}^{\theta}=\mathcal{I}^{\alpha}{}_{\tilde K\tilde K}{}^{\theta}=0,\quad \alpha=\Big\{\hat t, \hat r, \theta,\phi\Big\}.
	\end{split}
\end{align}
Hence,
\begin{align}\label{app14}
	\begin{split}
		& K_{NN}=\mathcal{I}^{\hat r}{}_{\tilde N\tilde N}{}^{\hat r},\quad J_{NN}=I_{NN}+\mathcal{I}^{\hat r}{}_{\tilde N\tilde N}{}^{\hat r}\partial_{\hat r},\\&
		I_{NN}=-\Big(\mathcal{Q}^{\hat r}_{\tilde N\tilde N}-2\,i\,\hat\omega\, \mathcal{I}^{\hat t}{}_{\tilde N\tilde N}{}^{\hat r}+ 2\,i\, m \,\mathcal{I}^{\hat t}{}_{\tilde N\tilde N}{}^{\phi}\Big)+\mathcal{I}^{\hat r}{}_{\tilde N\tilde N}{}^{\hat r}\partial_{\hat r}
	\end{split}
\end{align}
and 
\begin{align}
	\begin{split}\label{app15}
		O_{NN}=&\, \mathcal{P}_{\tilde N \tilde N}-i\,\hat \omega\, \mathcal{Q}^{\hat t}{}_{\tilde N \tilde N}-\hat \omega^2\,\mathcal{I}^{\hat t}_{\tilde N\tilde N}{}^{\hat t}+i\,m\Big(\mathcal{Q}^{\phi}{}_{\tilde{N}\tilde{N}}-2\,i\,\hat{\omega}\,\mathcal{I}^{\hat{t}}{}_{\tilde N\tilde N}{}^{\phi}\Big)\\&
		-m^2\,\mathcal{I}^{\phi}{}_{\tilde N\tilde N}{}^{\phi}+I_{NN}\partial_{\hat r}.
	\end{split}
\end{align}
Then expressions in \autoref{app12} get simplified. 
\end{widetext}
		\section{Comparison with previous works}\label{App:Comparison}
		\begin{widetext}
	In this section, we provide a comparison between our results with the ones existing in the literature. Ref.~\cite{Pani} calculated the gravitational wave flux for an EMRI system with spinning (but not deformed) secondary. The authors provided the data for the flux for $q=3\times 10^{-5}$ in \cite{Pani_code}. Note that, the value of $\mathcal{F}^{(0)}$, $\mathcal{F}^{(1)}$ and $\mathcal{F}^{(2)}$ does not depend on the $q$ and $\chi$. In \autoref{tab:flux0_com} and \autoref{tab:flux1_com}, we present a comparison between  our result and \cite{Pani} for the values of $\mathcal{F}^{(0)}$ and $\mathcal{F}^{(1)}$ for certain values of primary spin $\ha$ and orbital radius $\hr$. 
 \par
	\begin{table*}[!htb]
	\def\arraystretch{1.3}      	
	\setlength{\tabcolsep}{1.5em}
\begin{adjustwidth}{}{}
\begin{equation*}
\begin{array}{lll|lll|lll}
\hline
\hline
\multicolumn{3}{c|}{\hat{a}=0}  & \multicolumn{3}{|c|}{\hat{a}=0.6}  &\multicolumn{3}{|c}{\hat{a}=0.99} 
\\
\hline
\hr &  \mathcal{F}^{(0)} & \mathcal{F}^{(0)} \textrm{in \cite{Pani}}& \hr &  \mathcal{F}^{(0)} & \mathcal{F}^{(0)} \textrm{in \cite{Pani}}&\hr &  \mathcal{F}^{(0)} & \mathcal{F}^{(0)} \textrm{in \cite{Pani}} \\
\hline
10. & 0.00006133  & 0.00006152 & 10. & 0.0000533997 & 0.00005354 & 10. & 0.0000493777 & 0.00004950 \\
9.  & 0.000105496 & 0.0001059   & 8.  & 0.000158493  & 0.0001593 & 8.  & 0.00014153   & 0.0001422  \\
8.  & 0.000194952 & 0.0001961   & 6.  & 0.000643599  & 0.0006505   & 6.  & 0.000534583  & 0.0005396  \\
7.  & 0.000396622 & 0.0004002    & 5.  & 0.00156892   & 0.001597   & 4.  & 0.00317882   & 0.003260   \\
6.1  & 0.000926716 & 0.0009403  & 4.  & 0.00474159   & 0.004905   & 2.  & 0.0414704    & 0.0430138  \\
\hline
\hline
\end{array}
\end{equation*}
\end{adjustwidth}
\caption{Comparison between our result and \cite{Pani} for the value of  $\mathcal{F}^{(0)}$ for different values of orbital radius and  $\hat{a}$.}
\label{tab:flux0_com}
\end{table*}
		\begin{table*}[!htb]
	\def\arraystretch{1.3}      	
	\setlength{\tabcolsep}{1.5em}
\begin{adjustwidth}{}{}
\begin{equation*}
\begin{array}{lll|lll|lll}
\hline
\hline
\multicolumn{3}{c|}{\hat{a}=0}  & \multicolumn{3}{|c|}{\hat{a}=0.6}  &\multicolumn{3}{|c}{\hat{a}=0.99} 
\\
\hline
\hr &  \mathcal{F}^{(1)} & \mathcal{F}^{(1)} \textrm{in \cite{Pani}}& \hr &  \mathcal{F}^{(1)} & \mathcal{F}^{(1)} \textrm{in \cite{Pani}}&\hr &  \mathcal{F}^{(1)} & \mathcal{F}^{(1)} \textrm{in \cite{Pani}} \\
\hline
10. &
  0.000013444 &
  0.0000135324 &
  10. &
  8.36333\times 10^{-6} &
  8.41351\times 10^{-6} &
  10. &
  5.8153\times 10^{-6}&
  5.84959\times 10^{-6} \\
9.  & 0.000027533  & 0.0000277788 & 8. & 0.0000340719 & 0.0000344465 & 8. & 0.0000215679 & 0.0000217974 \\
8.  & 0.0000620701 & 0.000062854  & 6. & 0.000208892  & 0.00021392   & 6. & 0.000110702  & 0.000113198  \\
7.  & 0.000159224  & 0.000162234  & 5. & 0.000665169  & 0.000691733  & 4. & 0.000937339  & 0.00099818   \\
6.1 & 0.000434991  & 0.000447657  & 4. & 0.00283884   & 0.00305542   & 2. & 0.0116319    & 0.0179791 \\
\hline
\hline
\bottomrule
\end{array}
\end{equation*}
\end{adjustwidth}
\caption{Comparison between our result and \cite{Pani} for the values of  $\mathcal{F}^{(1)}$ for different values of orbital radius and  $\hat{a}$.} 
\label{tab:flux1_com}
\end{table*}
	\begin{figure*}[b!]
		\minipage{0.46\textwidth}
		\includegraphics[width=\linewidth]{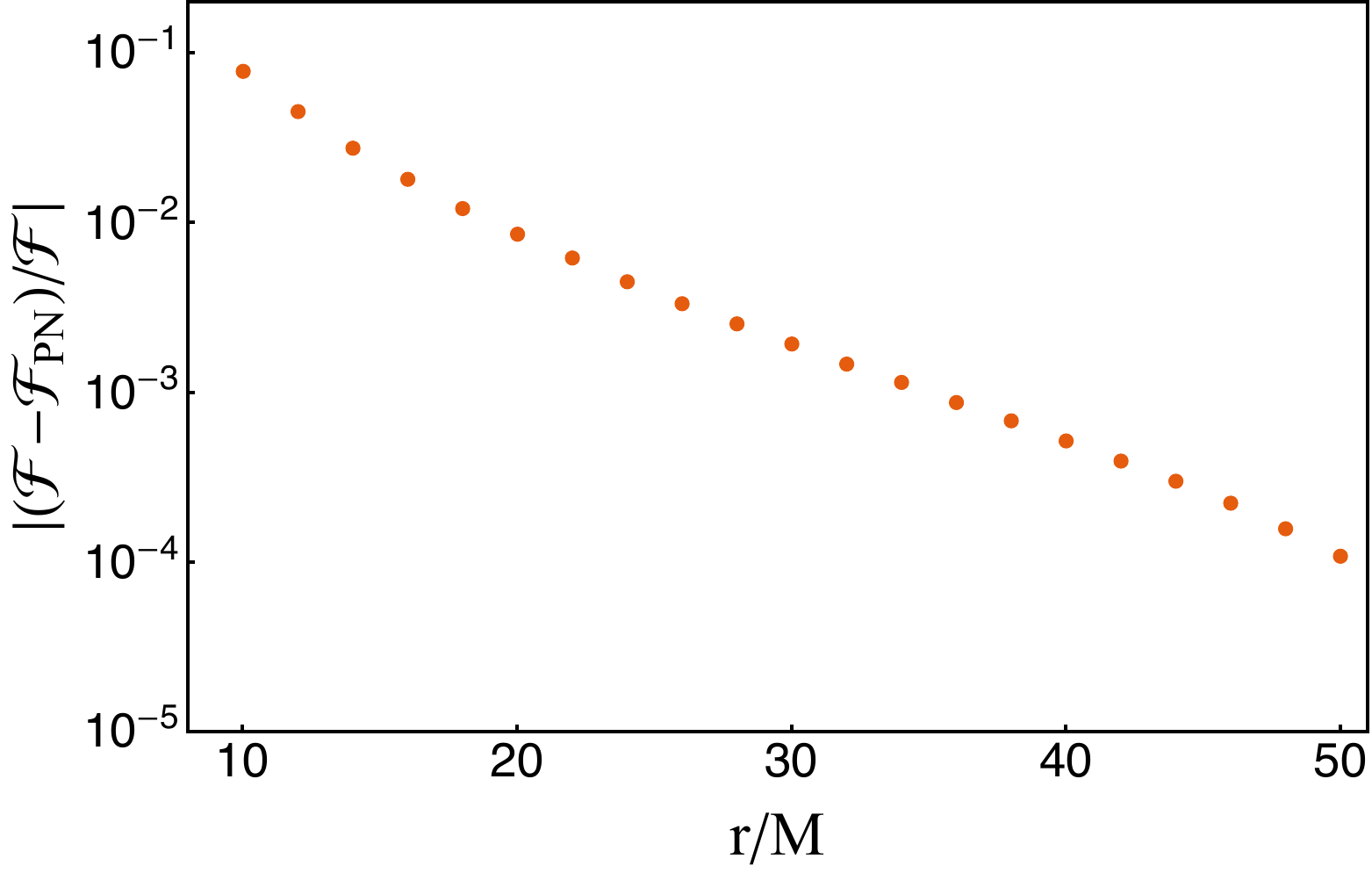}
		\endminipage
		\caption{Relative error between the numerical and PN results as a function orbital radius $\hr$. Here, we take $\ha=0.6$, $q=10^{-4}$, $\chi=1$ and $C_Q=10$.
		}\label{fig_error_PN}
	\end{figure*}
 Furthermore, we compare our results for gravitational wave flux with those obtained in post-Newtonian (PN) theory. Within the context of PN theory, gravitational wave flux can be written as 
 \begin{equation}
     \mathcal{F}_{\textrm{PN}}=\mathcal{F}_{\textrm{NS}}+\mathcal{F}_{\textrm{SO}}+\mathcal{F}_{\textrm{SS}}+...
 \end{equation}
where $\mathcal{F}_{\textrm{NS}}$ represents the non-spinning contribution to the flux, the expression of which is given in Eq.~(2.8) of Ref.~\cite{PhysRevD.98.124033}. The linear order correction term $\mathcal{F}_{\textrm{SO}}$ arises due to spin-orbital coupling (see Eq.~(2.6) in Ref.~\cite{PhysRevD.100.044007}) while the quadratic order correction term $\mathcal{F}_{\textrm{SS}}$ originates from spin-spin interactions (see Eq.~(2.7) in Ref.~\cite{PhysRevD.100.044007} or Eq.~(4.12) in Ref.~\cite{Bohe:2015ana}). In \autoref{fig_error_PN}, we plot the relative error between the numerical and PN results $\Delta_{\textrm{RE}}=|(\mathcal{F}-\mathcal{F}_{\textrm{PN}})/\mathcal{F}|$ at different values orbital radius. Here, we take $\ha=0.6$, $q=10^{-4}$, $\chi=1$ and $C_Q=10$.
As expected, our numerical results agree quite well with PN results when the orbital separation is large ($\Delta_{\textrm{RE}}\approx 10^{-4} $ at $\hr=50$). However, when orbital separation is small, there is a mismatch between numerical and PN result ($\Delta_{\textrm{RE}}\approx 0.01$ at $\hr=10$). 
 \par
\end{widetext}
	\bibliography{Reference_1}
\end{document}